\documentclass[eqsecnum, aps, amsmath, amssymb, singlecolumn, superscriptaddress, 11pt ]{revtex4-2}
\usepackage{graphicx}
\usepackage{dcolumn}
\usepackage[colorlinks=true,
            linkcolor=blue,
            urlcolor=blue,
            citecolor=blue]{hyperref}
\usepackage{bm}
\usepackage{enumerate}
\usepackage{lipsum}
\usepackage{threeparttable}
\usepackage{epstopdf}
\usepackage{float}
\usepackage{xcolor,colortbl}
\usepackage[caption = false]{subfig}
\usepackage{tabularx}
\usepackage{chngpage}

\begin{document}


\title{Coupled-cluster theory for  trapped bosonic mixtures}
\author{Anal Bhowmik}
\email{anal.bhowmik-1@ou.edu}
\affiliation{Homer L. Dodge Department of Physics and Astronomy,
The University of Oklahoma, Norman, Oklahoma 73019, USA}
\affiliation{Center for Quantum Research and Technology, The University of Oklahoma, Norman, Oklahoma 73019, USA}
\affiliation{Department of Physics, University of Haifa, Haifa 3498838, Israel}
\affiliation{Haifa Research Center for Theoretical Physics and Astrophysics, University of Haifa, Haifa 3498838, Israel}
\author{Ofir E. Alon}
\affiliation{Department of Physics, University of Haifa, Haifa 3498838, Israel}
 \affiliation{Haifa Research Center for Theoretical Physics and Astrophysics, University of Haifa, Haifa 3498838, Israel}

\date{\today}





\begin{abstract}
We develop a  coupled-cluster theory for  bosonic mixtures of binary species in external traps, providing a promising theoretical approach to demonstrate highly accurately the many-body physics of  mixtures of Bose-Einstein condensates. The coupled-cluster wavefunction for the binary species is obtained when an exponential cluster operator $e^T$, where $T=T^{(1)}+T^{(2)}+T^{(12)}$ and $T^{(1)}$ accounts for excitations in species-1, $T^{(2)}$ for excitations in species-2, and $T^{(12)}$ for combined excitations in both species, acts on the ground state configuration prepared by accumulating all bosons in a single orbital for each species.  We have explicitly derived the working equations for the bosonic mixtures by truncating the cluster operator upto the single and double excitations and using an arbitrary sets of orthonormal orbitals for each of the species. Further, the comparatively simplified version of the working equations are formulated using the Fock-like operators. Finally, using  an exactly solvable many-body  model for bosonic mixtures that exists in the literature allows us to implement and test the performance and accuracy of the coupled-cluster theory for situations with balanced as well as imbalanced boson numbers and for weak to moderately strong intra- and inter-species interaction strengths. The comparison between  our computed results using coupled-cluster theory with the respective analytical exact results displays remarkable agreement exhibiting excellent success of the coupled-cluster theory for bosonic mixtures. All in all, the correlation exhaustive  coupled-cluster theory shows  encouraging results  and it could be a promising approach in paving the way for high-accuracy modelling  of various bosonic mixture systems. 
\end{abstract}

\maketitle
\section{INTRODUCTION}

Mixtures of  bosonic species created from ultracold quantum gasses are highly investigated topic, providing  more degrees-of-freedom compared to single species, due to the  advanced  controllability of the system's parameters by the state of the art experiments. Excellent experimental control of the strengths of inter- and intra-species interactions and of the external confinement establishes  mixtures of bosonic species as a rich model  to investigate many-body quantum physics. A great variety of physical phenomena involving  mixtures of bosonic species have attracted much attention, such as, the phase separation \cite{Timmermans1998}, condensate depletion \cite{Sakhel2008}, fermionization \cite{Zollnerl2008}, quantum phase transition in waveguides \cite{Girardeau2009}, persistent currents \cite{Anoshkin2013}, quantum mechanical stabilization \cite{Petrov2015}, entanglement induced interactions \cite{Chen2018},   the miscible to immiscible phase transition \cite{Ticknor2013, Nicklas2015}, spin-charge separation  \cite{Kleine2008}, emergence of spin-roton \cite{Lee2022}, ferrodark solitons \cite{Yu2022}, and quantum droplets \cite{Smith2021}.

The ground-state properties of trapped bosonic mixtures have been  extensively investigated  both theoretically and experimentally \cite{Hall1998,StamperKurn1998,Chui2001,Thalhammer2008,Gautam2010,McCarron2011,
McCarron2015,Alon2017,Bhowmik2018,Trautmann2018,Alon2020,Bai2020,Bisset2021}.
In terms of theoretical description, although many observations  of bosonic mixtures so far have been explained by applying the standard mean-field theory, namely,  Gross–Pitaevskii  theory,  the many-body  modelling is an indispensable tool to capture  fundamental understanding and outline new schemes of variety of quantum phenomena for applications. Presently, the  popularly available  many-body theories are the quantum Monte Carlo method \cite{Cikojevic2018, Dzelalija2020}, the Bose-Hubbard model \cite{Lewenstein2012,Morera2020}, self-consistent many-body theory \cite{Leveque2018} for mixtures, and the most successful in accounting for quantum correlations, the multilayer multiconfigurational time-dependent Hartree method \cite{Alon2007,Cao2013,Cao2017,Manthe2017}.

Since  the coupled-cluster theory was reformulated and introduced to electron-structure theory in \cite{Cizek1966,Paldus1975},    it has  become one of the most successful  methods of choice in quantum chemistry    when  accuracy is concerned \cite{Lindgren1985, Bartlett2007}.  For many-fermion systems,     the coupled-cluster theory has already been proved  to be  a very reliable and accurate approach \cite{Bishop2003,CarskyJosef2010,White2018}, also in the relativistic regime of interest \cite{Nataraj2008,Bhowmik2017, Bhowmik2020,Chakraborty2022}. In Ref \cite{Cederbaum2006}, a coupled-cluster theory for trapped interacting indistinguishable  bosons was derived and a few numerical applications were shown upto $10^4$ particles with sufficiently strong inter-boson interaction strength. Their investigation suggests that the coupled-cluster theory would be a practical approach to apply beyond single-component  bosonic systems  and thereby study further the bosonic mixtures of different species. Here, we would like to mention some of the recent advancements of  coupled-cluster theory, such as, for electron-phonon systems \cite{White2020}, polarons \cite{Mordovina2020,Haugland2020}, and  the investigation of multiexcitonic interactions \cite{Ellis2016}

In this work we develop a theoretical framework of the coupled-cluster theory for  systems  of trapped binary bosonic mixtures.  The general theory includes all kinds of excitations  and we call it coupled-cluster theory for mixtures, or, briefly, CC-M. To begin with, we include  the single and double excitations in the mixture of two species, and thereby, according to the order of excitations included the theory becomes the coupled-cluster singles doubles theory for bosonic mixtures, CCSD-M.  We derive the working equations of the coupled-cluster theory with an  arbitrary sets of orthonormal orbitals and also  using Fock-like operators.  We implement the theory and illustrate the potential usage by comparing it  to an exactly solvable many-body model. This enables us to get precise results on how accurate we are in scenarios that occur for  binary mixtures  with  balanced and imbalanced particle numbers  and having  weak to strong intra- and inter-species interaction strengths.

 The structure of the paper is as follows: In section~\ref{Section II}, we present the CC-M ansatz and  provide the basic formulation of the cluster operators for the excitations for the mixture.  Section~\ref{Section III} presents the details of Fock-like operators for the bosonic mixture of binary species. In Section~\ref{Section IV}, the coupled-cluster theory for the bosonic mixtures is developed and the detailed derivation is shown. In Section~\ref{Section V}, we provide the potential of our coupled-cluster theory by comparing the results for  scenarios  when the mixtures have balanced and imbalanced particle numbers and for  different strengths of intra- and inter-species interaction  to the analytical results from an exactly solvable model. Finally, in Section~\ref{Section VI}, we summarize and conclude our work. The appendices provide additional details on the derivation of the theory and its implementation.

\section{Theoretical framework for coupled-cluster theory for bosonic mixtures}\label{Section II}
In order to develop  the coupled-cluster theory for binary  mixture  of bosons, we consider the number of bosons in species-1 and species-2 to be $N_1$ and $N_2$, respectively,  with the total number of bosons $N=N_1+N_2$.  For simplicity, we consider the bosons to be spinless.  We introduce the number of one-particle functions or orbitals  for species-1 as $M_1$  and for species-2 as  $M_2$.  We start from the  Schr\"odinger equation,   $H|\Psi\rangle=E_0|\Psi\rangle$, where $E_0$ is the exact energy and $\Psi$ is  the coupled-cluster wavefunction, to be defined below. The Hamiltonian $H$ for the mixture of two species can be written as
\begin{eqnarray}\label{eqn_2.1}
&&{H}={H}^{(1)}+{H}^{(2)}+{H}^{(12)},\nonumber \\
&&{H}^{(1)}=\sum_{p,q}h_{\bar{p}\bar{q}}^{(1)}{a}_{p}^\dagger {a}_{q}+\dfrac{1}{2}\sum_{p,q,r,s}V_{\bar{p}\bar{q}\bar{r}\bar{s}}^{(1)} {a}_{p}^\dagger {a}_{q}^\dagger {a}_{r} {a}_{s},\nonumber \\
&&{H}^{(2)}=\sum_{i,j}h_{ij}^{(2)}{b}_{i}^\dagger {b}_{j}+\dfrac{1}{2}\sum_{i,j,k,l}V_{ijkl}^{(2)} {b}_{i}^\dagger {b}_{j}^\dagger {b}_{k} {b}_{l},\nonumber \\
&& {H}^{(12)}=\sum_{p,q,i,j}V_{\bar{p}i\bar{q}j}^{(12)}{a}_{p}^\dagger {b}_i^\dagger {a}_{q}{b}_j. 
\end{eqnarray}
Here,  ${a}_p^\dagger$ and ${a}_p$, where $p=1,2,...,M_1$, are the bosonic creation and destruction operators, respectively, corresponding to the orbitals  $\varphi_p(\textbf{r}_1)$ of species-1  and similarly for species-2, the corresponding operators are ${b}_i^\dagger$ and ${b}_i$, where $i=1,2,...,M_2$, for the orbitals $\psi_i(\textbf{r}_2)$.  {The  symbols $\{pqrs\}$ and $\{ijkl\}$ are used for the species-1 and species-2, respectively,  until specified differently.}  Furthermore, to ease the reading of equations, whenever we have indices in matrix elements for species-1, we  write a bar on top. For instance in ${H}^{(12)}$, the matrix element is $V_{\bar{p}i\bar{q}j}^{(12)}$ where the  variables are $p$ and $q$ for species-1. The creation and destruction operators follow the usual bosonic commutator relations which read

 \begin{eqnarray}\label{eqn_2.2}
 &&[{a}_p,{a}_q^\dagger ]=\delta_{pq}, [{a}_p,{a}_q ]=0, [{a}_p^\dagger,{a}_q^\dagger ]=0,\nonumber\\
  &&[{b}_i,{b}_j^\dagger ]=\delta_{ij}, [{b}_i,{b}_j ]=0, [{b}_i^\dagger,{b}_j^\dagger ]=0,\nonumber\\
   &&[{a}_p,{b}_i^\dagger ]=0, [{a}_p,{b}_i ]=0, [{a}_p^\dagger,{b}_i^\dagger ]=0.
 \end{eqnarray}
In the Hamiltonian Eq.~\ref{eqn_2.1}, the one-particle operators for species-1 and species-2 are
\begin{eqnarray}\label{eqn_2.3}
&&h_{\bar{p}\bar{q}}^{(1)}=\int \varphi_p^*\hat{h}^{(1)}\varphi_qd\textbf{r}_1,\nonumber \\
&&h_{ij}^{(2)}=\int \psi_i^*\hat{h}^{(2)}\psi_jd\textbf{r}_2,
\end{eqnarray}
respectively, and the two-particle interactions in the Hamiltonian   Eq.~\ref{eqn_2.1} are given by
\begin{eqnarray}\label{eqn_2.4}
&&V_{\bar{p}\bar{q}\bar{r}\bar{s}}^{(1)}=\int \int \varphi_p^* \varphi_q^* {V}^{(1)}(\textbf{r}_1-\textbf{r}_1^\prime)\varphi_r \varphi_s d\textbf{r}_1 d\textbf{r}_1^\prime ,\nonumber\\
&&V_{ijkl}^{(2)}=\int \int \psi_i^*\psi_j^*{V}^{(2)}(\textbf{r}_2-\textbf{r}_2^\prime)\psi_k\psi_l d\textbf{r}_2d\textbf{r}_2^\prime,\nonumber \\
&&V_{\bar{p}i\bar{q}j}^{(12)}=\int \int \varphi_p^*\psi_i^*{V}^{(12)}(\textbf{r}_1-\textbf{r}_2)\varphi_q\psi_jd\textbf{r}_1d\textbf{r}_2.
\end{eqnarray}
Here the first and second terms stand for the two-particle interactions for species-1 and species-2, respectively. The last term indicates the two-particle interaction between the bosons of species-1 and species-2. 

{In order to solve the Schr\"{o}dinger equation $H|\Psi\rangle=E_0|\Psi\rangle$ with the Hamiltonian  presented  in Eq.~\ref{eqn_2.1},} using the coupled-cluster theory, we start with the definition of the exact wavefunction of  bosonic mixture which is obtained by employing an exponential operator onto the ground configuration

\begin{equation}\label{eqn_2.5}
|\Psi\rangle= e^T|\phi_0\rangle.
\end{equation}
The ground configuration  {takes the form of}  product state,  $\phi_0=\prod_{n=1}^{N_1}\varphi_1(r_{1n})\prod_{m=1}^{N_2}\psi_1(r_{2m})$,  which defines the standard mean-field for the bosonic mixture, also see below. We note that other mean-fields are possible for bosons \cite{Cederbaum2003,Alon2006}, but we will not pursue such a choice here.

For the bosonic mixture, the cluster operator $T$ is  conveniently divided into three parts,

\begin{equation}\label{eqn_2.6}
T= T^{(1)}+T^{(2)}+T^{(12)}.
\end{equation}
Here, the cluster operators, $T^{(1)}$, $T^{(2)}$, and $T^{(12)}$ correspond the superposition of excitation operators. Moreover, $T^{(1)}$  are identical to the single-component cluster operator and operates only on the Hilbert space of species-1. The same holds true for  $T^{(2)}$, but it operates only  on the Hilbert space of species-2. In contrast, $T^{(12)}$ yields the simultaneous excitations in the Hilbert spaces of both species-1 and species-2. In terms of excitations, the cluster operators can be written as 
\begin{equation}\label{eqn_2.7}
T^{(1)}=\sum_{n=1}^{N_1}T_{n}^{(1)}=\sum_{n=1}^{N_1}t_n^{(1)}(a_{1})^n=\sum_{n=1}^{N_1}\sum_{p_1,...,p_n=2}^{M_1} c_{p_1 p_2...p_n} a_{p_1}^\dagger a_{p_2}^\dagger ...a_{p_n}^\dagger (a_{1})^n,
\end{equation}

\begin{equation}\label{eqn_2.8}
T^{(2)}=\sum_{m=1}^{N_2}T_{m}^{(2)}=\sum_{m=1}^{N_2}t_m^{(2)}(b_1)^m=\sum_{m=1}^{N_2}\sum_{i_1,...,i_m=2}^{M_2} d_{i_1 i_2...i_m} b_{i_1}^\dagger b_{i_2}^\dagger ...b_{i_m}^\dagger(b_1)^m.
\end{equation}
Note that for each species the first orbital is occupied and the second orbital and onward are the virtual orbitals.  In Eq.~\ref{eqn_2.7}, $t_n^{(1)}=\sum_{p_1,...,p_n=2}^{M_1} c_{p_1 p_2...p_n} a_{p_1}^\dagger a_{p_2}^\dagger ...a_{p_n}^\dagger$ creates the excitations in the virtual orbitals of species-1. Similarly, in Eq.~\ref{eqn_2.8},  $t_m^{(2)}=\sum_{i_1,...,i_m=2}^{M_2} d_{i_1 i_2...i_m} b_{i_1}^\dagger b_{i_2}^\dagger ...b_{i_m}^\dagger$ generates the excitations in the virtual orbitals of species-2. $T^{(12)}$ has the form

\begin{eqnarray}\label{eqn_2.9}
T^{(12)}&=&\sum_{n^{\prime}=1}^{N_1} \sum_{m^{\prime}=1}^{N_2} T_{n^{\prime}m^{\prime}}^{(12)}=\sum_{n^{\prime}=1}^{N_1} \sum_{m^{\prime}=1}^{N_2}t_{n^\prime m^\prime}^{(12)}(a_{{1}})^{n^\prime}(b_1)^{m^\prime}\nonumber \\&=&\sum_{n^{\prime}=1}^{N_1} \sum_{m^{\prime}=1}^{N_2}\sum_{p_1,...,p_{n^\prime}=2}^{M_1}\sum_{i_1,...,i_{m^\prime}=2}^{M_2}e_{p_1...p_{n^\prime}i_1...i_{m^\prime}}a_{p_1}^\dagger ...a_{p_{n^\prime}}^\dagger b_{i_1}^\dagger b_{i_2}^\dagger ...b_{i_{m^\prime}}^\dagger(a_{{1}})^{n^\prime}(b_1)^{m^\prime}.
\end{eqnarray}
In the expression of  $T^{(12)}$, one can find that  $t_{n^\prime m^\prime}^{(12)}=\sum_{p_1,...,p_{n^\prime}=2}^{M_1}\sum_{i_1,...,i_{m^\prime}=2}^{M_2}e_{p_1...p_{n^\prime}i_1...i_{m^\prime}}a_{p_1}^\dagger ...a_{p_{n^\prime}}^\dagger\\ b_{i_1}^\dagger b_{i_2}^\dagger ...b_{i_{m^\prime}}^\dagger$ is responsible for the simultaneous excitations in both species.  In Eqs.~\ref{eqn_2.7} and ~\ref{eqn_2.8}, the expansions of $T^{(1)}=T_1^{(1)}+T_2^{(1)}+T_3^{(1)}+...$ and  $T^{(2)}=T_1^{(2)}+T_2^{(2)}+T_3^{(2)}+...$ include the singles, doubles, triples, ... and so on excitations. While for $T^{(12)}= T_{11}^{(12)}+T_{12}^{(12)}+T_{21}^{(12)}+...$, see Eq.~\ref{eqn_2.9}, the excitations start from doubles, each for one species. { We shall find out the  unknown coefficients  $c_{p_1 p_2...p_n}$, $d_{i_1 i_2...i_m}$, and $e_{p_1...p_{n^\prime}i_1...i_{m^\prime}}$ when we will explicitly develop below the coupled-cluster theory for bosonic mixtures.}  Since the summations in Eqs.~\ref{eqn_2.7} to ~\ref{eqn_2.9} are unrestricted, the coefficients do not depend on the ordering of the subscripts.  {Also, it is convenient to  find the commutation relations among the cluster operators,} $[T_{n}^{(1)},T_{n^\prime}^{(1)}]=[T_{m}^{(2)},T_{m^\prime}^{(2)}]=[T_{n}^{(1)},T_{m}^{(2)}]=[T_{n}^{(1)},T_{n^{\prime}m^{\prime}}^{(12)}]=[T_{m^\prime}^{(2)},T_{n^{\prime}m^{\prime}}^{(12)}]=[T_{nm}^{(12)},T_{n^{\prime}m^{\prime}}^{(12)}]=0$. The implication of the commutation relations among the cluster operators is that the excitation operator is breakable to a product of all the partial excitation operators, i.e.,  $e^T|\phi_0\rangle=e^{T_1}e^{T_2}e^{T_{12}}|\phi_0\rangle$. Note that, as $T$ is the excitation operator, the wavefunction in Eq.~\ref{eqn_2.5} satisfies the intermediate normalization $\langle \phi_0|\Psi\rangle=1$.

\section{Fock-like operators for a bosonic mixture of binary species}\label{Section III}
For the study of the coupled-cluster theory for bosonic mixtures, we will first derive the working equations when the orbitals are arbitrary or unspecified to demonstrate a general perspective of the theory. However, we will also utilize the  orbitals arising from the Fock-like operators to give an illustrative example.   To derive the Fock-like operators, we start from the mean-field  energy functional for a mixture which reads 

\begin{eqnarray}\label{eqn_3.1}
E_{\text{MF}}&=&N_1\Bigg[\int d\textbf{r}_1\varphi_1^*(\textbf{r}_1)\Big({h}^{(1)}+\dfrac{\Lambda_1}{2}{J}_{11}+\dfrac{\Lambda_{21}}{2}{{J}_{21}}\Big)\varphi_1(\textbf{r}_1)\Bigg]\nonumber\\
&+&N_2\Bigg[\int d\textbf{r}_2\psi_1^*(\textbf{r}_2)\Big({h}^{(2)}+\dfrac{\Lambda_2}{2}{J}_{22}+\dfrac{\Lambda_{12}}{2}{{J}_{12}}\Big)\psi_1(\textbf{r}_2)\Bigg],
\end{eqnarray}
where the  interaction parameters are given by $\Lambda_1=\lambda_1(N_1-1)$, $\Lambda_2=\lambda_2(N_2-1)$, $\Lambda_{21}=\lambda_{12}N_2$, and $\Lambda_{12}=\lambda_{12}N_1$, and satisfy $N_1\Lambda_{21}=N_2\Lambda_{12}$. Here $\lambda_1$ and $\lambda_2$ are the  intra-species interaction strengths of species-1 and species-2, respectively, and $\lambda_{12}$ is the  inter-species interaction strength between species-1 and species-2. Also, the direct interaction operators ${J}_{11}$,  ${J}_{22}$, ${J}_{12}$, and ${J}_{21}$ are local operators and are defined by

\begin{eqnarray}\label{eqn_3.2}
&&{J}_{11}=\int \varphi_1^*(\textbf{r}_1^{\prime}){V}^{(1)}(\textbf{r}_1-\textbf{r}_1^\prime)\varphi_1(\textbf{r}_1^\prime)d\textbf{r}_1^\prime,\nonumber\\
&&{J}_{22}=\int \psi_1^*(\textbf{r}_2^\prime){V}^{(2)}(\textbf{r}_2-\textbf{r}_2^\prime)\psi_1(\textbf{r}_2^\prime)d\textbf{r}_2^\prime,\nonumber\\
&&{J}_{21}=\int \psi_1^*(\textbf{r}_2){V}^{(12)}(\textbf{r}_1-\textbf{r}_2)\psi_1(\textbf{r}_2)d\textbf{r}_2,\nonumber\\
&&{J}_{12}=\int \varphi_1^*(\textbf{r}_1){V}^{(12)}(\textbf{r}_1-\textbf{r}_2)\varphi_1(\textbf{r}_1)d\textbf{r}_1.
\end{eqnarray}
Minimizing the mean-field energy functional  with respect to the  orbitals $\varphi_1(\textbf{r}_1)$ and $\psi_1(\textbf{r}_2)$, the two-coupled mean-field equations of the mixture are derived,

\begin{eqnarray}\label{eqn_3.3}
&&[{h}^{(1)}+\Lambda_1{J}_{11}+\Lambda_{21}{J}_{21}]\varphi_1(\textbf{r}_1)=\mu_{1}^{(1)}\varphi_1(\textbf{r}_1),\nonumber \\\
&&[{h}^{(2)}+\Lambda_2{J}_{22}+\Lambda_{12}{J}_{12}]\psi_1(\textbf{r}_2)=\mu_{1}^{(2)}\psi_1(\textbf{r}_2)
\end{eqnarray}
Here $\mu_{1}^{(1)}$ and $\mu_{1}^{(2)}$ are the chemical potentials of the ground state of species-1 and species-2, respectively. Eq.~\ref{eqn_3.3}  presents the  Hermitian Fock-like operators
\begin{eqnarray}\label{eqn_3.4}
&&{F}_1={h}^{(1)}+\Lambda_1{J}_{11}+\Lambda_{21}{J}_{21},\nonumber \\
&&{F}_2={h}^{(2)}+\Lambda_2{J}_{22}+\Lambda_{12}{J}_{12}.
\end{eqnarray}
Eq.~\ref{eqn_3.3} {determines the ground state of the bosonic mixtures} and the ground state orbitals define the Fock-like operators. Now,  the eigenfunctions of the Fock-like operators gives us the complete orthonormal basis sets
\begin{eqnarray}\label{eqn_3.5}
&&{F}_1\varphi_i= \mu_{i}^{(1)}\varphi_i,\nonumber\\
&&{F}_2\psi_i= \mu_{i}^{(2)}\psi_i.
\end{eqnarray}
From Eq.~\ref{eqn_3.5}, the eigenvalues are computed as $\langle\varphi_i|{F}_1|\varphi_i\rangle=\mu_{i}^{(1)}$ and $\langle\psi_i|{F}_2|\psi_i\rangle=\mu_{i}^{(2)}$.  Using the Fock-like  operators,  one can simplify the general working equations  { by utilizing the relations between the one-body  and two-body matrix elements,} see details in  Appendix A.

\section{DERIVATION OF THE WORKING EQUATIONS}\label{Section IV}

{Starting from the Schr\"odinger equation $H|\Psi\rangle=E_0|\Psi\rangle$, where $H$ is given in Eq.~\ref{eqn_2.1}, and plugging the coupled-cluster ansatz Eq.~\ref{eqn_2.5} on both the left and right hand sides, the exact energy of the  ground state can be expressed as an expectation value,}

\begin{equation}\label{eqn_4.11}
E_0=\langle\phi_0|e^{-T}He^T|\phi_0\rangle.
\end{equation} 
{The transformed Hamiltonian entering the expression for the energy Eq.~\ref{eqn_4.11} can be expanded using the relation} $\dot{A}=e^{-T}Ae^{T}=A+\dfrac{1}{1!}[A,T]+\dfrac{1}{2!}[[A,T],T]+ ...$. Now we break the transformed Hamiltonian in accordance with the number of operators depending on  the occupied orbitals $\varphi_1$ and $\psi_1$. Here,  the transformed one-particle operator is found to be
\begin{eqnarray}\label{eqn_4.12}
\dot{H_0}&=&h_{\bar{1}\bar{1}}^{(1)} \dot{a}_{1}^\dagger \dot{a}_{1}+\sum_{s=2}^{M_1}h_{\bar{1}\bar{s}}^{(1)}\dot{a}_{1}^\dagger \dot{a}_{s}+\sum_{s=2}^{M_1}h_{\bar{s}\bar{1}}^{(1)} \dot{a}_{s}^\dagger \dot{a}_{1} +\sum_{r,s=2}^{M_1}h_{\bar{r}\bar{s}}^{(1)} \dot{a}_{r}^\dagger \dot{a}_{s}\nonumber\\
&+& h_{11}^{(2)} \dot{b}_1^\dagger \dot{b}_1+\sum_{l=2}^{M_2}h_{1l}^{(2)}\dot{b}_1^\dagger \dot{b}_l+\sum_{l=2}^{M_2}h_{l1}^{(2)} \dot{b}_l^\dagger \dot{b}_1 +\sum_{k,l=2}^{M_2}h_{kl}^{(2)} \dot{b}_k^\dagger \dot{b}_l,
\end{eqnarray}
 which includes a total of eight terms for two species, and among them the second and the sixth terms are the most intricate {in evaluating} the exact energy.  The transformed two-body operator of the Hamiltonian is divided into three parts,  $\dot{V}^{(1)}$, $\dot{V}^{(2)}$, and $\dot{V}^{(12)}$.  The first, $\dot{V}^{(1)}$, and the second, $\dot{V}^{(2)}$, parts correspond to excitations in species-1 and species-2, respectively, and  consist of nine terms each. The third  part $\dot{V}^{(12)}$ is the most involved one as it contains the  excitations in both species and it has sixteen terms.  All terms are explicitly shown in  Appendix C.  {As shown in Eqs.~\ref{eqn_4.12}, \ref{Potential_1}, \ref{Potential_2}, and \ref{Potential_12},  the transformed Hamiltonian contains transformed creation and destruction operators corresponding to the occupied and unoccupied orbitals. Now, in the following subsection, we present the expansions of the transformed  creation and destruction operators. }
 
\subsection{General relations}
In this section, we derive and discuss the working equations of the coupled-cluster theory for the bosonic mixture. At first, we transform the bosonic destruction and creation operators for both the species using the expansion $\dot{A}\equiv e^{-T}Ae^{T} $. The  {similarity transformations of the} destruction and creation operators  are needed eventually to construct the transformed Hamiltonian under investigation. One can readily find, for both  species, that the destruction operator corresponding to the occupied orbitals, $\varphi_1$ and $\psi_1$, and the creation operators of the virtual orbitals, $\varphi_p$ and $\psi_i$ where $p\geq 2$ and $i\geq 2$, are invariant to the coupled-cluster transformation:
\begin{eqnarray}\label{eqn_4.1}
&&\dot{a}_{{1}}=a_{{1}},\nonumber\\
&&\dot{a}_{{p}}^\dagger=a_{{p}}^\dagger, \hspace{0.2cm} p=2, 3,..., M_1,\nonumber\\
&&\dot{b}_1=b_1,\nonumber\\
&&\dot{b}_i^\dagger=b_i^\dagger, \hspace{0.2cm} i=2, 3,..., M_2.
\end{eqnarray}
In contrast, the creation operators corresponding to the orbitals  occupied in $\phi_0$ for each species, and the destruction operators of the virtual orbitals are modified due to the  {similarity transformation}  for both species and one can readily find the relations
\begin{eqnarray}\label{eqn_4.2}
&&\dot{a}_{{1}}^\dagger=a_{{1}}^\dagger-\mathcal{K}_{{1}},\nonumber\\
&&\dot{a}_{p}=a_{p}+\mathcal{K}_{p},\nonumber\\
&&\dot{b}_1^\dagger=b_1^\dagger-\mathcal{L}_1,\nonumber\\
&&\dot{b}_i=b_i+\mathcal{L}_i,
\end{eqnarray}
where $\mathcal{K}_{{1}}$ and $\mathcal{K}_{{p}}$ can be expressed as
\begin{equation}\label{eqn_4.3}
\mathcal{K}_{{1}}=\sum_{n=1}^{N_1} nt_{n}^{(1)}{(a_{{1}})}^{n-1}+\sum_{n=1}^{N_1}\sum_{m=1}^{N_2}nt_{nm}^{(12)}{(a_{{1}})}^{n-1}b_1^m,
\end{equation}

\begin{equation}\label{eqn_4.4}
\mathcal{K}_{{p}}=\sum_{n=1}^{N_1} nt_{n}^{(1)(p)}{(a_{{1}})}^{n}+\sum_{n=1}^{N_1}\sum_{m=1}^{N_2}n\eta_{nm}^{(12)(p)}{a_{{1}}}^nb_1^m.
\end{equation}
Here, the operators $t_{n}^{(1)}$ and $t_{nm}^{(12)}$ are defined before, see Eqs.~\ref{eqn_2.7} and ~\ref{eqn_2.9}.  The operators $t_{n}^{(1)(p)}$  operate  in the virtual space of species-1 and create excitations, while  $\eta_{nm}^{(12)(p)}$ are also responsible for excitations but  acting in the virtual space of both species.  $t_{n}^{(1)(p)}$  and $\eta_{nm}^{(12)(p)}$ read

\begin{equation}\label{eqn_4.5}
t_{n}^{(1)(p)}=\sum_{p_2, p_3,...,p_n=2}^{M_1} c_{pp_2...p_n}  a_{p_2}^\dagger a_{p_3}^\dagger ...a_{p_n}^\dagger,
\end{equation}

\begin{equation}\label{eqn_4.6}
\eta_{nm}^{(12)(p)}=\sum_{p_2, p_3,...,p_n=2}^{M_1} \sum_{i_1,i_2,...,i_{m}=2}^{M_2}e_{pp_2...p_{n}i_1i_2...i_{m}}  a_{p_2}^\dagger a_{p_3}^\dagger ...a_{p_n}^\dagger b_{i_1}^\dagger b_{i_2}^\dagger ...b_{i_m}^\dagger.
\end{equation}
Similarly, for species-2, $\mathcal{L}_1$ and $\mathcal{L}_i$ are expressed as

\begin{equation}\label{eqn_4.7}
\mathcal{L}_1=\sum_{m=1}^{N_2} mt_{m}^{(2)}{(b_1)}^{m-1}+\sum_{n=1}^{N_1}\sum_{m=1}^{N_2}mt_{nm}^{(12)}a_1^n b_1^{m-1},
\end{equation}

\begin{equation}\label{eqn_4.8}
\mathcal{L}_i=\sum_{m=1}^{N_2} mt_{m}^{(2)(i)}{(b_1)}^{m}+\sum_{n=1}^{N_1}\sum_{m=1}^{N_2}m \tau_{nm}^{(12)(i)}a_1^n b_1^m.
\end{equation}
The operators $t_{m}^{(2)(i)}$ generate excitations and operate  in the virtual space of species-2 but  the operators $\tau_{nm}^{(12)(i)}$  excite bosons in both species simultaneously.   $t_{m}^{(2)(i)}$  and $\tau_{nm}^{(12)(i)}$ can be represented as

\begin{equation}\label{eqn_4.9}
t_{m}^{(2)(i)}=\sum_{i_2, i_3,...,i_m=2}^{M_2} d_{ii_2...i_m}  b_{i_2}^\dagger b_{i_3}^\dagger ...b_{i_m}^\dagger,
\end{equation}

\begin{equation}\label{eqn_4.10}
\tau_{nm}^{(12)(i)}=\sum_{p_1, p_2,...,p_n=2}^{M_1} \sum_{i_2, i_3,...,i_m=2}^{M_2} e_{p_1p_2...p_nii_2...i_m}  a_{p_1}^\dagger a_{p_2}^\dagger ...a_{p_n}^\dagger b_{i_2}^\dagger b_{i_3}^\dagger ...b_{i_m}^\dagger,
\end{equation}
To determine the unknown coefficients, $c_{p_1 p_2...p_n}$, $d_{i_1 i_2...i_m}$, and $e_{p_1...p_{n^\prime}i_1...i_{m^\prime}}$, it is recommended to expand the first few terms of $\mathcal{K}_{{1}}$, $\mathcal{K}_{{p}}$, $\mathcal{L}_1$, and $\mathcal{L}_i$. The expansions of $\mathcal{K}_{{1}}$, $\mathcal{K}_{{p}}$, $\mathcal{L}_1$, and $\mathcal{L}_i$, which are defined in Eqs.~\ref{eqn_4.2}, are explicitly presented  in  Appendix B.

In the following calculations, it is noted that the operators $\mathcal{K}$ and $\mathcal{L}$ fulfil the commutation relation  $[\mathcal{K}_{{p}},\mathcal{K}_{{q}}]=[\mathcal{K}_{{1}},\mathcal{K}_{{p}}]=0$, $[\mathcal{L}_i,\mathcal{L}_j]=[\mathcal{L}_1,\mathcal{L}_j]=0$, and $[\mathcal{L}_1,\mathcal{K}_{{p}}]=[\mathcal{K}_{{1}},\mathcal{L}_j]=[\mathcal{L}_i,\mathcal{K}_{{p}}]= 0$. Also, the actions of the $\mathcal{K}$ and $\mathcal{L}$ on $\langle\phi_0|$ from the right are $\langle\phi_0|\mathcal{K}_{{1}}= 0, \langle\phi_0|(a_{{1}})^n\mathcal{K}_{{1}}= 0, \langle\phi_0|\mathcal{K}_{{i}}= c_{{i}}\langle\phi_0|a_{{1}}$ and $\langle\phi_0|\mathcal{L}_1= 0, \langle\phi_0|(b_1)^m\mathcal{L}_1= 0, \langle\phi_0|\mathcal{L}_j= d_j\langle\phi_0|b_1$.

\subsection{The energy and its structure}

We now calculate the energy using $E_0=\langle\phi_0|\dot{H}|\phi_0\rangle$. {We  find that the energy is given by the} combination of three parts 
\begin{equation}\label{eqn_4.13}
E_0=E_1+E_2+E_{12},
\end{equation}
where after some intricate algebra, one can  readily find $E_1$, $E_2$, and $E_{12}$ as  

\begin{eqnarray}\label{eqn_4.14}
E_1&=&N_1\left[h_{\bar{1}\bar{1}}^{(1)}+\dfrac{N_1-1}{2}V_{\bar{1}\bar{1}\bar{1}\bar{1}}^{(1)}\right]\nonumber\\
&+& N_1\Biggl[\sum_{s=2}^{M_1}[h_{\bar{1}\bar{s}}^{(1)}+(N_1-1)V_{\bar{1}\bar{1}\bar{1}\bar{s}}^{(1)}]c_{s}+\dfrac{N_1-1}{2}\sum_{r,s=2}^{M_1}V_{\bar{1}\bar{1}\bar{r}\bar{s}}^{(1)}(2c_{rs}+c_{r} c_{s})\Biggr],
\end{eqnarray}

\begin{eqnarray}\label{eqn_4.15}
E_2&=&N_2\left[h_{11}^{(2)}+\dfrac{N_2-1}{2}V_{1111}^{(2)}\right]\nonumber\\
&+& N_2\Biggl[\sum_{l=2}^{M_2}[h_{1l}^{(2)}+(N_2-1)V_{111l}^{(2)}]d_l+\dfrac{N_2-1}{2}\sum_{k,l=2}^{M_2}V_{11kl}^{(2)}(2d_{kl}+d_k d_l)\Biggr],
\end{eqnarray}

\begin{equation}\label{eqn_4.16}
E_{12}=N_1 N_2 \Biggl[V_{\bar{1}1\bar{1}1}^{(12)}+\sum_{s=2}^{M_1} V_{\bar{1}1\bar{s}1}^{(12)}c_{s}+\sum_{l=2}^{M_2} V_{\bar{1}1\bar{1}l}^{(12)}d_l +\sum_{s=2}^{M_1}\sum_{l=2}^{M_2}V_{\bar{1}1\bar{s}l}^{(12)} (c_{s} d_l+e_{ls})\Biggr].
\end{equation}
We notice that, in order to calculate  $E_1$, the first and second terms of $\dot{H}_0$  and the first, second, and fourth terms of $\dot{V}^{(1)}$ contribute. While for $E_2$, the fifth and sixth terms of $\dot{H}_0$ and the first, second, and fourth terms of $\dot{V}^{(2)}$ contribute. To determine $E_{12}$, only the first, second, third, and sixth terms of $\dot{V}^{(12)}$ contribute, also see Appendix C. 

Here, $E_1$ depends on the singles and doubles coefficients of species-1  and, analogously, $E_2$ depends on the singles and doubles coefficients of species-2. $E_1$ and $E_2$ have the analog form of single species coupled-cluster theory,  whereas    $E_{12}$ does not have an analog  in  single species theory as it is generated due to the inter-species interaction. If there is no inter-species interaction, $E_0$ boils down to the energy of two independent single species bosonic system. Note that the equation of $E_0$ presented here, Eqs.~\ref{eqn_4.13} to ~\ref{eqn_4.16}, is valid for all orders of coupled-cluster theory for bosonic mixtures, CC-M.

The mean-field energy, $E_{\text{MF}}=\langle \phi_0|{H}|\phi_0\rangle$, is contained in the first and fifth terms of $\dot{H}_0$ and the first terms of $\dot{V}^{(1)}$, $\dot{V}^{(2)}$, and $\dot{V}^{(12)}$.  $E_{\text{MF}}$ is found to be
\begin{equation}\label{eqn_4.17}
E_{\text{MF}}=N_1\left[h_{\bar{1}\bar{1}}^{(1)}+\dfrac{N_1-1}{2}V_{\bar{1}\bar{1}\bar{1}\bar{1}}^{(1)}\right]+N_2\left[h_{11}^{(2)}+\dfrac{N_2-1}{2}V_{1111}^{(2)}\right]+N_1 N_2 V_{\bar{1}1\bar{1}1}^{(12)}.
\end{equation}
{ The definition of the correlation energy of the ground state is the difference between many-body and mean-field energies.  For the binary bosonic mixtures, the correlation  energy is $(E_1+E_2+E_{12})-E_{\text{MF}}$ which can be explicitly  written as}

\begin{eqnarray}\label{eqn_4.18}
E_{\text{cor}}&=&N_1\Biggl[\sum_{s=2}^{M_1}[h_{\bar{1}\bar{s}}^{(1)}+(N_1-1)V_{\bar{1}\bar{1}\bar{1}\bar{s}}^{(1)}]c_{s}+\dfrac{N_1-1}{2}\sum_{r,s=2}^{M_1}V_{\bar{1}\bar{1}\bar{r}\bar{s}}^{(1)}(2c_{rs}+c_{r} c_{s})\Biggr]\nonumber \\
&+& N_2\Biggl[\sum_{l=2}^{M_2}[h_{1l}^{(2)}+(N_2-1)V_{111l}^{(2)}]d_l+\dfrac{N_2-1}{2}\sum_{k,l=2}^{M_2}V_{11kl}^{(2)}(2d_{kl}+d_k d_l)\Biggr]\nonumber \\
&+&N_1 N_2 \Biggl[\sum_{s=2}^{M_1} V_{\bar{1}1\bar{s}1}^{(12)}c_{s}+\sum_{l=2}^{M_2} V_{\bar{1}1\bar{1}l}^{(12)}d_l +\sum_{s=2}^{M_1}\sum_{l=2}^{M_2}V_{\bar{1}1\bar{s}l}^{(12)} (c_{s} d_l+e_{ls})\Biggr].
\end{eqnarray}
In the calculation of $E_{\text{cor}}$, we have used arbitrary sets of orthonormal orbitals. If {one would make use of}  the  orbitals generating from the Fock-like operators ${F}_1$ and ${F}_2$, see Eqs.~\ref{eqn_3.4} and ~\ref{eqn_3.5} discussed in last section, and Eq.~\ref{eqn_A.1}  in Appendix A, the correlation energy  simplifies and reads

\begin{eqnarray}\label{eqn_4.19}
E_{\text{cor}}&=&\dfrac{N_1(N_1-1)}{2}\sum_{r,s=2}^{M_1}V_{\bar{1}\bar{1}\bar{r}\bar{s}}^{(1)}(2c_{rs}+c_{r} c_{s})+\dfrac{N_2(N_2-1)}{2}\sum_{k,l=2}^{M_2}V_{11kl}^{(2)}(2d_{kl}+d_k d_l)\nonumber\\
&&+N_1N_2\sum_{s=2}^{M_1}\sum_{l=2}^{M_2}V_{\bar{1}1\bar{s}l}^{(12)} (c_{s} d_l+e_{ls}).
\end{eqnarray}
The other terms  disappear due to the facts that $\langle\varphi_s|{F}_1|\varphi_1\rangle=0$ and $\langle\psi_l|{F}_2|\psi_1\rangle=0$. All in all, the total energy of the mixture is modified to

\begin{equation}
E_0=E_{\text{MF}}-E_{\text{cor}},
\end{equation}
where $E_{\text{cor}}$ is {the correlation energy} for the binary mixture of bosons and presented in Eq.~\ref{eqn_4.19}.
\subsection{Equations for the coefficients}
The correlation energy due to the  state dressed from $\phi_0$ to $\Psi$ is expressed in terms of coefficients  $\{c_s\}$, $\{d_l\}$, $\{c_{rs}\}$, $\{d_{kl}\}$, and $\{e_{ls}\}$. The orbitals $\phi_i$ and $\psi_j$ can be conveniently chosen to simplify  $E_{\text{cor}}$, details are in the next section.  Note  that even the simplified form of the  equations can not be utilized  to determine the unknown coefficients mentioned above since  $E_0=\langle\phi_0|e^{-T}He^T|\phi_0\rangle$ is not subject to a variational principle.  Therefore, one requires  approximations. For the coupled-cluster theory for bosonic mixtures we may chose approximations such as the combination of    keeping $M_1$ and $M_2$  small and truncating the coupled-cluster excitations. In the process of truncating the excitations, one may consider $\phi_0$ and all the singly and doubly  excited configurations analogous by the atomic structure calculations.

Here, we observe that $E_0|\phi_0\rangle=e^{-T}He^T|\phi_0\rangle$  holds, and hence projecting on any excited configuration of species-1 and species-2  gives us the required  equation for the coefficients. Therefore, the singly excited configurations of species-1 provide the $(M_1-1)$ equations
\begin{equation}\label{eqn_4.20}
\langle \phi_0|a_{1}^\dagger a_{\bar{i}}\dot{H}|\phi_0\rangle=0,  \hspace{0.2cm}\bar{i}=2,3,...,M_1. 
\end{equation}
Similarly,  the singly excited configurations of species-2 lead to the $(M_2-1)$ equations

\begin{equation}\label{eqn_4.21}
\langle \phi_0|b_1^\dagger b_i\dot{H}|\phi_0\rangle=0, \hspace{0.2cm}i=2,3,...,M_2. 
\end{equation}
The doubly excited configurations  generate  the $\dfrac{1}{2}M_1(M_1-1)$ equations for species-1

\begin{equation}\label{eqn_4.22}
\langle \phi_0|(a_{1}^\dagger)^2 a_{\bar{i}}a_{\bar{j}}\dot{H}|\phi_0\rangle=0, \hspace{0.2cm}\bar{i}\geq\bar{j}=2,3,...,M_1, 
\end{equation}
and the $\dfrac{1}{2}M_2(M_2-1)$ equations for species-2

\begin{equation}\label{eqn_4.23}
\langle \phi_0|(b_1^\dagger)^2 b_i b_j\dot{H}|\phi_0\rangle=0, \hspace{0.2cm} i\geq j=2,3,...,M_2. 
\end{equation}
In addition, two simultaneous single excitations, one for each species,  provide the $(M_1-1)(M_2-1)$ equations

\begin{equation}\label{eqn_4.24}
\langle \phi_0|a_{1}^\dagger b_1^\dagger a_{\bar{i}}b_i\dot{H}|\phi_0\rangle=0, \hspace{0.2cm}\bar{i}=2,3,...,M_1,   \hspace{0.2cm}i=2,3,...,M_2.
\end{equation}
Here the number of independent equations Eqs.~\ref{eqn_4.20} to ~\ref{eqn_4.24} is
exactly  equal to the number of distinct coefficients. Indeed,  there are $M_1-1$ coefficients of $c_{\bar{i}}$, $M_2-1$ coefficients of $d_i$, $M_1(M_1-1)/2$ coefficients of $c_{\bar{i}\bar{j}}$, $M_2(M_2-1)/2$ coefficients of  $d_{ij}$,  and $(M_1-1)(M_2-1)$ coefficients of  $e_{\bar{i}i}$. The equations presented
above are  coupled to each other by these unknown coefficients.

 It is necessary to discuss here what would be the equations if one would go beyond the second order coupled-cluster theory. To include  all triple excitations, one requires to  solve four more equations, and they are  for  $\langle \phi_0|(a_{1}^\dagger)^3 a_{\bar{i}}a_{\bar{j}}a_{\bar{k}}\dot{H}|\phi_0\rangle=0, \bar{i}\geq\bar{j}\geq\bar{k}=2,3,...,M_1$  {for species-1},  $\langle \phi_0|(b_1^\dagger)^3 b_i b_j b_k\dot{H}|\phi_0\rangle=0, i\geq j\geq k=2,3,...,M_2$ {for species-2}, and  the combined excitations $\langle \phi_0|(a_{1}^\dagger)^2 b_1^\dagger a_{\bar{i}}a_{\bar{j}}b_i\dot{H}|\phi_0\rangle=0$ and  $\langle \phi_0|a_{1}^\dagger (b_1^\dagger)^2 a_{\bar{i}}b_ib_j\dot{H}|\phi_0\rangle=0$  where $\bar{i}\geq \bar{j} =2,3,...,M_1,   i\geq j=2,3,...,M_2$ for both species. For triple excitations, additional four  different types of  unknown coefficients  emerge which are $\{c_{\bar{i}\bar{j}\bar{k}}\}$, $\{d_{ijk}\}$, $\{e_{\bar{i}\bar{j}i}\}$ and $\{e_{\bar{i}ij}\}$. Similarly for quadrupole excitations, one needs to solve five additional coupled equations.

Now, we evaluate the series of coupled  Eqs.~\ref{eqn_4.20} to ~\ref{eqn_4.24} in order to determine the unknown coefficients. The series contains sets of equation having excitation operators, $T_n^{(1)}$, $T_m^{(2)}$, and $T_{n^\prime m^\prime}^{(12)}$, where ideally $n$ and $n^\prime$ run from 1 to $N_1$ and $m$ and $m^\prime$ run from 1 to $N_2$, see Eqs.~\ref{eqn_2.7} to ~\ref{eqn_2.9}. In practice, one has to truncate the expansions  of $T_n^{(1)}$, $T_m^{(2)}$, and $T_{n^\prime m^\prime}^{(12)}$.  {If each of the excitation operators}, $T_n^{(1)}$ and  $T_m^{(2)}$,   consists of one excitation to the virtual orbital we call it coupled-cluster singles approach. When $T_n^{(1)}$ and  $T_m^{(2)}$ each contains two excitations to the virtual orbital, and $T_{n^\prime m^\prime}^{(12)}$ consists two simultaneous single excitations to the virtual orbitals, we call it coupled-cluster singles doubles for bosonic  mixture (CCSD-M).

\subsubsection{The intra-species coefficients}

We have derived  the equations ~\ref{eqn_4.20} to ~\ref{eqn_4.24} for an arbitrary
sets of orthonormal orbitals, see the details in Appendix D, but present here only the expressions found from  the Fock orbitals of ${F}_1$  and ${F}_2$, see Eq.~\ref{eqn_3.4}. The derivations are very lengthy and involved.  Now we start from solving  $\langle \phi_0|a_{1}^\dagger a_{\bar{i}}\dot{H}|\phi_0\rangle=0$ and $\langle \phi_0|b_1^\dagger b_i\dot{H}|\phi_0\rangle=0$. We obtain $M_1-1$ coupled equations with $\bar{i}=2,3,...,M_1$

\begin{eqnarray}\label{eqn_4.25}
\mu_1^{(1)}c_{\bar{i}}&=&-\sum_{s=2}^{M_1}V_{\bar{1}\bar{1}\bar{1}\bar{s}}^{(1)}(N_1-1)(2c_{s\bar{i}}+c_{\bar{i}}c_s)+(N_1-1)\sum_{s=2}^{M_1}V_{\bar{1}\bar{i}\bar{s}\bar{1}}^{(1)}c_{s}\nonumber \\
&&-\sum_{l=2}^{M_2}V_{\bar{1}1\bar{1}l}^{(12)}N_2(e_{\bar{i}l}+  c_{\bar{i}}d_l)+N_2\sum_{l=2}^{M_2}V_{\bar{i}1\bar{1}l}^{(12)}d_l\nonumber\\
&&+\sum_{r,s=2}^{M_1}V_{\bar{1}\bar{1}\bar{r}\bar{s}}^{(1)}\bar{\alpha}_{rs\bar{i}}+\sum_{r,s=2}^{M_1}V_{\bar{1}\bar{i}\bar{r}\bar{s}}^{(1)}[(N_1-1)(2c_{rs}+c_{r}c_{s})]\nonumber\\
&&+\dfrac{1}{2}N_2(N_2-1)\sum_{k,l=2}^{M_2}V_{11kl}^{(2)}(2e_{\bar{i}lk}+e_{\bar{i}k}d_l+e_{\bar{i}l}d_k)\nonumber\\
&&\nonumber \\
&&+\sum_{s=2}^{M_1}\sum_{l=2}^{M_2}V_{\bar{1}1\bar{s}l}^{(12)}[(N_1-1)N_2(2e_{\bar{i}sl}+c_s e_{\bar{i}l}+2c_{s\bar{i}}d_l)-N_2(c_{\bar{i}}e_{sl}+c_{\bar{i}}c_sd_l)]\nonumber \\
&&+N_2\sum_{s=2}^{M_1}\sum_{l=2}^{M_2}V_{\bar{i}1\bar{s}l}^{(12)}(e_{sl}+c_{s}d_l),
\end{eqnarray}
and $M_2-1$ coupled equation with $i=2,3,...,M_2$

\begin{eqnarray}\label{eqn_4.26}
\mu_1^{(2)}d_i&=&-\sum_{l=2}^{M_2}V_{111l}^{(2)}(N_2-1)(2d_{li}+d_id_l)+(N_2-1)\sum_{l=2}^{M_2}V_{1il1}^{(2)}d_l\nonumber \\
&&-\sum_{s=2}^{M_1}V_{\bar{1}1\bar{s}1}^{(12)}N_1(e_{si}+ c_sd_i)+N_1 \sum_{s=2}^{M_1}V_{\bar{1}i\bar{s}1}^{(12)}c_{s}  \nonumber\\
&&+\sum_{k,l=2}^{M_2}V_{11kl}^{(2)}\alpha_{kli}+\sum_{k,l=2}^{M_2}V_{1ikl}^{(2)}[(N_2-1)(2d_{lk}+d_kd_l)]\nonumber\\
&&+\dfrac{1}{2}N_1(N_1-1)\sum_{r,s=2}^{M_1}V_{\bar{1}\bar{1}\bar{r}\bar{s}}^{(1)}(2e_{sri}+e_{ri}c_s+e_{si}c_r)\nonumber \\
&&+\sum_{s=2}^{M_1}\sum_{l=2}^{M_2}V_{\bar{1}1\bar{s}l}^{(12)}[N_1(N_2-1)(2e_{sli}+2c_{s}d_{li}+d_l e_{si})-N_1(d_ie_{sl}+c_sd_id_l)]\nonumber \\
&&+N_1\sum_{s=2}^{M_1}\sum_{l=2}^{M_2}V_{\bar{1}i\bar{s}l}^{(12)}(e_{sl}+c_{s}d_l),
\end{eqnarray}
where the quantities $\bar{\alpha}_{rs\bar{i}}$ and $\alpha_{kli}$ are given by
\begin{eqnarray}\label{eqn_4.27}
&& \bar{\alpha}_{rs\bar{i}}=[(N_1-1)(N_1-2)(3c_{s\bar{i}r}+c_{s\bar{i}}c_{r}+c_{r\bar{i}}c_{s})-(N_1-1)(2c_{\bar{i}}c_{rs}+c_{\bar{i}}c_{r}c_{s})],\nonumber\\
&&\alpha_{kli}=(N_2-1)(N_2-2)(3d_{lik}+d_{li}d_{k}+d_{ki}d_l)-(N_2-1)(2d_id_{lk}+d_id_kd_l)].
\end{eqnarray}
In   Eqs.~\ref{eqn_4.25} and ~\ref{eqn_4.26}, there are terms, explicitly, $c_{s\bar{i}r}$, $e_{\bar{i}lk}$, $e_{\bar{i}sl}$, $d_{lik}$, $e_{sri}$, and $e_{sli}$, which  contain triple excitations. These coefficients have to be put equal to zero if coupled-cluster singles doubles is to be used. Similarly,  when we take only the single excitations which leads to the coupled-cluster singles approach,  the general equations for the  arbitrary sets of orthonormal orbitals are presented in the Appendix D.

Now, to evaluate the working equations of the CCSD-M, we need to determine  the set of $M_1(M_1-1)/2$, $M_2(M_2-1)/2$, and $(M_1-1)(M_2-1)$ distinct coupled equations resulting from $\langle \phi_0|(a_{1}^\dagger)^2 a_{\bar{i}}a_{\bar{j}}\dot{H}|\phi_0\rangle=0$, $\langle \phi_0|(b_1^\dagger)^2 b_i b_j\dot{H}|\phi_0\rangle=0$, and  $\langle \phi_0|a_{1}^\dagger b_1^\dagger a_{\bar{i}}b_i\dot{H}|\phi_0\rangle=0$, respectively.  To remind, we only present here the working equations generating from the Fock orbitals of ${F}_1$ and ${F}_2$. The sets of coupled-equations for the double excitations with $\bar{i},\bar{j}=2,3,...,M_1$ read

\begin{eqnarray}\label{eqn_4.28}
(4\mu_1^{(1)}&-&2\mu_{\bar{i}}^{(1)}-2\mu_{\bar{j}}^{(1)})c_{\bar{i}\bar{j}}=V_{\bar{i}\bar{j}\bar{1}\bar{1}}^{(1)}+V_{\bar{1}\bar{1}\bar{1}\bar{1}}^{(1)}(2c_{\bar{i}\bar{j}}+c_{\bar{i}}c_{\bar{j}})\nonumber\\
&&-[V_{\bar{i}\bar{1}\bar{1}\bar{1}}^{(1)}c_{\bar{j}}+V_{\bar{j}\bar{1}\bar{1}\bar{1}}^{(1)}c_{\bar{i}}]-\sum_{s=2}^{M_1}V_{\bar{1}\bar{1}\bar{1}\bar{s}}^{(1)}\bar{\beta}_{s\bar{i}\bar{j}}\nonumber \\
&&-\sum_{s=2}^{M_1}V_{\bar{i}\bar{1}\bar{s}\bar{1}}^{(1)}(2c_{s\bar{j}}+c_{\bar{j}}c_s)+\sum_{s=2}^{M_1}V_{\bar{1}\bar{i}\bar{s}\bar{1}}^{(1)}[2(N_1-2)c_{s\bar{j}}-c_{\bar{j}}c_{s}]\nonumber\\
&&-\sum_{s=2}^{M_1}V_{\bar{j}\bar{1}\bar{s}\bar{1}}^{(1)}(2c_{s\bar{i}}+c_{\bar{i}}c_s)+\sum_{s=2}^{M_1}V_{\bar{1}\bar{j}\bar{s}\bar{1}}^{(1)}[2(N_1-2)c_{s\bar{i}}-c_{\bar{i}}c_{s}]\nonumber \\
&&+\sum_{s=2}^{M_1}V_{\bar{i}\bar{j}\bar{s}\bar{1}}^{(1)}c_{s}+\sum_{s=2}^{M_1}V_{\bar{j}\bar{i}\bar{s}\bar{1}}^{(1)}c_{s}+\sum_{r,s=2}^{M_1}V_{\bar{i}\bar{j}\bar{r}\bar{s}}^{(1)}[2c_{sr}+c_{r}c_{s}]+\sum_{r,s=2}^{M_1}V_{\bar{1}\bar{1}\bar{r}\bar{s}}^{(1)}\bar{\gamma}_{rs\bar{i}\bar{j}}\nonumber \\
&&+\sum_{r,s=2}^{M_1}V_{\bar{1}\bar{i}\bar{r}\bar{s}}^{(1)}[2(N_1-2)c_{r\bar{j}}c_{s}+2(N_1-2)c_{s\bar{j}}c_{r}-2c_{sr}c_{\bar{j}}-c_{\bar{j}}c_{r}c_{s}]\nonumber \\
&&+\sum_{r,s=2}^{M_1}V_{\bar{1}\bar{j}\bar{r}\bar{s}}^{(1)}[2(N_1-2)c_{r\bar{i}}c_{s}+2(N_1-2)c_{s\bar{i}}c_{r}-2c_{sr}c_{\bar{i}}-c_{\bar{i}}c_{r}c_{s}]\nonumber \\
&&+N_2\sum_{l=2}^{M_2}[V_{\bar{i}1\bar{1}l}^{(12)}e_{\bar{j}l}+V_{\bar{j}1\bar{1}l}^{(12)}e_{\bar{i}l}]-N_2\sum_{l=2}^{M_2}V_{\bar{1}1\bar{1}l}^{(12)}(4c_{\bar{i}\bar{j}}d_l+c_{\bar{i}}e_{\bar{j}l}+c_{\bar{j}}e_{\bar{i}l})\nonumber \\
&&+\dfrac{1}{2}N_2(N_2-1)\sum_{k,l=2}^{M_2}V_{11kl}^{(2)}[e_{\bar{i}k}e_{\bar{j}l}+e_{\bar{j}k}e_{\bar{i}l}]+\sum_{s=2}^{M_1}\sum_{l=2}^{M_2}V_{\bar{1}1\bar{s}l}^{(12)}\bar{\delta}_{sl\bar{i}\bar{j}}\nonumber \\
&&\nonumber \\
&&+N_2\sum_{s=2}^{M_1}\sum_{l=2}^{M_2}[V_{\bar{i}1\bar{s}l}^{(12)}(2c_{s\bar{j}}d_l+c_se_{\bar{j}l})+V_{\bar{j}1\bar{s}l}^{(12)}(2c_{s\bar{i}}d_l+c_se_{\bar{i}l})].
\end{eqnarray}
The quantities $\bar{\beta}_{s\bar{i}\bar{j}}$, $\bar{\gamma}_{rs\bar{i}\bar{j}}$, and $\bar{\delta}_{sl\bar{i}\bar{j}}$ have the form

\begin{eqnarray}\label{eqn_4.29}
&&\bar{\beta}_{s\bar{i}\bar{j}}=[2(N_1-3)(c_{\bar{i}}c_{s\bar{j}}+c_{\bar{j}}c_{s\bar{i}})+4(N_1+2)c_{s}c_{\bar{i}\bar{j}}+2c_{\bar{i}}c_{\bar{j}}c_{s}], \nonumber \\
&&\bar{\gamma}_{rs\bar{i}\bar{j}}=4c_{r\bar{i}}c_{s\bar{j}}(N_1-2)(N_1-3)-2(N_1-2)\big[2c_{\bar{i}\bar{j}}(2c_{rs}+c_{r}c_{s})\nonumber \\
&& \hspace{1cm}+c_{r\bar{i}}c_{s}c_{\bar{j}}+c_{s\bar{i}}c_{r}c_{\bar{j}}+c_{r\bar{j}}c_{s}c_{\bar{i}}+c_{s\bar{j}}c_{r}c_{\bar{i}}\big]-(2c_{rs}+c_{r}c_{s})(2c_{\bar{i}\bar{j}}-c_{\bar{i}}c_{\bar{j}}),\nonumber \\
&&\bar{\delta}_{sl\bar{i}\bar{j}}=2(N_1-2)N_2(c_{s\bar{i}}e_{\bar{j}l}+c_{s\bar{j}}e_{\bar{i}l})\nonumber \\
&&\hspace{1cm}-N_2(4c_{\bar{i}\bar{j}}e_{sl}+4c_{\bar{i}\bar{j}}c_sd_l+2c_{s\bar{j}}c_{\bar{i}}d_l+2c_{\bar{j}}d_lc_{s\bar{i}}+c_{\bar{i}}c_se_{\bar{j}l}+c_{\bar{j}}c_se_{\bar{i}l}).
\end{eqnarray}
Analogously, the $M_2(M_2-1)/2$ set of coupled equations generated from $\langle \phi_0|(b_1^\dagger)^2 b_i b_j\dot{H}|\phi_0\rangle=0$   with $i,j=2,3,...,M_2$ can be written as

\begin{eqnarray}\label{eqn_4.30}
(4\mu_1^{(2)}&-&2\mu_{i}^{(2)}-2\mu_{j}^{(2)})d_{ij}=V_{ij11}^{(2)}+V_{1111}^{(2)}(2d_{ij}+d_id_j)\nonumber\\
&&-[V_{i111}^{(2)}d_j+V_{j111}^{(2)}d_i]-\sum_{l=2}^{M_2}V_{111l}^{(2)}\beta_{lij}\nonumber \\
&&-\sum_{l=2}^{M_2}V_{i1l1}^{(2)}(2d_{lj}+d_jd_l)+\sum_{l=2}^{M_2}V_{1il1}^{(2)}[2(N_2-2)d_{lj}-d_jd_l]\nonumber \\
&&-\sum_{l=2}^{M_2}V_{j1l1}^{(2)}(2d_{li}+d_id_l)+\sum_{l=2}^{M_2}V_{1jl1}^{(2)}[2(N_2-2)d_{li}-d_id_l]\nonumber \\
&&+\sum_{l=2}^{M_2}V_{ijl1}^{(2)}d_l+\sum_{l=2}^{M_2}V_{jil1}^{(2)}d_l+\sum_{k,l=2}^{M_2}V_{ijkl}^{(2)}[2d_{lk}+d_kd_l]+\sum_{k,l=2}^{M_2}V_{11kl}^{(2)}\gamma_{klij}\nonumber \\
&&+\sum_{k,l=2}^{M_2}V_{1ikl}^{(2)}[2(N_2-2)d_{kj}d_l+2(N_2-2)d_{lj}d_k-2d_{lk}d_j-d_jd_kd_l]\nonumber \\
&&+\sum_{k,l=2}^{M_2}V_{1jkl}^{(2)}[2(N_2-2)d_{ki}d_l+2(N_2-2)d_{li}d_k-2d_{lk}d_i-d_id_kd_l]\nonumber \\
&&+N_1\sum_{s=2}^{M_1}[V_{\bar{1}i\bar{s}1}^{(12)}e_{sj}+V_{\bar{1}j\bar{s}1}^{(12)}e_{si}]-N_1\sum_{s=2}^{M_1}V_{\bar{1}1\bar{s}1}^{(12)}(4d_{ij}c_{s}+e_{sj}d_i+e_{si}d_j)\nonumber \\
&&+\dfrac{1}{2}N_1(N_1-1)\sum_{r,s=2}^{M_1}V_{\bar{1}\bar{1}\bar{r}\bar{s}}^{(1)}[e_{ri}e_{sj}+e_{rj}e_{si}]+\sum_{s=2}^{M_1}\sum_{l=2}^{M_2}V_{\bar{1}1\bar{s}l}^{(12)}\delta_{slij}
\nonumber \\
&&+N_1\sum_{s=2}^{M_1}\sum_{l=2}^{M_2}[V_{\bar{1}i\bar{s}l}^{(12)}(2d_{lj}c_{s}+e_{sj}d_l)+V_{\bar{1}j\bar{s}l}^{(12)}(2d_{li}c_{s}+e_{si}d_l)],
\end{eqnarray}
where the quantities, $\beta_{lij}$, $\gamma_{klij}$, and  $\delta_{slij}$, appearing in Eq.~\ref{eqn_4.30} are

\begin{eqnarray}\label{eqn_4.31}
&&\beta_{lij}=[2(N_2-3)(d_id_{lj}+d_jd_{li})+4(N_2+2)d_ld_{ij}+2d_id_jd_l],\nonumber\\
&&\gamma_{klij}=4d_{ki}d_{lj}(N_2-2)(N_2-3)-2(N_2-2)\big[2d_{ij}(2d_{kl}+d_kd_l)\nonumber \\
&&\hspace{1cm}+d_{ki}d_ld_j+d_{li}d_kd_j+d_{kj}d_ld_i+d_{lj}d_kd_i\big]-(2d_{kl}+d_kd_l)(2d_{ij}-d_id_j),\nonumber \\
&&\delta_{slij}=2N_1(N_2-2)(e_{si}d_{lj}+e_{sj}d_{li})\nonumber \\
&&\hspace{1cm}-N_1(4d_{ij}e_{sl}+4d_{ij}c_sd_l+2c_sd_id_{lj}+2c_sd_jd_{li}+d_id_le_{sj}+d_jd_le_{si}).
\end{eqnarray}
Note that the general forms of the Eqs.~\ref{eqn_4.28} and ~\ref{eqn_4.30} are presented in Appendix D.

\subsubsection{The inter-species coefficients}
In this subsection we show the final form of the $(M_1-1)(M_2-1)$ equations resulting from $\langle \phi_0|a_{1}^\dagger b_1^\dagger a_{\bar{i}}b_i\dot{H}|\phi_0\rangle=0$ where $\bar{i}=2,3,...,M_1$ and   $i=2,3,...,M_2$ using Fock-like operators and Appendix A. The general form of this expansion is presented in Appendix D. Using the Fock-like operators, one can readily find

\begin{eqnarray}\label{eqn_4.32}
(\mu_1^{(1)}+\mu_1^{(2)}&-&\mu_{\bar{i}}^{(1)}-\mu_i^{(2)})e_{\bar{i}i}=V_{\bar{i}i\bar{1}1}^{(12)}-V_{\bar{1}i\bar{1}1}^{(12)}c_{\bar{i}}-V_{\bar{i}1\bar{1}1}^{(12)} d_i\nonumber \\
&&+V_{\bar{1}1\bar{1}1}^{(12)}(e_{\bar{i}i}+c_{\bar{i}}d_i)+\sum_{s=2}^{M_1}V_{\bar{i}i\bar{s}1}^{(12)}c_s+\sum_{l=2}^{M_2}V_{\bar{i}i\bar{1}l}^{(12)}d_l\nonumber \\
&&-\sum_{s=2}^{M_1}V_{\bar{1}\bar{1}\bar{1}\bar{s}}^{(1)}[(N_1+1)(c_se_{\bar{i}i}+c_{\bar{i}}e_{si})]+\sum_{s=2}^{M_1}V_{\bar{1}1\bar{s}1}^{(12)}[N_2(c_se_{\bar{i}i}+c_{\bar{i}}e_{si})-\bar{\chi}_{s\bar{i}i}]\nonumber \\
&&-\sum_{l=2}^{M_2}V_{111l}^{(2)}[(N_2+1)(d_le_{\bar{i}i}+d_ie_{\bar{i}l})]+\sum_{l=2}^{M_2}V_{\bar{1}1\bar{1}l}^{(12)}[N_1(d_le_{\bar{i}i}+d_ie_{\bar{i}l})-\chi_{l\bar{i}i}]\nonumber \\
&&+\sum_{s=2}^{M_1}V_{\bar{1}\bar{i}\bar{s}\bar{1}}^{(1)}(N_1-1)e_{si}-\sum_{s=2}^{M_1}V_{\bar{i}1\bar{s}1}^{(12)}(e_{si}+c_sd_i)\nonumber \\
&&+\sum_{l=2}^{M_2}V_{1il1}^{(2)}(N_2-1)e_{\bar{i}l}-\sum_{l=2}^{M_2}V_{\bar{1}i\bar{1}l}^{(12)}(e_{\bar{i}l}+c_{\bar{i}}d_l)\nonumber \\
&&+\sum_{s=2}^{M_1}V_{\bar{1}i\bar{s}1}^{(12)}[2(N_1-1)c_{s\bar{i}}-c_{\bar{i}}c_s]+\sum_{l=2}^{M_2}V_{\bar{i}1\bar{1}l}^{(12)}[2(N_2-1)d_{li}-d_{i}d_l]\nonumber \\
&&+\sum_{r,s=2}^{M_1}V_{\bar{1}\bar{1}\bar{r}\bar{s}}^{(1)}\bar{\xi}_{rs\bar{i}i}+(N_1-1)\sum_{r,s=2}^{M_1}V_{\bar{1}\bar{i}\bar{r}\bar{s}}^{(1)}(c_se_{ri}+c_re_{si})\nonumber \\
&&+\sum_{k,l=2}^{M_2}V_{11kl}^{(2)}\xi_{kl\bar{i}i}+(N_2-1)\sum_{k,l=2}^{M_2}V_{1ikl}^{(2)}(d_le_{\bar{i}k}+d_ke_{\bar{i}l})\nonumber \\
&&+\sum_{s=2}^{M_1}\sum_{l=2}^{M_2}V_{\bar{1}i\bar{s}l}^{(12)}[(N_1-1)(2c_{si}d_l+c_se_{\bar{i}l})-(c_{\bar{i}}e_{sl}+c_{\bar{i}}c_sd_l)]\nonumber \\
&&+\sum_{s=2}^{M_1}\sum_{l=2}^{M_2}V_{\bar{i}1\bar{s}l}^{(12)}[(N_2-1)(2c_sd_{li}+d_le_{si})-(d_ie_{sl}+c_sd_id_l)]\nonumber \\
&&+\sum_{s=2}^{M_1}\sum_{l=2}^{M_2}V_{\bar{1}1\bar{s}l}^{(12)}\zeta_{sl\bar{i}i}+\sum_{s=2}^{M_1}\sum_{l=2}^{M_2}V_{\bar{i}i\bar{s}l}^{(12)}(e_{sl}+c_sd_l),
\end{eqnarray}
where $\mu_1^{(1)}$ and $\mu_1^{(2)}$ are the chemical potentials for the ground orbitals of species-1 and species-2, respectively.  $\mu_{\bar{i}}^{(1)}$ and $\mu_i^{(2)}$ are the chemical potential of the $\bar{i}$-th and $i$-th orbital of species-1 and species-2, respectively. The parameters $\bar{\xi}_{rs\bar{i}i}$, $\xi_{kl\bar{i}i}$, $\bar{\chi}_{s\bar{i}i}$, $\chi_{l\bar{i}i}$, and $\zeta_{sl\bar{i}i}$ appearing in Eq.~\ref{eqn_4.32} are given by

\begin{eqnarray}\label{eqn_4.33}
&&\bar{\xi}_{rs\bar{i}i}=(N_1-1)[(N_1-2)(c_{r\bar{i}}e_{si}+c_{si}e_{r\bar{i}})-2c_{rs}e_{\bar{i}i}-c_rc_se_{\bar{i}i}-c_{\bar{i}}c_se_{ri}-c_{\bar{i}}c_re_{si}],\nonumber\\
&&\xi_{kl\bar{i}i}=(N_2-1)[(N_2-2)(d_{ki}e_{\bar{i}l}+d_{li}e_{\bar{i}k})-2d_{lk}e_{\bar{i}i}-d_kd_le_{\bar{i}i}-d_id_le_{\bar{i}k}-d_id_ke_{\bar{i}l}],\nonumber\\
&&\bar{\chi}_{s\bar{i}i}=(N_1+N_2-1)c_se_{\bar{i}i}+2(N_1-1)d_ic_{s\bar{i}}+(N_2-1)c_{\bar{i}}e_{si}-c_{\bar{i}}c_sd_i,\nonumber\\
&&\chi_{l\bar{i}i}=(N_1+N_2-1)d_le_{\bar{i}i}+(N_1-1)d_ie_{\bar{i}l}+2(N_2-1)c_{\bar{i}}d_{li}-c_{\bar{i}}d_id_l,\nonumber\\
&&\zeta_{sl\bar{i}i}=4(N_1-1)(N_2-1)c_{s\bar{i}}d_{li}-(N_1+N_2-1)(e_{\bar{i}i}e_{sl}+c_sd_le_{\bar{i}i})\nonumber \\
&&\hspace{1cm}-(N_1-1)(2c_{s\bar{i}}d_id_l+c_sd_ie_{\bar{i}l})-(N_2-1)(2c_{\bar{i}}c_sd_{li}+c_{\bar{i}}d_le_{si})\nonumber \\
&&\hspace{1cm}+c_{\bar{i}}d_ie_{sl}+c_{\bar{i}}d_ic_sd_l.
\end{eqnarray}
{Hitherto},  we have presented the details of  derivation of the working equations of CCSD-M. In the next section, we would like to apply and check their accuracy. Our strategy is to recruit a solvable many-body model for a mixture and study the performance for different interactions, attractive and repulsive, with balanced and  imbalanced numbers of bosons.

\section{ILLUSTRATIVE EXAMPLES}\label{Section V}
Until now, {we} have formulated the working equations for CCSD-M using the arbitrary sets of orbitals and Fock-like operators. Here,  we aim at implementing and benchmarking the new theory by utilizing those working equations. We elaborate a few  illustrative examples in case of the  harmonic-interaction model and compare the ground-state energy with the  corresponding analytical exact results. The exactly solvable many-body harmonic interaction  model  has been employed to compare various properties of a many-body system as this model is  one of the non-trivial scenarios with an analytical solution for the ground state of a many-particle system \cite{Leveque2018, Lode2012, Fasshauer2016,Lode2020}. This model describes a many-body system when the particles are trapped in a harmonic potential as well as the inter-particle interaction has the form of harmonic  nature.  For bosonic mixtures, the harmonic interaction model is extended in Ref. \cite{Bouvrie2014,Klaiman2017, Alon2017}. 

  We consider a binary-species mixture having $N_1$ and $N_2$ numbers of bosons in species-1 and species-2, respectively,  where both species are trapped in a one-dimensional external harmonic potential. Here, the two species are  localized at the origin with the one-body term in the Hamiltonian for species-1 reads  $-\dfrac{1}{2}\dfrac{\partial^2}{\partial r_1^2}+\dfrac{1}{2}\omega^2r_1^2$  and that for species-2 $-\dfrac{1}{2}\dfrac{\partial^2}{\partial r_2^2}+\dfrac{1}{2}\omega^2r_2^2$,  where $\omega$ is the frequency of the trap and it is considered to be one throughout this work.  The intra-species interactions for species-1 and species-2 are   $\lambda_1(r_1-r_1^\prime)^2$ and  $\lambda_2(r_2-r_2^\prime)^2$, respectively, while the inter-species interaction is  $\lambda_{12}(r_1-r_2)^2$. Here,   we have the degrees-of-freedom of the number of bosons in each species and the strengths and signs of the intra- and inter-species interactions. If the numbers of bosons and intra-species interaction parameters in both species are same then the mixture is called balanced and  if otherwise, it is imbalanced.

For the CCSD-M calculation of the energy, we restrict the orbital space upto  $M_1=M_2=2$ orbitals, i.e., $\varphi_1$ and $\varphi_2$ for  species-1, and $\psi_1$ and $\psi_2$ for species-2. The quality of this truncation for bosonic mixtures will be assessed too. To use the coupled-cluster theory for an example of harmonic-interaction model, we (i) solve the mean-field equations to determine $\mu_1^{(1)}$ and $\mu_1^{(2)}$,  (ii) solve the Fock-like equations for the second orbitals of the species  to get the chemical potentials, $\mu_2^{(1)}$ and $\mu_2^{(2)}$, and (iii) calculate all matrix elements.   {In addition,} steps (i)-(iii) can all be carried out analytically.

Here, for each species, the ground orbital, $\varphi_1(\psi_1)$, possesses gerade symmetry and the excited orbital $\varphi_2(\psi_2)$ has ungerade symmetry. The normalized ground and first excited orbitals for the two species take on the form as 

\begin{eqnarray}\label{orbitals}
\varphi_1(r_1)=\dfrac{\Omega_1^{1/4}}{\pi^{1/4}}exp\bigg[-\dfrac{{\Omega_1}}{2}r_1^2\bigg],\nonumber\\
\varphi_2(r_1)=\dfrac{\sqrt{2}\Omega_1^{3/4}}{\pi^{1/4}}r_1exp\bigg[-\dfrac{{\Omega_1}}{2}r_1^2\bigg],\nonumber\\
\psi_1(r_2)=\dfrac{\Omega_2^{1/4}}{\pi^{1/4}}exp\bigg[-\dfrac{{\Omega_2}}{2}r_2^2\bigg],\nonumber\\
\psi_2(r_2)=\dfrac{\sqrt{2}\Omega_2^{3/4}}{\pi^{1/4}}r_2exp\bigg[-\dfrac{{\Omega_2}}{2}r_2^2\bigg],
\end{eqnarray}
where $\Omega_1$ and $\Omega_2$ are dressed frequencies that depend on   the inter- and intra-species interaction parameters  as follows $\Omega_1=\sqrt{\omega^2+2(\Lambda_1+\Lambda_{21})}$ and  $\Omega_2=\sqrt{\omega^2+2(\Lambda_2+\Lambda_{12})}$. As we consider only two orbitals for each species and they have different spatial  symmetries, therefore  in case of harmonic-interaction model, one  can readily find that the singles  coefficients vanish, $c_s=d_l=0$ when $s=2$ and $l=2$. Moreover,  Eqs.~\ref{eqn_4.28}, ~\ref{eqn_4.30}, and ~\ref{eqn_4.32} for the specific case of $M_1=M_2=2$ take the  simplified form of three quadratic coupled equations for the $c_{22}$, $d_{22}$, and $e_{22}$ coefficients which can be  {written as}

\begin{eqnarray}\label{eqn_5.1}
(N_1^2&&-7N_1+9)V_{\bar{1}\bar{1}\bar{2}\bar{2}}^{(1)}c_{22}^2+\dfrac{N_2(N_2-1)}{4}V_{1122}^{(2)}e_{22}^2\nonumber \\
&&+\bigg[(\mu_2^{(1)}-\mu_1^{(1)})+\dfrac{1}{2}(V_{\bar{1}\bar{1}\bar{1}\bar{1}}^{(1)}+V_{\bar{2}\bar{2}\bar{2}\bar{2}}^{(1)})+(N_1-2)V_{\bar{1}\bar{2}\bar{2}\bar{1}}^{(1)}-V_{\bar{2}\bar{1}\bar{2}\bar{1}}^{(1)}\bigg]c_{22}\nonumber \\
&&+\dfrac{N_2}{2}V_{\bar{2}1\bar{1}2}^{(12)}e_{22}+N_2(N_1-3)V_{\bar{1}1\bar{2}2}^{(12)}c_{22}e_{22}+\dfrac{V_{\bar{2}\bar{2}\bar{1}\bar{1}}^{(1)}}{4}=0,
\end{eqnarray}
where Eq.~\ref{eqn_5.1} couples  the coefficients $c_{22}$ and   $e_{22}$,

\begin{eqnarray}\label{eqn_5.2}
(N_2^2&&-7N_2+9)V_{1122}^{(2)}d_{22}^2+\dfrac{N_1(N_1-1)}{4}V_{\bar{1}\bar{1}\bar{2}\bar{2}}^{(1)}e_{22}^2\nonumber \\
&&+\bigg[(\mu_2^{(2)}-\mu_1^{(2)})+\dfrac{1}{2}(V_{1111}^{(2)}+V_{2222}^{(2)})+(N_2-2)V_{1221}^{(2)}-V_{2121}^{(2)}\bigg]d_{22}\nonumber \\
&&+\dfrac{N_1}{2}V_{\bar{1}2\bar{2}1}^{(12)}e_{22}+N_1(N_2-3)V_{\bar{1}1\bar{2}2}^{(12)}d_{22}e_{22}+\dfrac{V_{2211}^{(2)}}{4}=0,
\end{eqnarray}
which couples the coefficients $d_{22}$ and   $e_{22}$, and

\begin{eqnarray}\label{eqn_5.3}
(N_1+&&N_2-1)V_{\bar{1}1\bar{2}2}^{(12)}e_{22}^2+\bigg[(\mu_1^{(1)}+\mu_1^{(2)}-\mu_2^{(1)}-\mu_2^{(2)})\nonumber \\
&&+V_{\bar{2}1\bar{2}1}^{(12)}+V_{\bar{1}2\bar{1}2}^{(12)}-V_{\bar{1}1\bar{1}1}^{(12)}-V_{\bar{2}2\bar{2}2}^{(12)}-(N_1-1)V_{\bar{1}\bar{2}\bar{2}\bar{1}}^{(1)}-(N_2-1)V_{1221}^{(2)}\bigg]e_{22}\nonumber \\
&&-2(N_1-1)V_{\bar{1}2\bar{2}1}^{(12)}c_{22}-2(N_2-1)V_{\bar{2}1\bar{1}2}^{(12)}d_{22}-4(N_1-1)(N_2-1)V_{\bar{1}1\bar{2}2}^{(12)}c_{22}d_{22}\nonumber \\
&&-(N_1^2-7N_1+6)V_{\bar{1}\bar{1}\bar{2}\bar{2}}^{(1)}c_{22}e_{22}-(N_2^2-7N_2+6)V_{1122}^{(2)}d_{22}e_{22}-V_{\bar{2}2\bar{1}1}^{(12)}=0,
\end{eqnarray}
which  accommodates explicitly the three unknown coefficients, $c_{22}$, $d_{22}$, and $e_{22}$.

 Before moving to the solution, let us briefly discuss what would be the general structure of the theory if one would use more virtual orbitals.  If we  simplify Eqs.~\ref{eqn_4.25}, ~\ref{eqn_4.26}, ~\ref{eqn_4.29}, ~\ref{eqn_4.31}, and ~\ref{eqn_4.33} for the higher orbital numbers, i.e., $M_1,M_2>2$, then we would obtain additional  coupled equations according to Eqs.~\ref{eqn_4.20} to ~\ref{eqn_4.24}. For example, when $M_1=M_2=3$, Eqs.~\ref{eqn_4.25}, ~\ref{eqn_4.26}, ~\ref{eqn_4.29}, ~\ref{eqn_4.31}, and ~\ref{eqn_4.33} boil down to two, two, three,  three, and four equations, respectively, and they are coupled to each other.  Also for obvious reasons, we require additional unknown coefficients, in this case, $c_3$, $d_3$, $c_{23}$, $d_{23}$, $e_{23}$, and $e_{32}$,  to determine  the energy.

In Eqs.~\ref{eqn_5.1} to ~\ref{eqn_5.3}, $\mu_1^{(1)}(\mu_1^{(2)})$ and $\mu_2^{(1)}(\mu_2^{(2)})$ are the respective chemical potentials of the ground and excited orbitals for species-1 (species-2), respectively. Given the orbitals $\varphi_1$, $\varphi_2$, $\psi_1$, and $\psi_2$ in Eq.~\ref{orbitals}, the chemical potentials can be determined analytically,
\begin{eqnarray}
\mu_1^{(1)}=\dfrac{{\Omega_1}}{2}+\dfrac{1}{2}\bigg(\dfrac{\Lambda_1}{{\Omega_1}}+\dfrac{\Lambda_{21}}{{\Omega_2}}\bigg),\nonumber \\
\mu_2^{(1)}=\dfrac{3{\Omega_1}}{2}+\dfrac{1}{2}\bigg(\dfrac{\Lambda_1}{{\Omega_1}}+\dfrac{\Lambda_{21}}{{\Omega_2}}\bigg),\nonumber \\
\mu_1^{(2)}=\dfrac{{\Omega_2}}{2}+\dfrac{1}{2}\bigg(\dfrac{\Lambda_2}{{\Omega_2}}+\dfrac{\Lambda_{12}}{{\Omega_1}}\bigg),\nonumber \\
\mu_2^{(2)}=\dfrac{3{\Omega_2}}{2}+\dfrac{1}{2}\bigg(\dfrac{\Lambda_2}{{\Omega_2}}+\dfrac{\Lambda_{12}}{{\Omega_1}}\bigg).
\end{eqnarray}
{Now, using Eq.~\ref{eqn_4.19},   one can  find  the correlation energy for the ground state  for $M_1=M_2=2$ as}

\begin{equation}\label{eqn_5.4}
E_{\text{cor}}=N_1(N_1-1)V_{\bar{1}\bar{1}\bar{2}\bar{2}}^{(1)}c_{22}+N_2(N_2-1)V_{1122}^{(2)}d_{22}+N_1N_2V_{\bar{1}1\bar{2}2}^{(12)} e_{22}.
\end{equation}
Here,   $c_s$ and $d_l$ are zero due to symmetry and, hence, do not contribute.  By solving the coupled Eqs.~\ref{eqn_5.1} to ~\ref{eqn_5.3}, we find the coefficients $c_{22}$, $d_{22}$, and $e_{22}$.  We would like to examine  the performance of the CCSD-M theory and,  thereby, we compare our results with analytical data found for the exact energy of a binary-species mixture.  The exact energy of the bosonic mixture of two species reads \cite{Alon2022} 

\begin{eqnarray}\label{eqn_5.5}
E_{\text{exact}}=\dfrac{1}{2}\Bigg[(N_1-1)\sqrt{\omega^2+2(N_1\lambda_1+N_2\lambda_{12})}&+&(N_2-1)\sqrt{\omega^2+2(N_2\lambda_2+N_1\lambda_{12})}\nonumber\\
&+&\sqrt{\omega^2+2(N_1+N_2)\lambda_{12}}+\omega\Bigg].
\end{eqnarray}
The mean-field energy for the binary-mixture is  found from the energy functional Eq.~\ref{eqn_3.1} and Gross-Pitaeveskii coupled equations  for mixture Eq.~\ref{eqn_3.3}, see also \cite{Alon2022}

\begin{eqnarray}\label{eqn_5.6}
E_{\text{MF}}&=&\dfrac{1}{2}\Bigg[N_1\Omega_1+N_2\Omega_2\Bigg].
\end{eqnarray}
Now, we show our numerical examples by calculating the  difference between the CCSD-M energy per particle and the corresponding analytical exact energy per particle, $\dfrac{E_{\text{cc}}-E_{\text{exact}}}{N}$, for various strengths of the intra- and inter-species interaction parameters. The  solutions of the three coupled equations, Eqs.~\ref{eqn_5.1} to ~\ref{eqn_5.3}, for  fixed inter- and intra-species interactions yield eight sets of results for $c_{22}$, $d_{22}$, and $e_{22}$, which eventually generate eight values of the ground state energy. The task is to determine the correct and accurate ground state energy from the eight sets of data. Among the eight sets of ground state energies, we notice that some values are complex and some are larger as well as  smaller compared to the mean-field energy. We can discard those energy values which are complex and,  {of course}, those which are larger compared to the mean-field energy,  as we expect any many-body approximation using the coupled-cluster theory to give us  lower  energy compared to the corresponding mean-field energy. Among the remaining energy values which are smaller compared to the mean-field energy, we observe that, apart from one value, the other energy values and their corresponding coefficients are fluctuating when we slowly change the interaction strength from repulsive to attractive. Finally, we notice that the correct energy value is the one when its respective coefficients ($c_{22}$ and $d_{22}$) have a monotonous feature with the increase of interaction strength and, moreover,   this particular energy value is the closest to the mean-field energy. See  Appendix E for the evaluation of the correct values of the coupled-cluster  coefficients [Fig.~\ref{FigA1}, ~\ref{FigA2}]  and energy [Fig.~\ref{FigA3}].
\begin{figure}[!h]
{\includegraphics[ scale=.28]{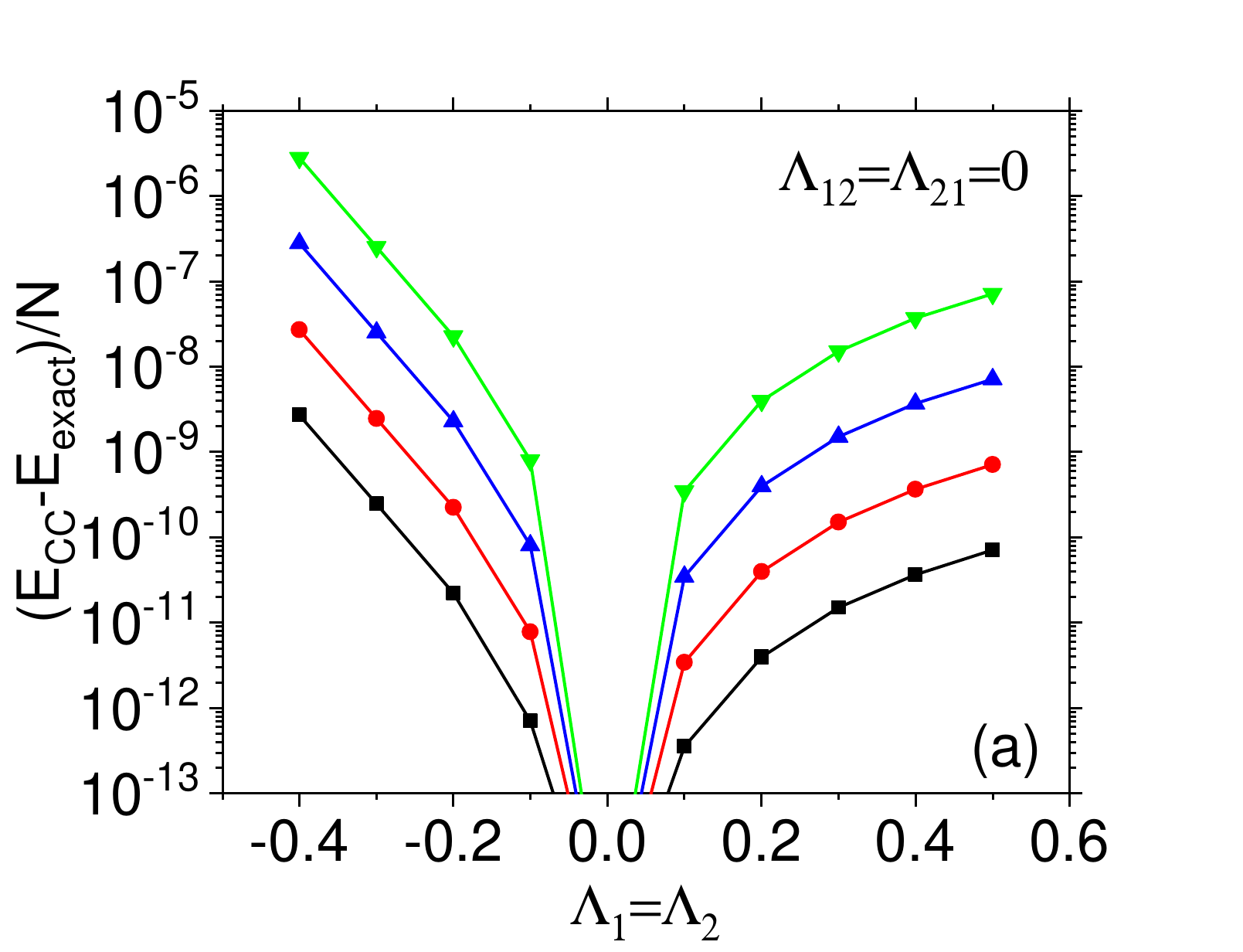}}
{\includegraphics[ scale=.28]{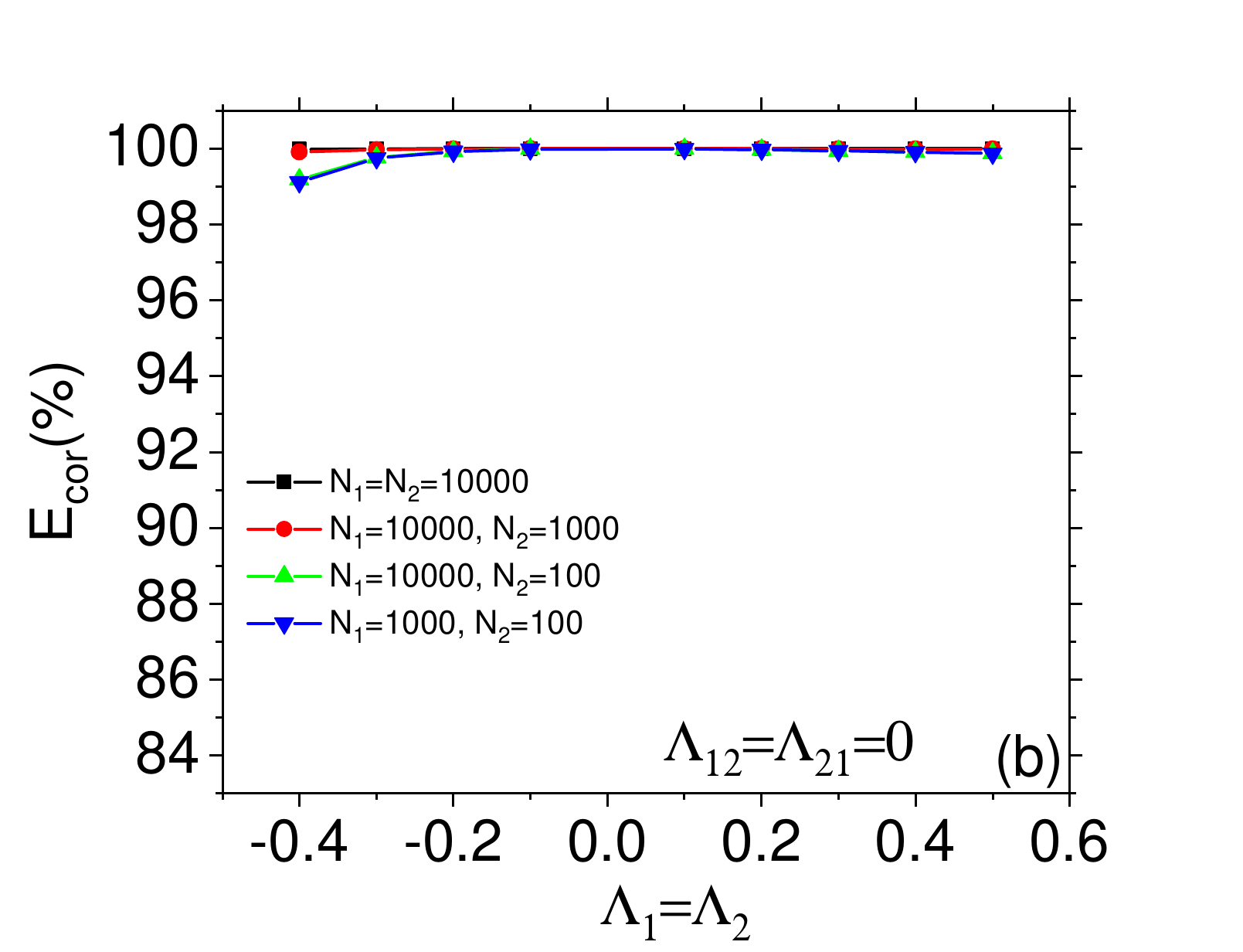}}\\
{\includegraphics[ scale=.28]{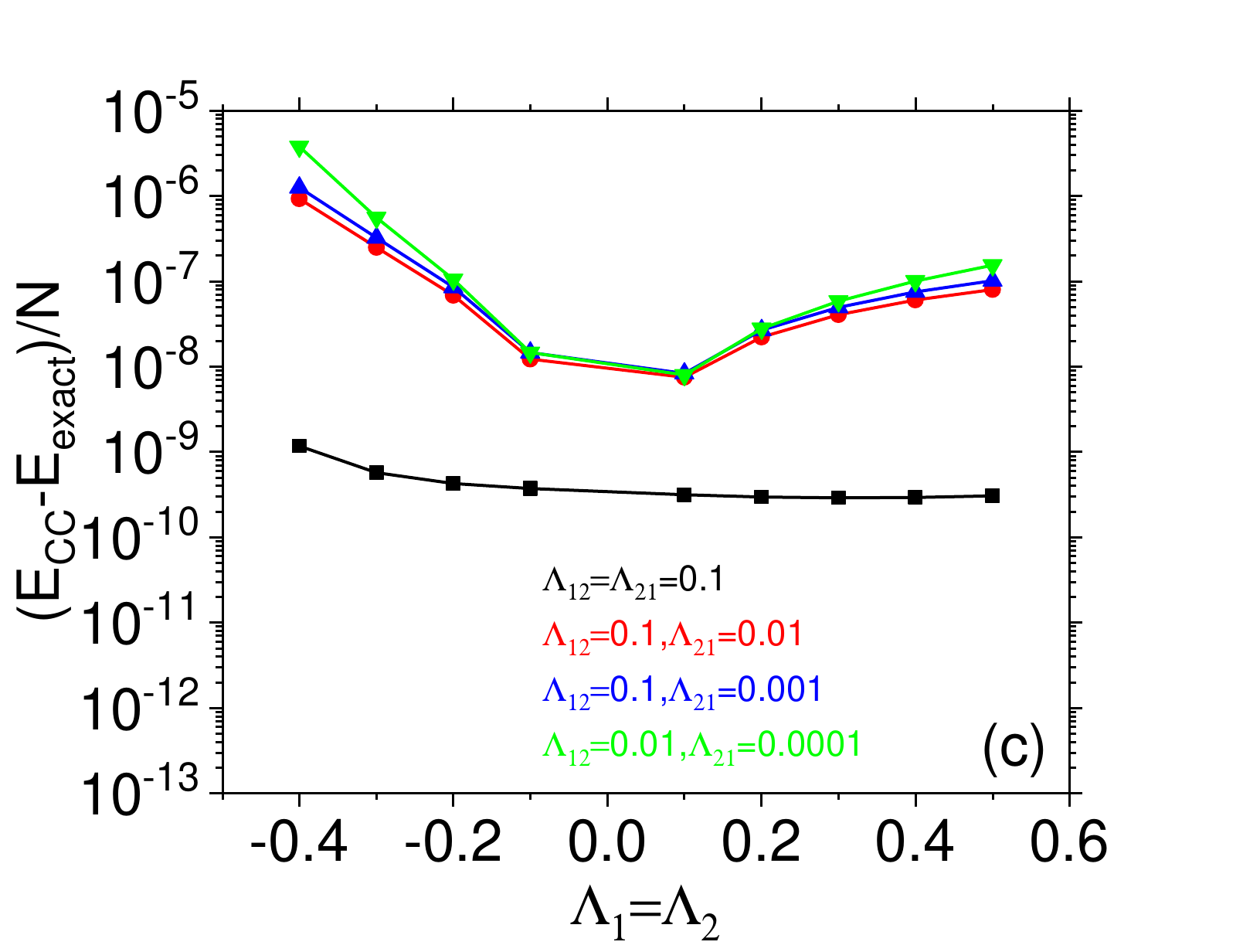}}
{\includegraphics[ scale=.28]{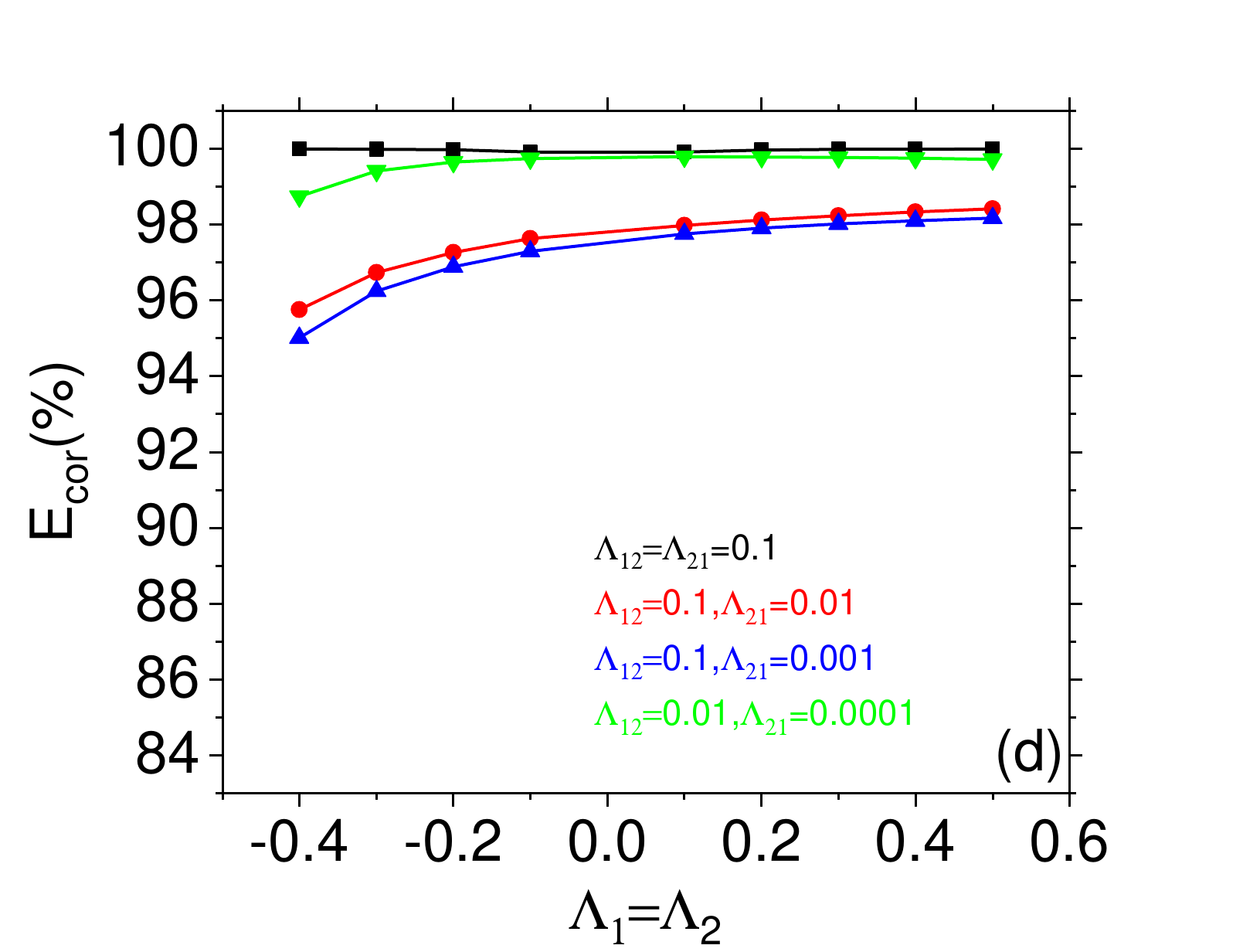}}\\
{\includegraphics[ scale=.28]{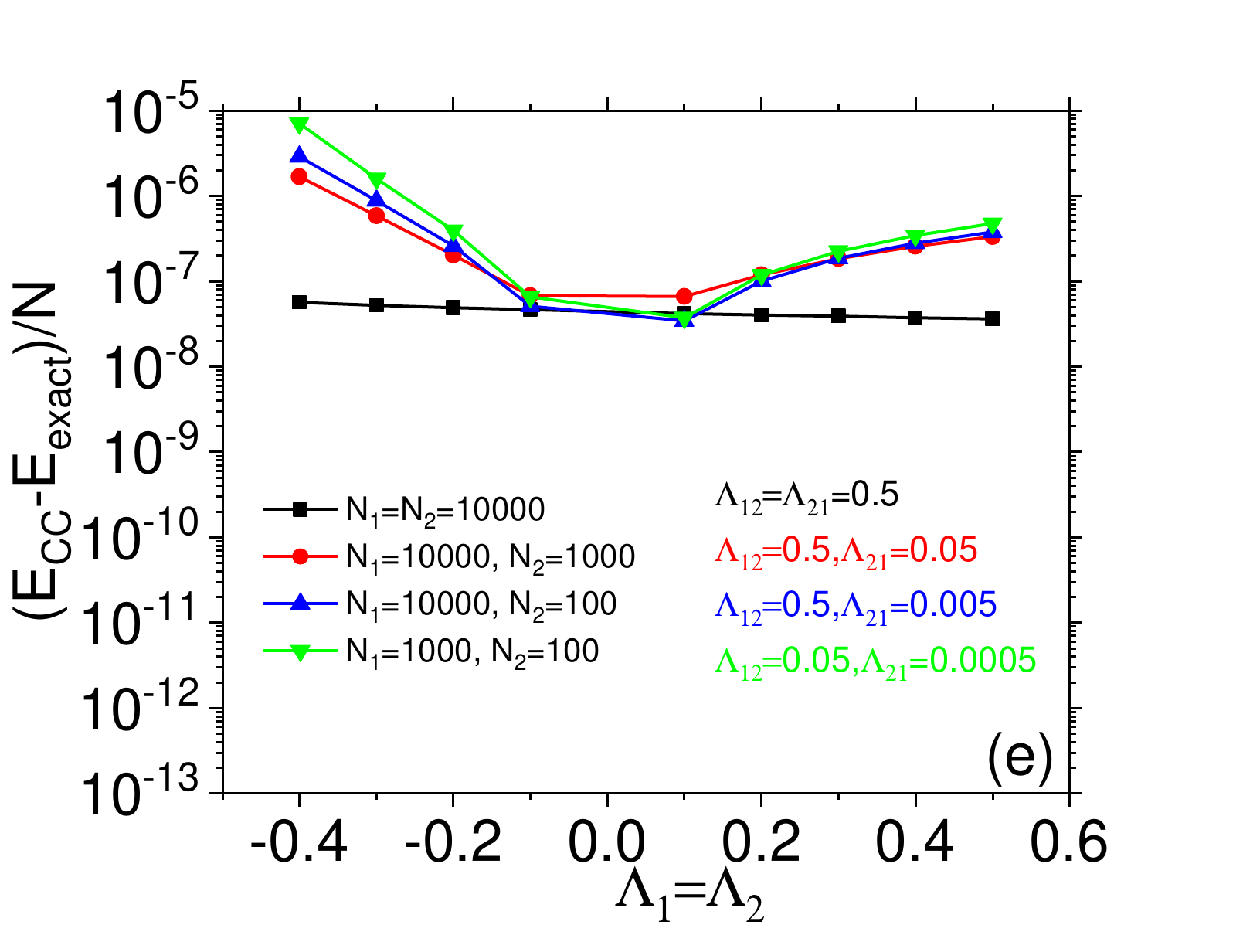}}
{\includegraphics[ scale=.28]{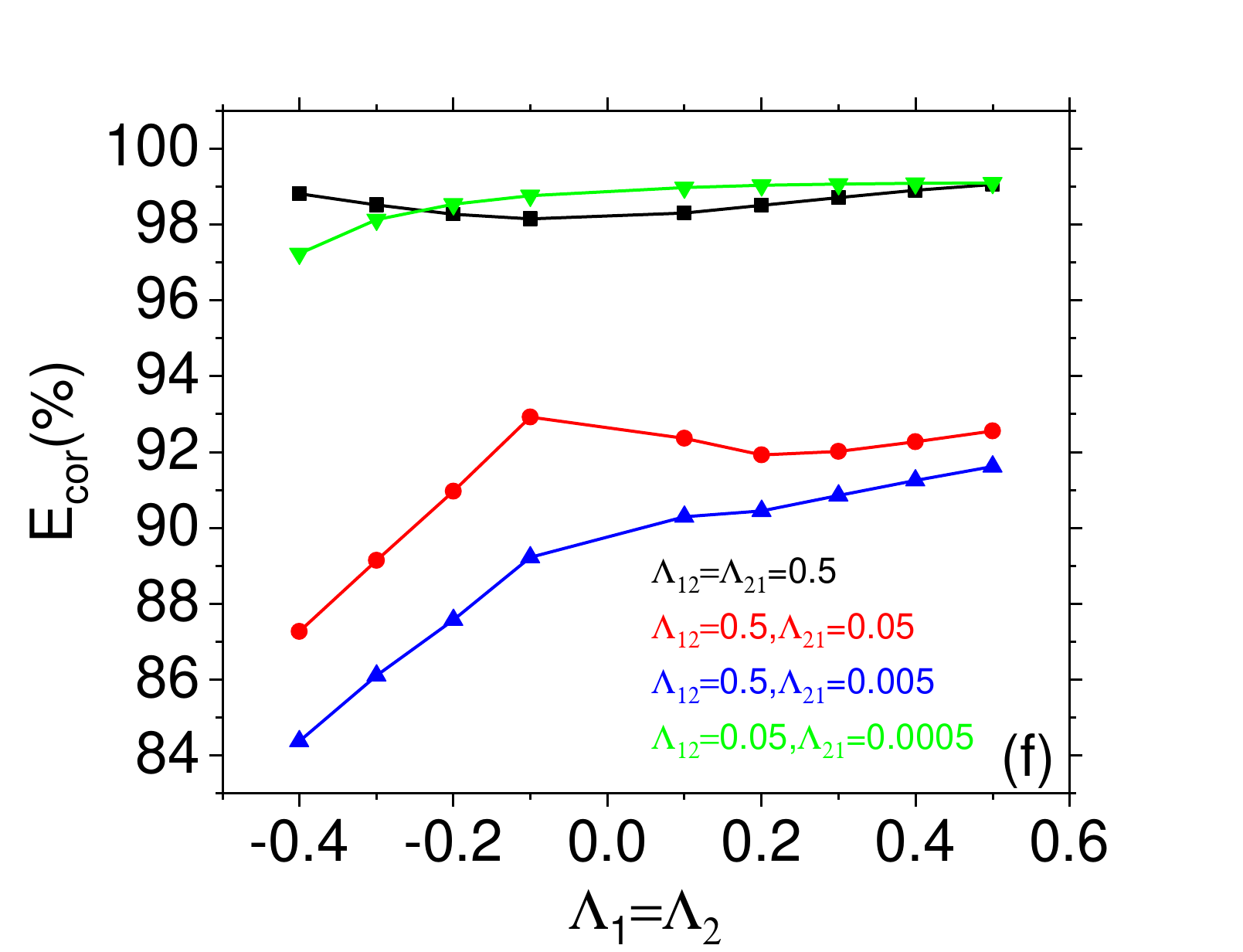}}\\
\caption{Variation of the difference between the coupled-cluster energy per particle and the corresponding analytical exact energy per particle, $\Delta_{\text{cc}}=\dfrac{E_{\text{cc}}-E_{\text{exact}}}{N}$,  with respect to the intra-species interaction parameters $\Lambda_1=\Lambda_2$ is shown in panels (a), (c), and (e). Panels (b), (d), and (f) present $E_{\text{cor}}(\%)=\dfrac{E_{\text{MF}}-E_{\text{cc}}}{E_{\text{MF}}-E_{\text{exact}}}\times 100$ as a function of the intra-species interaction parameters.  The coupled-cluster energy is calculated for the combinations of equal and unequal numbers of bosons of the two species: $N_1=N_2=10000$ {(black squares)}, $N_1=10000$ and $N_2=1000$  {(red circles)},  $N_1=10000$ and $N_2=100$ {(blue triangles)}, and $N_1=1000$ and $N_2=100$  {(green triangles)}. Captions are shown in panel (b) and (e). For panels in the first row, the two species are non-interacting to each other. Panels (a) and (b)  serve for benchmarking the coupled-cluster theory for bosonic single species. In panels (c) to (f), $\Lambda_{12}$ and $\Lambda_{21}$ are demonstrated in the same color as their corresponding curves.  Continuous curves are to guide the eye only.   All the quantities are dimensionless.  }
\label{Fig1}
\end{figure}

In this work we concentrate on the energy and we examine two features, the variation of the difference between the CCSD-M energy per particle and the corresponding analytical exact energy per particle, $\Delta_{\text{cc}}=\dfrac{E_{\text{cc}}-E_{\text{exact}}}{N}$ (left column of Fig.~\ref{Fig1})  and  the percentage of correlation energy, $E_{\text{cor}}(\%)=\dfrac{E_{\text{MF}}-E_{\text{cc}}}{E_{\text{MF}}-E_{\text{exact}}}\times 100$ (right column of Fig.~\ref{Fig1})  as a function of the intra-species interaction parameter.  The figure presents  the combinations of a balanced  number of bosons of the two species  $N_1=N_2=10000$, and imbalanced numbers of bosons, $N_1=10000$ and $N_2=1000$,  $N_1=10000$ and $N_2=100$, and $N_1=1000$ and $N_2=100$. The inter-species interaction parameters, $\Lambda_{12}$ and $\Lambda_{21}$, are displayed in each panel. Note that the left and right columns correspond to each other with the same numbers of bosons and interaction parameters. Let us define the strength of the intra- and inter-species interactions.  If  $\lambda_1$, $\lambda_2$, and $\lambda_{12}$  are $<0.01$ the system is referred to as weakly interacting, in between $0.011$ and $0.05$ defines a medium strength interaction,  and $>0.05$ is strong interaction. To remind the reader $\Lambda_1=\lambda_1(N_1-1)$, $\Lambda_2=\lambda_2(N_2-1)$, $\Lambda_{21}=\lambda_{12}N_2$, and $\Lambda_{12}=\lambda_{12}N_1$. Thus, the largest interaction parameter used in the examples below are:  $\Lambda_1=10^4$, $\Lambda_2=10^4$, $\Lambda_{12}=50$, and $\Lambda_{21}=50$.

 Before we are going to analyze the performance of the CCSD-M, let us take  two species which are non-interacting to each other, $\Lambda_{12}=\Lambda_{21}=0$,  and check the performance of the coupled cluster theory. For $\Lambda_{12}=\Lambda_{21}=0$, the coefficient $e_{22}=0$ and the performance of the coupled-cluster for mixtures boils down to the coupled-cluster for single species \cite{Cederbaum2006}, see Eq.~\ref{eqn_5.4} for correlation energy.  The benchmark presented here is very promising.

Fig.~\ref{Fig1} (a) and (b) present $\Delta_{\text{cc}}$ and $E_{\text{cor}}(\%)$, respectively,  for a basic situation  when the  two species are non-interacting to each other for several balanced and imbalanced combinations of boson numbers.  It is found that the deviation $\Delta_{\text{cc}}$ is always less than $10^{-5}$ with, obviously, no deviation   when $\Lambda_1=\Lambda_2=0$. Moreover, apart from $\Lambda_1=\Lambda_2=0$, for a fixed combination of bosons, balanced or imbalanced, the minimal $\Delta_{\text{cc}}$ occurs for the attractive inter-species interaction $\Lambda_1=\Lambda_2=+0.1$. Fig.~\ref{Fig1} (b) shows that, for  $\Lambda_{12}=\Lambda_{21}=0$, CCSD-M theory with two orbitals in each species can capture more than $99\%$ of the correlation energy, for the considered intra-species interaction parameters, which is very promising.

 Now, when we switch on the inter-species interaction parameters, see Figs.~\ref{Fig1}  (c) and (d), $\Delta_{\text{cc}}$ varies approximately from $10^{-5}$ to $10^{-9}$,  and the corresponding $E_{\text{cor}}(\%)$ deviates between $0.01\%$ to $5\%$ from the exact correlation energy. As we further increase of  $\Lambda_{12}$ and $\Lambda_{21}$ five times for each cases, $\Delta_{\text{cc}}$  is around $10^{-5}-10^{-7}$, see Fig.~\ref{Fig1} (e). Moreover, the coupled-cluster theory with $M_1=M_2=2$ can capture more than $84\%$ of correlation energy.   Naturally,  for strong inter-species interaction, one requires more orbitals to incorporate the accurate correlation energy. All in all, Fig.~\ref{Fig1} exhibits that  we require either an additional number of orbitals, or a higher order of excitations,  or the combination of both  to get the full correlation energy in the  particle imbalanced cases with repulsive intra-species interaction parameter.

\begin{figure}[!h]
{\includegraphics[ scale=.30]{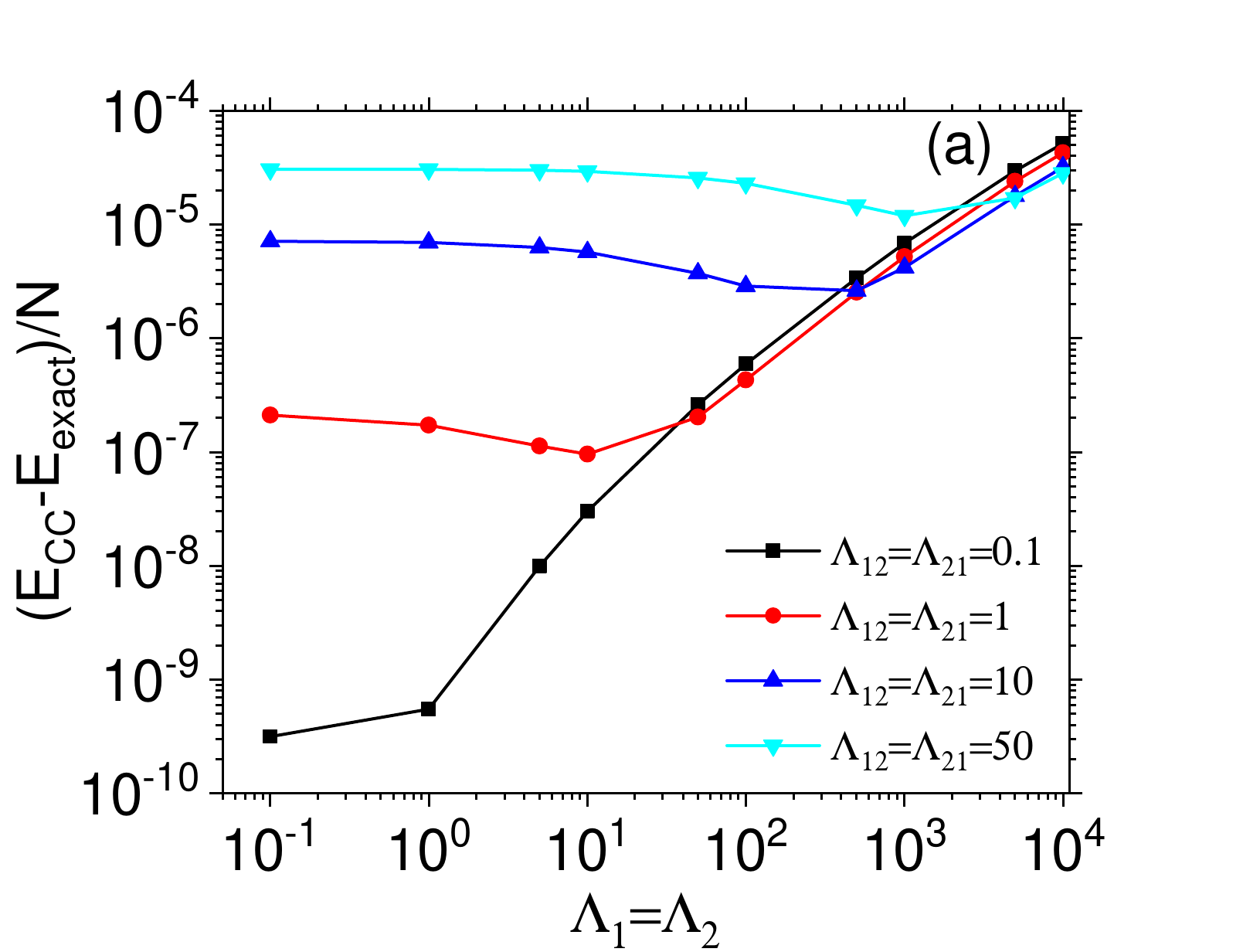}}
{\includegraphics[ scale=.30]{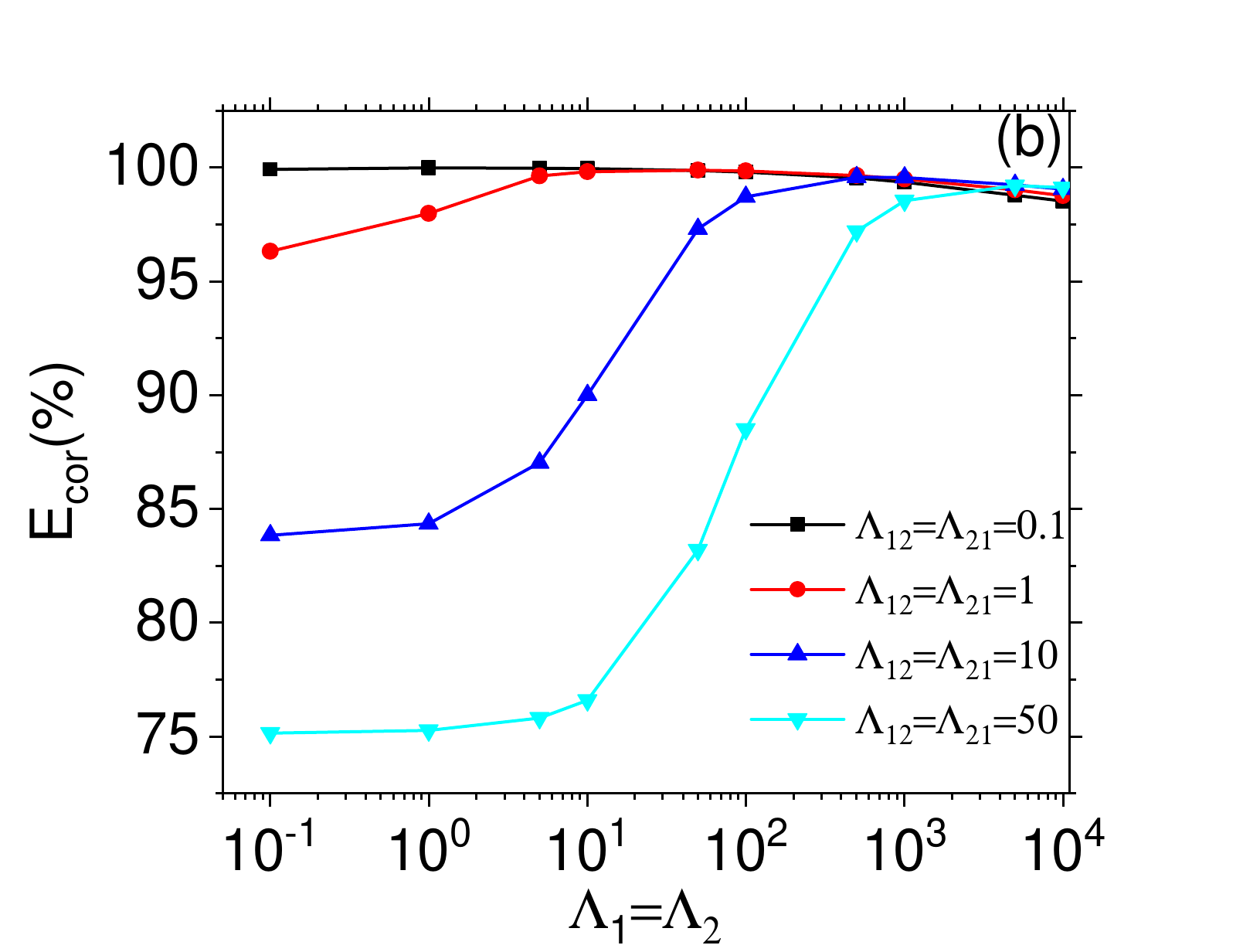}}
\caption{  Variation of  (a) the difference between the coupled-cluster energy per particle and the corresponding analytical exact energy per particle, $\Delta_{\text{cc}}=\dfrac{E_{\text{cc}}-E_{\text{exact}}}{N}$,  and (b) $E_{\text{cor}}(\%)=\dfrac{E_{\text{MF}}-E_{\text{cc}}}{E_{\text{MF}}-E_{\text{exact}}}\times 100$ as a function of intra-species interaction parameters $\Lambda_1=\Lambda_2$. The coupled-cluster energy is calculated for the  numbers of bosons  $N_1=N_2=10000$.  Results are presented  for the inter-species interaction parameters, $\Lambda_{12}$ and $\Lambda_{21}$,  0.1, 1, 10, and 50. It is found that if we increase both intra- and inter-species interaction parameters, CCSD-M can  capture higher fractions of the correlation energy.   All the quantities are dimensionless. }
\label{Fig2}
\end{figure}
Till now, the CCSD-M approach shows excellent success for weakly interacting bosonic mixtures. Here, we check the applicability  of the CCSD-M theory for a strongly interacting system with a large number of bosons in the two species. Fig.~\ref{Fig2}  exhibits (a) $\Delta_{\text{cc}}=\dfrac{E_{\text{cc}}-E_{\text{exact}}}{N}$ and (b) $E_{\text{cor}}(\%)$ for $N_1=N_2=10000$ with various inter-species interaction parameters  ranging from weak to  medium, namely, 0.1, 1, 10, and 50,   while the intra-species interaction parameters  vary from weak $10^{-2}$ to strong $10^4$.  For weak $\Lambda_{12}=\Lambda_{21}=0.1$, $\Delta_{\text{cc}}$ increases with the values of $\Lambda_1$ and $\Lambda_2$. Interestingly, for comparatively stronger values  $\Lambda_{12}=\Lambda_{21}=1$, $10$, and $50$, we observe that $\Delta_{\text{cc}}$ decreases at the beginning and then increases with the intra-species interaction parameter. This change in nature and the crossing of  $\Delta_{\text{cc}}$  curves correlates with the  $E_{\text{cc}}-E_{\text{MF}}$.  On the whole, we observe that $\Delta_{\text{cc}}<10^{-4}$ even for strong intra- and inter-species interaction strengths with a large number  {of} bosons in each species which proves the remarkable success of CCSD-M theory.

Now we discuss how much of the correlation energy is captured by the CCSD-M with the number of orbitals $M_1=M_2=2$, see Fig.~\ref{Fig2} (b). When the inter-species interaction is weak, $\Lambda_{12}=\Lambda_{21}=0.1$, the coupled-cluster theory can determine more than $98\%$ of the correlation energy for the weak to strong intra-species interaction parameters.  The correlation energy deviates from the exact results  as one increases the  inter-species interaction.  For  inter-species interaction  satisfying  $\Lambda_{12}=\Lambda_{21}\geq 1$, it is interesting to see that,  when the inter- and intra-species interactions are of the same order,  the CCSD-M theory with $M_1=M_2=2$ orbitals captures more than  about $85\%$ of correlation energy.  Moreover, we observe that for  strong values of the intra-species interactions  $\Lambda_1=\Lambda_2$, two orbitals for each species can produce more than $97.5\%$ of the correlation energy, which exhibits the potential of coupled-cluster theory for mixtures.

\section{Conclusions}\label{Section VI}

In this work, we present the theoretical development and
implementation details of the coupled-cluster theory for the bosonic mixture of binary species, with the numbers of bosons $N_1$ and $N_2$ for species-1 and species-2, respectively, in external trap potentials. In the coupled-cluster theory for mixtures, the ansatz of the many-body wavefunction is obtained when the three exponential cluster operators $e^{T^{(1)}}$,  $e^{T^{(2)}}$, and $e^{T^{(12)}}$, are applied onto   the ground configuration. Since $T^{(1)}$, $T^{(2)}$, and $T^{(12)}$ commute with each other,  their exponents can be separated. Here, $T^{(12)}=0$ implies that  there is no inter-species interaction between the two bosonic species and the theory derived here boils down to a single-species coupled-cluster ansatz for each of the species.   $T^{(1)}$ and $T^{(2)}$ incorporate the single, double, triple, ... excitations in  each species, while $T^{(12)}$ starts from the double excitations, one for each species.   As per the bosons statistics,  there is no restriction in occupying a particular orbital for bosons.  Our starting point for building up correlations is the standard mean-field for which $N_1$ bosons occupy one orbital of species-1 and $N_2$ bosons are sitting in another orbital of species-2. These orbitals are obtained by the solution of the coupled Gross-Pitaeveskii equations of the mixtures.

Next, we have derived the involved  working  equations for the  unknown coefficients in the coupled-cluster theory for  bosonic mixtures using an arbitrary sets of orthonormal orbitals.    Also, the comparatively simplified version of the working equations are derived for the  coefficients  using Fock-like operators.  The  working equations with Fock-like operators as well as those for an arbitrary sets of orthonormal orbitals consist of $M_1-1$ and $M_2-1$ equations for single excitations in species-1 and species-2, respectively, $M_1(M_1-1)/2$ and $M_2(M_2-1)/2$ equations for double excitations in species-1 and species-2, respectively, and $(M_1-1)(M_2-1)$ equations for simultaneous excitations in the two species, and so on. Utilizing orthonormal orbitals, we find that the correlation energy for the mixture depends on the  different coefficients, namely, $\{c_s\}$, $\{d_l\}$, $\{c_{rs}\}$, $\{d_{kl}\}$, and $\{e_{ls}\}$, which originate for the single and double excitations only.

Furthermore, we have implemented our developed coupled-cluster theory on the harmonic interaction model for mixtures and compared the results with this exactly solvable many-body model for the different strengths of the inter-species and intra-species interactions. The investigation serves to check the theory, implementation, and the usage of the theory, that is the truncation to CCSD-M and the inclusion of $M_1=M_2=2$, for the studied examples.  We have calculated  the energy for the mixture with  $N_1$ and $N_2$ ranging  from 100 to 10000 bosons, and  {in various scenarios for balanced and imbalanced numbers  of bosons}  between the two species.  To check the performance of our theory, we calculated the difference between the coupled-cluster energy per particle and the corresponding analytical exact energy per particle, $\Delta_{\text{cc}}=\dfrac{E_{\text{cc}}-E_{\text{exact}}}{N}$.  We have shown  how much of the exact correlation energy can be captured by the coupled-cluster theory by calculating $E_{\text{cor}}(\%)=\dfrac{E_{\text{MF}}-E_{\text{cc}}}{E_{\text{MF}}-E_{\text{exact}}}\times 100$. We found that even for rather strong intra- and inter-species interactions  and relatively large numbers of bosons for  each species $N_1=N_2=10^4$,  the CCSD-M provides remarkable success.  

 {Before ending this section, it is worthwhile to mention the computational scaling of the CCSD-M approach in comparison to the configuration interaction which employs a basis set expansion \cite{Haugset1998}. For example, the largest system we have in our work is 10000 bosons in  each species which produces  $10001^2 \approx 10^8$ coefficients for configuration interaction with two orbitals per species. While, there are only three coefficients for CCSD-M. For more number of bosons, the number of  coefficients increases for configuration interaction accordingly. This analysis is  impressive in terms of saving computational resources with respect to configuration interaction and it  shows the excellent  applicability of CCSD-M approach.   }  The quality of coupled-cluster theory opens the way to investigate few- to many-boson binary mixture up to fairly  strong interaction strengths  where several orbitals,  or higher order excitations,  or both of them are required to describe the physics accurately. 

All in all, it is found that the  coupled-cluster theory for bosonic mixtures is a promising many-body approach. As an outlook, one could  be interested to investigate various properties of bosonic mixtures for other external potentials and inter-bosons interactions \cite{Cikojevic2018}.  {For instance, one can compute the reduced one-particle density matrix of each species from the CC-M wavefunction, and from them the respective eigenvalues. This will tell one how the condensate fraction of each species is modified by the inter-species and intra-species interactions. Also, it would be interesting to extract from the CC-M wavefunction the entanglement between species-1 and species-2. Examining the CC-M ansatz 
Eqs.~\ref{eqn_2.5} and \ref{eqn_2.6}, it is clear that the cluster operator $T^{(12)}$ alone governs the entanglement between the two species.  Moreover, one can determine the expectation value of an operator, $\Omega$, using the so-called $\Lambda$-equation, $\langle\Omega\rangle=\langle\phi_0|(1+\Lambda)e^{-T}\Omega e^T|\phi_0\rangle$, where $\Lambda$ is the Lagrange multiplier \cite{Matthews2020}. For that, the equation of motion for $\Lambda$  would have to be derived.}  Based on our CCSD-M, one can anticipate the development of a time-dependent coupled-cluster theory for bosons first, and then one could expect to expand and explore the dynamics of bosonic mixtures using time-dependent coupled-cluster.  This is a challenge to be undertaken in future research. We believe that fermionic time-dependent coupled-cluster theory will be helpful in this direction \cite{Huber2011,Ofstad2023}.

\section*{Acknowledgement}

This research was supported by the Israel Science Foundation (Grant No.  1516/19).

\section*{Data Availability}
The data that support the findings of this work are available from the corresponding author upon reasonable request.
\clearpage

\section*{Appendix}
\subsection*{A. Relations between one-body and two-body terms}\label{Appendix A}
\setcounter{equation}{0}
\renewcommand{\theequation}{A.\arabic{equation}}
Using the eigenvalue-equations of the Fock-like operators, ${F}_1\varphi_i= \mu_{i}^{(1)}\varphi_i$ and  ${F}_2\psi_i= \mu_{i}^{(2)}\psi_i$, described in the main text [Eqs.~\ref{eqn_3.4} and ~\ref{eqn_3.5} ], one can find a few relations between one-body and two-body terms which can assist in simplifying the working equations of the coupled-cluster theory for bosonic mixtures.  The relations are as follows

\begin{eqnarray}\label{eqn_A.1}
&&h_{\bar{1}\bar{1}}^{(1)}+(N_1-1)V_{\bar{1}\bar{1}\bar{1}\bar{1}}^{(1)}+N_2V_{\bar{1}1\bar{1}1}^{(12)}=\mu_1^{(1)},\nonumber \\
&&h_{11}^{(2)}+(N_2-1)V_{1111}^{(2)}+N_1V_{\bar{1}1\bar{1}1}^{(12)}=\mu_1^{(2)},\nonumber \\
&&\sum_{s=2}^{M_1}\Big[h_{\bar{s}\bar{s}}^{(1)}+(N_1-1)V_{\bar{s}\bar{1}\bar{s}\bar{1}}^{(1)}+N_2V_{\bar{s}1\bar{s}1}^{(12)}\Big]=\mu_s^{(1)},\nonumber \\
&&\sum_{l=2}^{M_2}\Big[h_{ll}^{(2)}+(N_2-1)V_{l1l1}^{(2)}+N_1V_{\bar{1}l\bar{1}l}^{(12)}\Big]=\mu_l^{(2)},\nonumber \\
&&\sum_{s=2}^{M_1}\Big[h_{\bar{1}\bar{s}}^{(1)}+(N_1-1)V_{\bar{1}\bar{1}\bar{s}\bar{1}}^{(1)}+N_2V_{\bar{1}1\bar{s}1}^{(12)}\Big]=0,\nonumber \\
&&\sum_{l=2}^{M_2}\Big[h_{1l}^{(2)}+(N_2-1)V_{11l1}^{(2)}+N_1V_{\bar{1}1\bar{1}l}^{(12)}\Big]=0,\nonumber \\
&&\sum_{s=2}^{M_1}\Big[h_{\bar{s}\bar{1}}^{(1)}+(N_1-1)V_{\bar{s}\bar{1}\bar{1}\bar{1}}^{(1)}+N_2V_{\bar{s}1\bar{1}1}^{(12)}\Big]=0,\nonumber \\
&&\sum_{l=2}^{M_2}\Big[h_{l1}^{(2)}+(N_2-1)V_{l111}^{(2)}+N_1V_{\bar{1}l\bar{1}1}^{(12)}\Big]=0.
\end{eqnarray}
The first two relations are obtained when the ground orbital is sandwiching in Eq.~\ref{eqn_3.3}  the Fock operator. Similarly, the third and fourth relations are found when the same excited orbitals act on the Fock operator from the left and right sides. We get  the other relations when ground and excited orbitals  are  sandwiching the Fock operator in Eq.~\ref{eqn_3.5}. The first four relations involve chemical potentials for the ground and virtual orbitals as the Hamiltonian is sandwiched between the same two orbitals.

\subsection*{B. Transformed creation and destruction operators}\label{Appendix B}
The derivation of the working equations of the coupled-cluster theory for bosonic  mixtures demands the transformation of the creation operators of the ground orbitals, $\dot{a}_{{1}}^\dagger=a_{{1}}^\dagger-\mathcal{K}_{{1}}$ and  $\dot{b}_1^\dagger=b_1^\dagger-\mathcal{L}_1$, and of the  destruction operators  of the virtual orbitals, $\dot{a}_{p}=a_{p}-\mathcal{K}_{p}$ and $\dot{b}_i=b_i-\mathcal{L}_i$, for both species.  The explicit expansions of the first few terms of $\mathcal{K}_{{1}}$, $\mathcal{K}_{{p}}$, $\mathcal{L}_1$, and $\mathcal{L}_i$ read

\setcounter{equation}{0}
\renewcommand{\theequation}{B.\arabic{equation}}
\begin{eqnarray}
\mathcal{K}_{{1}}&=&\sum_{p=2}^{M_1}c_{p} a_{p}^\dagger+2\sum_{p,q=2}^{M_1}c_{pq}a_{p}^\dagger a_{q}^\dagger a_{1}  +3\sum_{p,q,r=2}^{M_1}c_{pqr}a_{p}^\dagger a_{q}^\dagger a_{r}^\dagger a_{1}^2+...\nonumber \\&+&\sum_{p=2}^{M_1}\sum_{i=2}^{M_2}e_{pi}a_p^\dagger b_i^\dagger b_1+2\sum_{p,q=2}^{M_1}\sum_{i=2}^{M_2}e_{pqi}a_p^\dagger a_q^\dagger b_i^\dagger a_1 b_1+\sum_{p=2}^{M_1}\sum_{i,j=2}^{M_2}e_{pij}a_p^\dagger b_i^\dagger b_j^\dagger b_1^2+...,
\end{eqnarray}

\begin{eqnarray}
\mathcal{K}_{p}&=&c_{p} a_{1}+2\sum_{q=2}^{M_1}c_{pq}a_{q}^\dagger  a_{1}^2+3\sum_{q,r=2}^{M_1}c_{pqr}a_{q}^\dagger a_{r}^\dagger  a_{1}^3+... \nonumber \\ &+& \sum_{i=2}^{M_2}e_{pi}b_i^\dagger a_1 b_1+2\sum_{q=2}^{M_1}\sum_{i=2}^{M_2}e_{pqi}a_q^\dagger b_i^\dagger a_1^2 b_1+\sum_{i,j=2}^{M_2}e_{pij}b_i^\dagger b_j^\dagger a_1 b_1^2+...,
\end{eqnarray}

\begin{eqnarray}
\mathcal{L}_1&=&\sum_{i=2}^{M_2}d_i b_i^\dagger+2\sum_{i,j=2}^{M_2}d_{ij}b_i^\dagger b_j^\dagger b_1 +3\sum_{i,j,k=2}^{M_2}d_{ijk}b_i^\dagger b_j^\dagger b_k^\dagger b_1^2+...\nonumber \\
&+&\sum_{p=2}^{M_1}\sum_{i=2}^{M_2}e_{pi}a_{p}^\dagger b_i^\dagger a_1+2\sum_{p=2}^{M_1}\sum_{i,j=2}^{M_2}e_{pij}a_{p}^\dagger b_i^\dagger b_j^\dagger a_1 b_1+\sum_{p,q=2}^{M_1}\sum_{i=2}^{M_2}e_{pqi}a_{p}^\dagger a_{q}^\dagger b_i^\dagger a_1^2+...,
\end{eqnarray}

\begin{eqnarray}
\mathcal{L}_i&=&d_i b_1+2\sum_{j=2}^{M_2}d_{ij}b_j^\dagger  b_1^2+3\sum_{j,k=2}^{M_2}d_{ijk}b_j^\dagger b_k^\dagger  b_1^3+... \nonumber \\ &+& \sum_{p=2}^{M_1}e_{pi}a_p^\dagger a_1 b_1+2\sum_{p=2}^{M_1}\sum_{j=2}^{M_2} e_{pij}a_p^\dagger b_j^\dagger a_1 b_1^2+\sum_{p,q=2}^{M_1}e_{pqi}a_p^\dagger a_q^\dagger a_1^2 b_1+....
\end{eqnarray}
The above mentioned expansions are used to   find out the unknown coefficients, $c_{p_1 p_2...p_n}$, $d_{i_1 i_2...i_m}$, and $e_{p_1...p_{n^\prime}i_1...i_{m^\prime}}$.

\subsection*{C. Transformed two-body operators}\label{Appendix C}

The transformed intra-species and inter-species two-body operators $\dot{V}^{(1)}$, $\dot{V}^{(2)}$, and $\dot{V}^{(12)}$  are required to solve  the coupled-cluster energy, see Sec. IVB, by finding out the unknown coefficients. $\dot{V}^{(1)}$ consists of nine terms listed below when each term is a combination of the creation and destruction operators of species-1. The nine terms are as follows 
\setcounter{equation}{0}
\renewcommand{\theequation}{C.\arabic{equation}}

\begin{eqnarray}\label{Potential_1}
&&\dot{V}^{(1)}=\sum_{w=1}^9 \dot{V}^{(1)}(w),\nonumber\\
&& \dot{V}^{(1)}(1)=\dfrac{1}{2} V_{\bar{1}\bar{1}\bar{1}\bar{1}}^{(1)}\dot{a}_{1}^\dagger \dot{a}_{1}^\dagger a_{1} a_{1}, \hspace{1.7cm}\dot{V}^{(1)}(2)= \sum_{s=2}^{M_1}V_{\bar{1}\bar{1}\bar{1}\bar{s}}^{(1)}\dot{a}_{1}^\dagger \dot{a}_{1}^\dagger a_{1} \dot{a}_{s}, \nonumber\\
&& \dot{V}^{(1)}(3)= \sum_{s=2}^{M_1}V_{\bar{s}\bar{1}\bar{1}\bar{1}}^{(1)}a_{s}^\dagger \dot{a}_{1}^\dagger a_{1} a_{1}, \hspace{1.45cm} \dot{V}^{(1)}(4)= \dfrac{1}{2}\sum_{r,s=2}^{M_1}V_{\bar{1}\bar{1}\bar{r}\bar{s}}^{(1)}\dot{a}_{1}^\dagger \dot{a}_{1}^\dagger \dot{a}_{r} \dot{a}_{s}, \nonumber\\
&& \dot{V}^{(1)}(5)= \dfrac{1}{2}\sum_{r,s=2}^{M_1}V_{\bar{r}\bar{s}\bar{1}\bar{1}}^{(1)}a_{r}^\dagger a_{s}^\dagger a_{1} a_{1}, \hspace{1cm} \dot{V}^{(1)}(6)= \sum_{r,s=2}^{M_1}(V_{\bar{1}\bar{r}\bar{1}\bar{s}}^{(1)}+V_{\bar{1}\bar{r}\bar{s}\bar{1}}^{(1)})\dot{a}_{1}^\dagger a_{r}^\dagger a_{1} \dot{a}_{s},\nonumber\\
&& \dot{V}^{(1)}(7)= \sum_{q,r,s=2}^{M_1}V_{\bar{1}\bar{q}\bar{r}\bar{s}}^{(1)}\dot{a}_{1}^\dagger a_{q}^\dagger \dot{a}_{r} \dot{a}_{s}, \hspace{1.25cm}\dot{V}^{(1)}(8)= \sum_{q,r,s=2}^{M_1}V_{\bar{q}\bar{r}\bar{s}\bar{1}}^{(1)}a_{q}^\dagger a_{r}^\dagger \dot{a}_{s} a_{{1}},\nonumber\\
&& \dot{V}^{(1)}(9)= \dfrac{1}{2}\sum_{p,q,r,s=2}^{M_1}V_{\bar{p}\bar{q}\bar{r}\bar{s}}^{(1)}a_{p}^\dagger a_{q}^\dagger \dot{a}_{r} \dot{a}_{s}.
\end{eqnarray}
Similarly,  $\dot{V}^{(2)}$ has nine terms and each term carries the creation and destruction operators of species-2. The nine terms read

\begin{eqnarray}\label{Potential_2}
&&\dot{V}^{(2)}=\sum_{w=1}^9 \dot{V}^{(2)}(w),\nonumber\\
&& \dot{V}^{(2)}(1)=\dfrac{1}{2} V_{1111}^{(2)}\dot{b}_1^\dagger \dot{b}_1^\dagger b_1 b_1,  \hspace{1.65cm}\dot{V}^{(2)}(2)= \sum_{l=2}^{M_2}V_{111l}^{(2)}\dot{b}_1^\dagger \dot{b}_1^\dagger b_1 \dot{b}_l,\nonumber\\
&& \dot{V}^{(2)}(3)= \sum_{l=2}^{M_2}V_{l111}^{(2)}b_l^\dagger \dot{b}_1^\dagger b_1 b_1,\hspace{1.5cm} \dot{V}^{(2)}(4)= \dfrac{1}{2}\sum_{k,l=2}^{M_2}V_{11kl}^{(2)}\dot{b}_1^\dagger \dot{b}_1^\dagger \dot{b}_k \dot{b}_l,\nonumber\\
&& \dot{V}^{(2)}(5)= \dfrac{1}{2}\sum_{k,l=2}^{M_2}V_{kl11}^{(2)}b_k^\dagger b_l^\dagger b_1 b_1, \hspace{1cm} \dot{V}^{(2)}(6)= \sum_{k,l=2}^{M_2}(V_{1k1l}^{(2)}+V_{1kl1}^{(2)})\dot{b}_1^\dagger b_k^\dagger b_1 \dot{b}_l, \nonumber\\
&& \dot{V}^{(2)}(7)= \sum_{j,k,l=2}^{M_2}V_{1jkl}^{(2)}\dot{b}_1^\dagger b_j^\dagger \dot{b}_k \dot{b}_l, \hspace{1.3cm} \dot{V}^{(2)}(8)= \sum_{j,k,l=2}^{M_2}V_{jkl1}^{(2)}b_j^\dagger b_k^\dagger \dot{b}_l b_1,\nonumber\\
&& \dot{V}^{(2)}(9)= \dfrac{1}{2}\sum_{i,j,k,l=2}^{M_2}V_{ijkl}^{(2)}b_i^\dagger b_j^\dagger \dot{b}_k \dot{b}_l.
\end{eqnarray}
$\dot{V}^{(12)}$ is made of sixteen terms and each of the terms is a mixture of the creation and destruction operators of species-1 and species-2. All sixteen  terms can be readily found as

\begin{eqnarray}\label{Potential_12}
&&\dot{V}^{(12)}=\sum_{w=1}^{16} \dot{V}^{(12)}(w),\nonumber\\
&& \dot{V}^{(12)}(1)= V_{\bar{1}1\bar{1}1}^{(12)}\dot{a}_{1}^\dagger \dot{b}_1^\dagger a_{1} b_1, \hspace{2.2cm} \dot{V}^{(12)}(2)= \sum_{l=2}^{M_2}V_{\bar{1}1\bar{1}l}^{(12)}\dot{a}_{1}^\dagger \dot{b}_1^\dagger a_{1} \dot{b}_l,\nonumber\\
&& \dot{V}^{(12)}(3)= \sum_{s=2}^{M_1}V_{\bar{1}1\bar{s}1}^{(12)}\dot{a}_{1}^\dagger \dot{b}_1^\dagger \dot{a}_{s} b_1, \hspace{1.65cm}\dot{V}^{(12)}(4)= \sum_{l=2}^{M_2}V_{\bar{1}l\bar{1}1}^{(12)}\dot{a}_{1}^\dagger b_l^\dagger a_{1}  b_1,\nonumber\\
&& \dot{V}^{(12)}(5)= \sum_{s=2}^{M_1}V_{\bar{s}1\bar{1}1}^{(12)}a_{s}^\dagger \dot{b}_1^\dagger a_{1} b_1, \hspace{1.65cm} \dot{V}^{(12)}(6)= \sum_{s=2}^{M_1}\sum_{l=2}^{M_2}V_{\bar{1}1\bar{s}l}^{(12)}\dot{a}_{1}^\dagger \dot{b}_1^\dagger \dot{a}_{s} \dot{b}_l,\nonumber\\
&& \dot{V}^{(12)}(7)= \sum_{s=2}^{M_1}\sum_{l=2}^{M_2} V_{\bar{s}l\bar{1}1}^{(12)} a_{s}^\dagger b_l^\dagger a_{1}  b_1,  \hspace{1.1cm}\dot{V}^{(12)}(8)= \sum_{r,s=2}^{M_1}V_{\bar{r}1\bar{s}1}^{(12)}a_{r}^\dagger \dot{b}_1^\dagger \dot{a}_{s}  b_1,\nonumber\\
&& \dot{V}^{(12)}(9)= \sum_{k,l=2}^{M_2}V_{\bar{1}k\bar{1}l}^{(12)}\dot{a}_{1}^\dagger b_k^\dagger  a_{1} \dot{b}_l, \hspace{1.65cm} \dot{V}^{(12)}(10)= \sum_{s=2}^{M_1}\sum_{l=2}^{M_2}V_{\bar{s}1\bar{1}l}^{(12)} a_{s}^\dagger \dot{b}_1^\dagger  a_{1} \dot{b}_l, \nonumber\\
&& \dot{V}^{(12)}(11)= \sum_{s=2}^{M_1}\sum_{l=2}^{M_2}V_{\bar{1}l\bar{s}1}^{(12)} \dot{a}_{1}^\dagger b_l^\dagger  \dot{a}_{s} b_1, \hspace{1cm} \dot{V}^{(12)}(12)= \sum_{s=2}^{M_1}\sum_{k,l=2}^{M_2}V_{\bar{1}k\bar{s}l}^{(12)} \dot{a}_{1}^\dagger b_k^\dagger  \dot{a}_{s} \dot{b}_l, \nonumber\\
&& \dot{V}^{(12)}(13)= \sum_{r,s=2}^{M_1}\sum_{l=2}^{M_2}V_{\bar{r}1\bar{s}l}^{(12)} a_{r}^\dagger \dot{b}_1^\dagger  \dot{a}_{s} \dot{b}_l, \hspace{1cm} \dot{V}^{(12)}(14)= \sum_{s=2}^{M_1}\sum_{k,l=2}^{M_2}V_{\bar{s}k\bar{1}l}^{(12)} a_{s}^\dagger  b_k^\dagger  a_{1} \dot{b}_l,\nonumber\\
&& \dot{V}^{(12)}(15)= \sum_{r,s=2}^{M_1}\sum_{l=2}^{M_2}V_{\bar{r}l\bar{s}1}^{(12)} a_{r}^\dagger  b_l^\dagger  \dot{a}_{s}  b_1, \hspace{1cm} \dot{V}^{(12)}(16)= \sum_{r,s=2}^{M_1}\sum_{k,l=2}^{M_2}V_{\bar{r}k\bar{s}l}^{(12)} a_{r}^\dagger  b_k^\dagger  \dot{a}_{s}  \dot{b}_l. 
\end{eqnarray}
The working equations of the coupled-cluster theory for bosonic mixtures are developed using the transformed two-body operators along with $\dot{H}_0$, see Eq.~\ref{eqn_4.12}.

\clearpage

\subsection*{D. The general  working equations}\label{Appendix D}
Here we present the general forms of the working equations of the coupled-cluster theory for bosonic mixtures. The derivation of the
coupled working equations is very lengthy and involved. We start with the single excitation in both species by solving the equations  $\langle \phi_0|a_{1}^\dagger a_{\bar{i}}\dot{H}|\phi_0\rangle=0$ and $\langle \phi_0|b_1^\dagger b_i\dot{H}|\phi_0\rangle=0$  where  $\bar{i}=2,3,...,M_1$  and  $i=2,3,...,M_2$. The general form of the single excitation in  species-1 can be readily obtained as

\setcounter{equation}{0}
\renewcommand{\theequation}{D.\arabic{equation}}

\begin{eqnarray}\label{general_1}
0&=&-h_{\bar{1}\bar{1}}^{(1)}c_{\bar{i}}+(N_1-1)\sum_{s=2}^{M_1}h_{\bar{1}\bar{s}}^{(1)}(2c_{s\bar{i}})-\sum_{s=2}^{M_1}h_{\bar{1}\bar{s}}^{(1)} (c_{\bar{i}}c_{s})+h_{\bar{i}\bar{1}}^{(1)}+\sum_{s=2}^{M_1}h_{\bar{i}\bar{s}}^{(1)}c_{s}+N_2\sum_{l=2}^{M_2}h_{1l}^{(2)}e_{\bar{i}l}\nonumber \\
&&-(N_1-1)V_{\bar{1}\bar{1}\bar{1}\bar{1}}^{(1)}c_{\bar{i}}+(N_1-1)(N_1-2)\sum_{s=2}^{M_1}V_{\bar{1}\bar{1}\bar{1}\bar{s}}^{(1)}(2c_{s\bar{i}})-2(N_1-1)\sum_{s=2}^{M_1}V_{\bar{1}\bar{1}\bar{1}\bar{s}}^{(1)}(c_{\bar{i}}c_{s})+(N_1-1)V_{\bar{i}\bar{1}\bar{1}\bar{1}}^{(1)}\nonumber \\
&&+\sum_{r,s=2}^{M_1}V_{\bar{1}\bar{1}\bar{r}\bar{s}}^{(1)}\bar{\alpha}_{rs\bar{i}}+(N_1-1)\Bigg[\sum_{s=2}^{M_1}(V_{\bar{1}\bar{i}\bar{1}\bar{s}}^{(1)}+V_{\bar{1}\bar{i}\bar{s}\bar{1}}^{(1)})c_{s}\Bigg]+(N_1-1)\Bigg[\sum_{r,s=2}^{M_1}V_{\bar{1}\bar{i}\bar{r}\bar{s}}^{(1)}(2c_{rs}+c_{r}c_{s})\Bigg]\nonumber\\
&&+N_2 (N_2-1) \sum_{l=2}^{M_2}V_{111l}^{(2)}e_{\bar{i}l}+\dfrac{1}{2}N_2(N_2-1)\sum_{k,l=2}^{M_2}V_{11kl}^{(2)}(2e_{\bar{i}lk}+e_{\bar{i}k}d_l+e_{\bar{i}l}d_k)\nonumber \\
&&-N_2 V_{\bar{1}1\bar{1}1}^{(12)} c_{\bar{i}}+(N_1-1)N_2\sum_{l=2}^{M_2}V_{\bar{1}1\bar{1}l}^{(12)}e_{\bar{i}l}-N_2 \sum_{l=2}^{M_2} V_{\bar{1}1\bar{1}l}^{(12)} c_{\bar{i}}d_l\nonumber\\
&&+N_2\Bigg[(N_1-1)\sum_{s=2}^{M_1}V_{\bar{1}1\bar{s}1}^{(12)}(2c_{s\bar{i}})-\sum_{s=2}^{M_1}V_{\bar{1}1\bar{s}1}^{(12)}(c_{\bar{i}}c_{s})+V_{\bar{i}1\bar{1}1}^{(12)}\Bigg]\nonumber\\
&&+(N_1-1)N_2\sum_{s=2}^{M_1}\sum_{l=2}^{M_2}V_{\bar{1}1\bar{s}l}^{(12)}(2e_{\bar{i}sl}+c_s e_{\bar{i}l}+2c_{s\bar{i}}d_l)-N_2\sum_{s=2}^{M_1}\sum_{l=2}^{M_2}V_{\bar{1}1\bar{s}l}^{(12)}(c_{\bar{i}}e_{sl}+c_{\bar{i}}c_sd_l)\nonumber \\
&&+N_2\Bigg[\sum_{s=2}^{M_1}V_{\bar{i}1\bar{s}1}^{(12)}c_{s}+\sum_{l=2}^{M_2}V_{\bar{i}1\bar{1}l}^{(12)}d_l+\sum_{s=2}^{M_1}\sum_{l=2}^{M_2}V_{\bar{i}1\bar{s}l}^{(12)}(e_{sl}+c_{s}d_l)\Bigg].
\end{eqnarray}
To calculate Eq.~\ref{general_1}, contributions from the fifth, seventh, and eighth  terms of $\dot{H}_0$ in Eq.~\ref{eqn_4.12}, the second, third, fifth, sixth, seventh, eighth, and ninth terms of $\dot{V}^{(2)}$ in Eq.~\ref{Potential_2}, and from  the fourth, seventh, ninth, eleventh, twelfth, fourteenth, fifteenth, and sixteenth terms of $\dot{V}^{(12)}$ in Eq.~\ref{Potential_12}  are zero due to the creation operators for virtual orbitals in the species-2. Also the fifth, eighth, and ninth terms of $\dot{V}^{(1)}$ in Eq.~\ref{Potential_1} do not have any contribution as they all exhibit two creation operators for virtual orbitals of   species-1. 

The general form of the single excitation in species-2 reads

\begin{eqnarray}\label{general_2}
0&=&-h_{11}^{(2)}d_i+(N_2-1)\sum_{l=2}^{M_2}h_{1l}^{(2)}(2d_{li})-\sum_{l=2}^{M_2}h_{1l}^{(2)} (d_i d_l)+h_{i1}^{(2)}+\sum_{l=2}^{M_2}h_{il}^{(2)}d_{l}+N_1\sum_{s=2}^{M_1}h_{\bar{1}\bar{s}}^{(1)}e_{si}\nonumber \\
&&-(N_2-1)V_{1111}^{(2)}d_i+(N_2-1)(N_2-2)\sum_{l=2}^{M_2}V_{111l}^{(2)}(2d_{li})-2(N_2-1)\sum_{l=2}^{M_2}V_{111l}^{(2)}(d_id_l)+(N_2-1)V_{i111}^{(2)}\nonumber \\
&&+\sum_{k,l=2}^{M_2}V_{11kl}^{(2)}\alpha_{kli}+(N_2-1)\Bigg[\sum_{l=2}^{M_2}(V_{1i1l}^{(2)}+V_{1il1}^{(2)})d_l\Bigg]+(N_2-1)\Bigg[\sum_{k,l=2}^{M_2}V_{1ikl}^{(2)}(2d_{lk}+d_kd_l)\Bigg]\nonumber\\
&&+N_1(N_1-1)\sum_{s=2}^{M_1}V_{\bar{1}\bar{1}\bar{1}\bar{s}}^{(1)}e_{si}+\dfrac{1}{2}N_1(N_1-1)\sum_{r,s=2}^{M_1}V_{\bar{1}\bar{1}\bar{r}\bar{s}}^{(1)}(2e_{sri}+e_{ri}c_s+e_{si}c_r)\nonumber \\
&&-N_1 V_{\bar{1}1\bar{1}1}^{(12)} d_i+N_1(N_2-1)\sum_{s=2}^{M_1}V_{\bar{1}1\bar{s}1}^{(12)}e_{si}-N_1 \sum_{s=2}^{M_1} V_{\bar{1}1\bar{s}1}^{(12)} d_ic_{s}\nonumber\\
&&+N_1\Bigg[(N_2-1)\sum_{l=2}^{M_2}V_{\bar{1}1\bar{1}l}^{(12)}(2d_{li})-\sum_{l=2}^{M_2}V_{\bar{1}1\bar{1}l}^{(12)}(d_id_l)+V_{\bar{1}i\bar{1}1}^{(12)}\Bigg]\nonumber\\
&&+N_1(N_2-1)\sum_{s=2}^{M_1}\sum_{l=2}^{M_2}V_{\bar{1}1\bar{s}l}^{(12)}(2e_{sli}+2c_{s}d_{li}+d_l e_{si})-N_1\sum_{s=2}^{M_1}\sum_{l=2}^{M_2}V_{\bar{1}1\bar{s}l}^{(12)}(d_ie_{sl}+d_ic_s+d_l)\nonumber \\
&&+N_1\Bigg[\sum_{s=2}^{M_1}V_{\bar{1}i\bar{s}1}^{(12)}c_{s}+\sum_{l=2}^{M_2}V_{\bar{1}i\bar{1}l}^{(12)}d_l+\sum_{s=2}^{M_1}\sum_{l=2}^{M_2}V_{\bar{1}i\bar{s}l}^{(12)}(e_{sl}+c_{s}d_l)\Bigg].
\end{eqnarray}
In order to determine Eq.~\ref{general_2},   the first, third, and fourth  terms of $\dot{H}_0$ in Eq.~\ref{eqn_4.12}, the second, third, fifth, sixth, seventh, eighth, and ninth terms of $\dot{V}^{(1)}$ in Eq.~\ref{Potential_1}, and  the fifth, seventh, eighth, tenth, thirteenth, fourteenth, fifteenth, and sixteenth terms of $\dot{V}^{(12)}$ in Eq.~\ref{Potential_12}  give rise to  zero due to the creation operators for virtual orbitals in the species-1. In addition, the fifth, eighth, and ninth terms of  $\dot{V}^{(2)}$ in  Eq.~\ref{Potential_2} do not have any contribution  due to the presence of two creation operators for virtual orbitals of  species-2. Eqs.~\ref{general_1} and ~\ref{general_2} include  coefficients $\bar{\alpha}_{rs\bar{i}}$ and $\alpha_{kli}$ which are presented in the main text, see Eq.~\ref{eqn_4.27}. For calculation of CCSD-M, in Eqs.~\ref{general_1} and ~\ref{general_2}, one should set the coefficients $c_{s\bar{i}r}$, $d_{lik}$, $e_{\bar{i}lk}$, $e_{\bar{i}sl}$, $e_{sri}$, and $e_{sli}$, which involve triple excitations, as zero.

To deduce the unknown coefficients of CCSD-M, it is required to solve the coupled equations resulting from $\langle \phi_0|(a_{1}^\dagger)^2 a_{\bar{i}}a_{\bar{j}}\dot{H}|\phi_0\rangle=0$, $\langle \phi_0|(b_1^\dagger)^2 b_i b_j\dot{H}|\phi_0\rangle=0$, and  $\langle \phi_0|a_{1}^\dagger b_1^\dagger a_{\bar{i}}b_i\dot{H}|\phi_0\rangle=0$, for the double excitations in the coupled-cluster theory. Substituting $\dot{H}$ into the above mentioned coupled equations, one can find  the rather lengthy general form  for the coupled-cluster double excitations.  The explicit form of $\langle \phi_0|(a_{1}^\dagger)^2 a_{\bar{i}}a_{\bar{j}}\dot{H}|\phi_0\rangle=0$ is

\begin{eqnarray}\label{general_3}
0&=&-4h_{\bar{1}\bar{1}}^{(1)}(c_{\bar{i}\bar{j}})-\sum_{s=2}^{M_1}h_{\bar{1}\bar{s}}^{(1)}(2c_{\bar{i}}c_{s\bar{j}}+2c_{\bar{j}}c_{s\bar{i}}+4c_{s}c_{\bar{i}\bar{j}})+\sum_{s=2}^{M_1}\Bigg[2h_{\bar{i}\bar{s}}^{(1)}c_{s\bar{j}}+2h_{\bar{j}\bar{s}}^{(1)}c_{s\bar{i}}\Bigg]\nonumber\\
&&-2(2N_1-3)V_{\bar{1}\bar{1}\bar{1}\bar{1}}^{(1)}c_{\bar{i}\bar{j}}+V_{\bar{1}\bar{1}\bar{1}\bar{1}}^{(1)}c_{\bar{i}}c_{\bar{j}}-\bigg[V_{\bar{i}\bar{1}\bar{1}\bar{1}}^{(1)}c_{\bar{j}}+V_{\bar{j}\bar{1}\bar{1}\bar{1}}^{(1)}c_{\bar{i}}\bigg]\nonumber \\
&&-\sum_{s=2}^{M_1}V_{\bar{1}\bar{1}\bar{1}\bar{s}}^{(1)}\Bigg[4(N_1-2)(c_{\bar{i}}c_{s\bar{j}}+c_{\bar{j}}c_{s\bar{i}}+2c_{s}c_{\bar{i}\bar{j}})+(4c_{s}c_{\bar{i}\bar{j}}-2c_{\bar{i}}c_{\bar{j}}c_{k})\Bigg]\nonumber \\
&&+\sum_{r,s=2}^{M_1}V_{\bar{1}\bar{1}\bar{r}\bar{s}}^{(1)}\bar{\gamma}_{rs\bar{i}\bar{j}}+\sum_{s=2}^{M_1}\bigg[V_{\bar{1}\bar{i}\bar{1}\bar{s}}^{(1)}+V_{\bar{1}\bar{i}\bar{s}\bar{1}}^{(1)}\bigg][2(N_1-2)c_{s\bar{j}}-c_{\bar{j}}c_{s}]\nonumber\\
&&+V_{\bar{i}\bar{j}\bar{1}\bar{1}}^{(1)}+\sum_{s=2}^{M_1}\bigg[V_{\bar{1}\bar{j}\bar{1}\bar{s}}^{(1)}+V_{\bar{1}\bar{j}\bar{s}\bar{1}}^{(1)}\bigg][2(N_1-2)c_{s\bar{i}}-c_{\bar{i}}c_{s}]\nonumber \\
&&+\sum_{r,s=2}^{M_1}V_{\bar{1}\bar{i}\bar{r}\bar{s}}^{(1)}\bigg[2(N_1-2)c_{r\bar{j}}c_{s}+2(N_1-2)c_{s\bar{j}}c_{r}-2c_{sr}c_{\bar{j}}-c_{\bar{j}}c_{r}c_{s}\bigg]\nonumber \\
&&+\sum_{r,s=2}^{M_1}V_{\bar{1}\bar{j}\bar{r}\bar{s}}^{(1)}\bigg[2(N_1-2)c_{r\bar{i}}c_{s}+2(N_1-2)c_{s\bar{i}}c_{r}-2c_{sr}c_{\bar{i}}-c_{\bar{i}}c_{r}c_{s}\bigg]\nonumber \\
&&+\sum_{s=2}^{M_1}V_{\bar{i}\bar{j}\bar{s}\bar{1}}^{(1)}c_{s}+\sum_{s=2}^{M_1}V_{\bar{j}\bar{i}\bar{s}\bar{1}}^{(1)}c_{s}+\sum_{r,s=2}^{M_1}V_{\bar{i}\bar{j}\bar{r}\bar{s}}^{(1)}[2c_{sr}+c_{r}c_{s}]\nonumber \\
&&+\dfrac{1}{2}N_2(N_2-1)\sum_{k,l=2}^{M_2}V_{11kl}^{(2)}[e_{\bar{i}k}e_{\bar{j}l}+e_{\bar{j}k}e_{\bar{i}l}]-4N_2V_{\bar{1}1\bar{1}1}^{(12)}(c_{\bar{i}\bar{j}})\nonumber\\
&&-N_2\sum_{s=2}^{M_1}V_{\bar{1}1\bar{s}1}^{(12)}(4c_{\bar{i}\bar{j}}c_{s}+2c_{s\bar{j}}c_{\bar{i}}+2c_{s\bar{i}}c_{\bar{j}})\nonumber \\
&&-N_2\sum_{l=2}^{M_2}V_{\bar{1}1\bar{1}l}^{(12)}(4c_{\bar{i}\bar{j}}d_l+c_{\bar{i}}e_{\bar{j}l}+c_{\bar{j}}e_{\bar{i}l})+\sum_{s=2}^{M_1}\sum_{l=2}^{M_2}V_{\bar{1}1\bar{s}l}^{(12)}\bar{\delta}_{sl\bar{i}\bar{j}}\nonumber \\
&&+2N_2\sum_{s=2}^{M_1}\Bigg[V_{\bar{i}1\bar{s}1}^{(12)}c_{s\bar{j}}+V_{\bar{j}1\bar{s}1}^{(12)}c_{s\bar{i}}\Bigg]+N_2\sum_{l=2}^{M_2}\Bigg[V_{\bar{i}1\bar{1}l}^{(12)}e_{\bar{j}l}+V_{\bar{j}1\bar{1}l}^{(12)}e_{\bar{i}l}\Bigg]\nonumber \\
&&+N_2\sum_{s=2}^{M_1}\sum_{l=2}^{M_2}\Bigg[V_{\bar{i}1\bar{s}l}^{(12)}(2c_{s\bar{j}}d_l+c_se_{\bar{j}l})+V_{\bar{j}1\bar{s}l}^{(12)}(2c_{s\bar{i}}d_l+c_se_{\bar{i}l})\Bigg].
\end{eqnarray}
Here,  Eq.~\ref{general_3} involves the   first, second, and fourth terms of Eq.~\ref{eqn_4.12},  all terms of Eq.~\ref{Potential_1}, only the fourth term of Eq.~\ref{Potential_2}, and the first, second, third, sixth, eighth, tenth, and thirteenth terms of Eq.~\ref{Potential_12}. The remaining terms do not contribute. 

Now, writing down the transformed Hamiltonian in $\langle \phi_0|(b_1^\dagger)^2 b_i b_j\dot{H}|\phi_0\rangle=0$, one can find  for species-2

\begin{eqnarray}\label{general_4}
0&=&-4h_{11}^{(2)}d_{ij}-\sum_{l=2}^{M_2}h_{1l}^{(2)}(2d_id_{lj}+2d_jd_{li}+4d_ld_{ij})+\sum_{l=2}^{M_2}\Bigg[2h_{il}^{(2)}d_{lj}+2h_{jl}^{(2)}d_{li}\Bigg]\nonumber\\
&&-2(2N_2-3)V_{1111}^{(2)}d_{ij}+V_{1111}^{(2)}d_id_j-\bigg[V_{i111}^{(2)}d_j+V_{j111}^{(2)}d_i\bigg] \nonumber \\
&&-\sum_{l=2}^{M_2}V_{111l}^{(2)}\Bigg[4(N_2-2)(d_id_{lj}+d_jd_{li}+2d_ld_{ij})+(4d_ld_{ij}-2d_id_jd_l)\Bigg]\nonumber \\
&&+\sum_{k,l=2}^{M_2}V_{11kl}^{(2)}\gamma_{klij}+\sum_{l=2}^{M_2}\bigg[V_{1i1l}^{(2)}+V_{1il1}^{(2)}\bigg][2(N_2-2)d_{lj}-d_jd_l]\nonumber\\
&&+V_{ij11}^{(2)}+\sum_{l=2}^{M_2}\bigg[V_{1j1l}^{(2)}+V_{1jl1}^{(2)}\bigg][2(N_2-2)d_{li}-d_id_l]\nonumber \\
&&+\sum_{k,l=2}^{M_2}V_{1ikl}^{(2)}\bigg[2(N_2-2)d_{kj}d_l+2(N_2-2)d_{lj}d_k-2d_{lk}d_j-d_jd_kd_l\bigg]\nonumber \\
&&+\sum_{k,l=2}^{M_2}V_{1jkl}^{(2)}\bigg[2(N_2-2)d_{ki}d_l+2(N_2-2)d_{li}d_k-2d_{lk}d_i-d_id_kd_l\bigg]\nonumber \\
&&+\sum_{l=2}^{M_2}V_{ijl1}^{(2)}d_l+\sum_{l=2}^{M_2}V_{jil1}^{(2)}d_l+\sum_{k,l=2}^{M_2}V_{ijkl}^{(2)}[2d_{lk}+d_kd_l]\nonumber \\
&&+\dfrac{1}{2}N_1(N_1-1)\sum_{r,s=2}^{M_1}V_{\bar{1}\bar{1}\bar{r}\bar{s}}^{(1)}[e_{ri}e_{sj}+e_{rj}e_{si}]
-4N_1V_{\bar{1}1\bar{1}1}^{(12)}d_{ij}\nonumber\\
&&-N_1\sum_{l=2}^{M_2}V_{\bar{1}1\bar{1}l}^{(12)}(4d_{ij}d_l+2d_{lj}d_i+2d_{li}d_j)\nonumber \\
&&-N_1\sum_{s=2}^{M_1}V_{\bar{1}1\bar{s}1}^{(12)}(4d_{ij}c_{s}+e_{sj}d_i+e_{si}d_j)+\sum_{s=2}^{M_1}\sum_{l=2}^{M_2}V_{\bar{1}1\bar{s}l}^{(12)}\delta_{slij}\nonumber \\
&&+2N_1\sum_{l=2}^{M_2}\Bigg[V_{\bar{1}i\bar{1}l}^{(12)}d_{lj}+V_{\bar{1}j\bar{1}l}^{(12)}d_{li}\Bigg]+N_1\sum_{s=2}^{M_1}\Bigg[V_{\bar{1}i\bar{s}1}^{(12)}e_{sj}+V_{\bar{1}j\bar{s}1}^{(12)}e_{si}\Bigg]\nonumber \\
&&+N_1\sum_{s=2}^{M_1}\sum_{l=2}^{M_2}\Bigg[V_{\bar{1}i\bar{s}l}^{(12)}(2d_{lj}c_{s}+e_{sj}d_l)+V_{\bar{1}j\bar{s}l}^{(12)}(2d_{li}c_{s}+e_{si}d_l)\Bigg].
\end{eqnarray}
For calculating Eq.~\ref{general_4}, the fifth, sixth, and eighth terms of Eq.~\ref{eqn_4.12},   only  the fourth term of Eq.~\ref{Potential_2}, all terms of Eq.~\ref{Potential_1}, and  the first, second, third, sixth, ninth, eleventh, and twelfth terms of Eq.~\ref{Potential_12} contribute.

Now,  the general form of $\langle \phi_0|a_{1}^\dagger b_1^\dagger a_{\bar{i}}b_i\dot{H}|\phi_0\rangle=0$ with  {arbitrary} orbitals becomes

\begin{eqnarray}\label{general_5}
0&=&-[h_{\bar{1}\bar{1}}^{(1)}+h_{11}^{(2)}]e_{\bar{i}i}-\sum_{s=2}^{M_1}h_{\bar{1}\bar{s}}^{(1)}(e_{\bar{i}i}c_s+e_{si}c_{\bar{i}})-\sum_{l=2}^{M_2}h_{1l}^{(2)}(e_{\bar{i}i}d_l+e_{\bar{i}l}d_i)+\sum_{s=2}^{M_1}h_{\bar{i}\bar{s}}^{(1)}e_{si}+\sum_{l=2}^{M_2}h_{il}^{(2)}e_{\bar{i}l}\nonumber \\
&&-[(N_1-1)V_{\bar{1}\bar{1}\bar{1}\bar{1}}^{(1)}+(N_2-1)V_{1111}^{(2)}]e_{\bar{i}i}+\sum_{r,s=2}^{M_1}V_{\bar{1}\bar{1}\bar{r}\bar{s}}^{(1)}\bar{\xi}_{rs\bar{i}i}+\sum_{k,l=2}^{M_2}V_{11kl}^{(2)}\xi_{kl\bar{i}i}\nonumber\\
&&-(N_1-1)\sum_{s=2}^{M_1}V_{\bar{1}\bar{1}\bar{1}\bar{s}}^{(1)}(2c_se{\bar{i}i}+2c_{\bar{i}}e_{si})-(N_2-1)\sum_{l=2}^{M_2}V_{111l}^{(2)}(2d_le_{\bar{i}i}+2d_ie_{\bar{i}l})\nonumber \\
&&+(N_1-1)\sum_{s=2}^{M_1}(V_{\bar{1}\bar{i}\bar{1}\bar{s}}^{(1)}+V_{\bar{1}\bar{i}\bar{s}\bar{1}}^{(1)})e_{si}+(N_2-1)\sum_{l=2}^{M_2}(V_{1i1l}^{(2)}+V_{1il1}^{(2)})e_{\bar{i}l}\nonumber \\
&&+(N_1-1)\sum_{r,s=2}^{M_1}V_{\bar{1}\bar{i}\bar{r}\bar{s}}^{(1)}(c_se_{ri}+c_re_{si})+(N_2-1)\sum_{k,l=2}^{M_2}V_{1ikl}^{(2)}(d_le_{\bar{i}k}+d_ke_{\bar{i}l})\nonumber \\
&&-V_{\bar{1}1\bar{1}1}^{(12)}[(N_1+N_2-1)e_{\bar{i}i}-c_{\bar{i}}d_i]-\sum_{s=2}^{M_1}V_{\bar{1}1\bar{s}1}^{(12)}\bar{\chi}_{s\bar{i}i}-\sum_{l=2}^{M_2}V_{\bar{1}1\bar{1}l}^{(12)}\chi_{l\bar{i}i}\nonumber\\
&&-V_{\bar{1}i\bar{1}1}^{(12)}c_{\bar{i}}-V_{\bar{i}1\bar{1}1}^{(12)} d_i+\sum_{s=2}^{M_1}\sum_{l=2}^{M_2}V_{\bar{1}1\bar{s}l}^{(12)}\zeta_{sl\bar{i}i}+\sum_{s=2}^{M_1}\sum_{l=2}^{M_2}V_{\bar{i}i\bar{s}l}^{(12)}(e_{sl}+c_sd_l)\nonumber \\
&&+V_{\bar{i}i\bar{1}1}^{(12)}+\sum_{s=2}^{M_1}V_{\bar{i}1\bar{s}1}^{(12)}[(N_2-1)e_{si}-c_sd_i]+\sum_{l=2}^{M_2}V_{\bar{1}i\bar{1}l}^{(12)}[(N_1-1)e_{\bar{i}l}-c_{\bar{i}}d_l]\nonumber \\
&&+\sum_{s=2}^{M_1}V_{\bar{1}i\bar{s}1}^{(12)}[2(N_1-1)c_{s\bar{i}}-c_{\bar{i}}c_s]+\sum_{l=2}^{M_2}V_{\bar{i}1\bar{1}l}^{(12)}[2(N_2-1)d_{li}-d_{i}d_l]\nonumber \\
&&+\sum_{s=2}^{M_1}\sum_{l=2}^{M_2}V_{\bar{1}i\bar{s}l}^{(12)}[(N_1-1)(2c_{si}d_l+c_se_{\bar{i}l})-(c_{\bar{i}}e_{sl}+c_{\bar{i}}c_sd_l)]+\sum_{s=2}^{M_1}V_{\bar{i}i\bar{s}1}^{(12)}c_s\nonumber \\
&&+\sum_{s=2}^{M_1}\sum_{l=2}^{M_2}V_{\bar{i}1\bar{s}l}^{(12)}[(N_2-1)(2c_sd_{li}+d_le_{si})-(d_ie_{sl}+c_sd_id_l)]+\sum_{l=2}^{M_2}V_{\bar{i}i\bar{1}l}^{(12)}d_l.
\end{eqnarray}
In order to deduce  Eq.~\ref{general_5},  the first, second, fourth, fifth, sixth, and eighth terms of Eq.~\ref{eqn_4.12},  the first, second, fourth, sixth, and  seventh   terms of Eq.~\ref{Potential_1} and of  Eq.~\ref{Potential_2}, and  all terms of Eq.~\ref{Potential_12} contribute. The terms $\bar{\gamma}_{rs\bar{i}\bar{j}}$ and $\bar{\delta}_{sl\bar{i}\bar{j}}$ present in Eq.~\ref{general_3}, $\gamma_{klij}$ and $\delta_{slij}$ in Eq.~\ref{general_4}, and $\bar{\xi}_{rs\bar{i}i}$, $\xi_{kl\bar{i}i}$, $\bar{\chi}_{s\bar{i}i}$, $\chi_{l\bar{i}i}$, and $\zeta_{sl\bar{i}i}$ in Eq.~\ref{general_5} are displayed in the main text. Eqs.~\ref{general_1} to ~\ref{general_5} are determined for the arbitrary sets of orthonormal orbitals. Using the relations  extracting from the Fock-like operators, see Eq.~\ref{eqn_A.1}, and Eqs.~\ref{general_1} to ~\ref{general_5}, one can transform the one-body operators to two-body operators and find the working equations of the coupled-cluster theory for bosonic mixtures.

\subsection*{E. Finding the coefficients $c_{22}$, $d_{22}$, and $e_{22}$}\label{Appendix E}
\begin{figure}[!h]
{\includegraphics[ scale=.30]{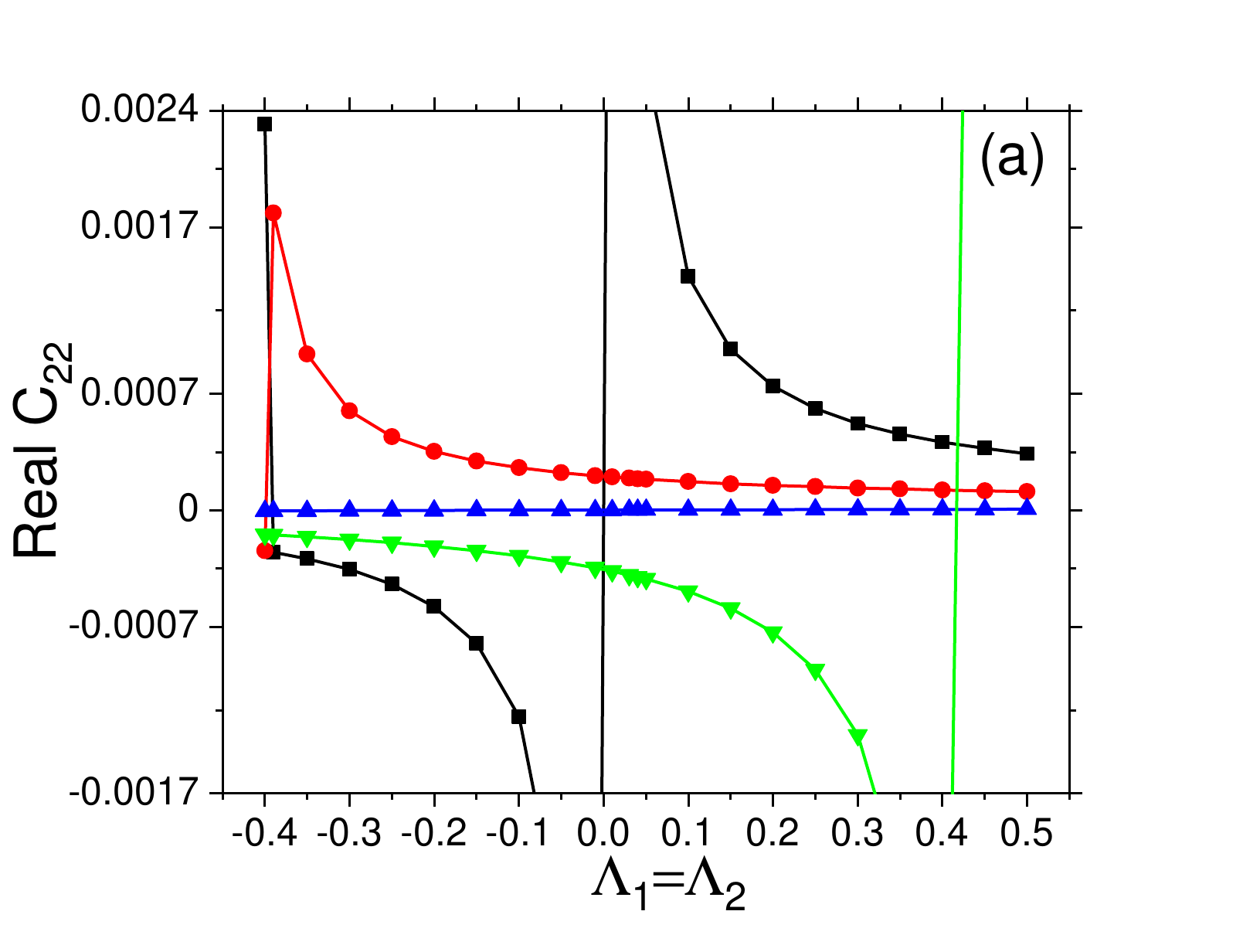}}
{\includegraphics[ scale=.30]{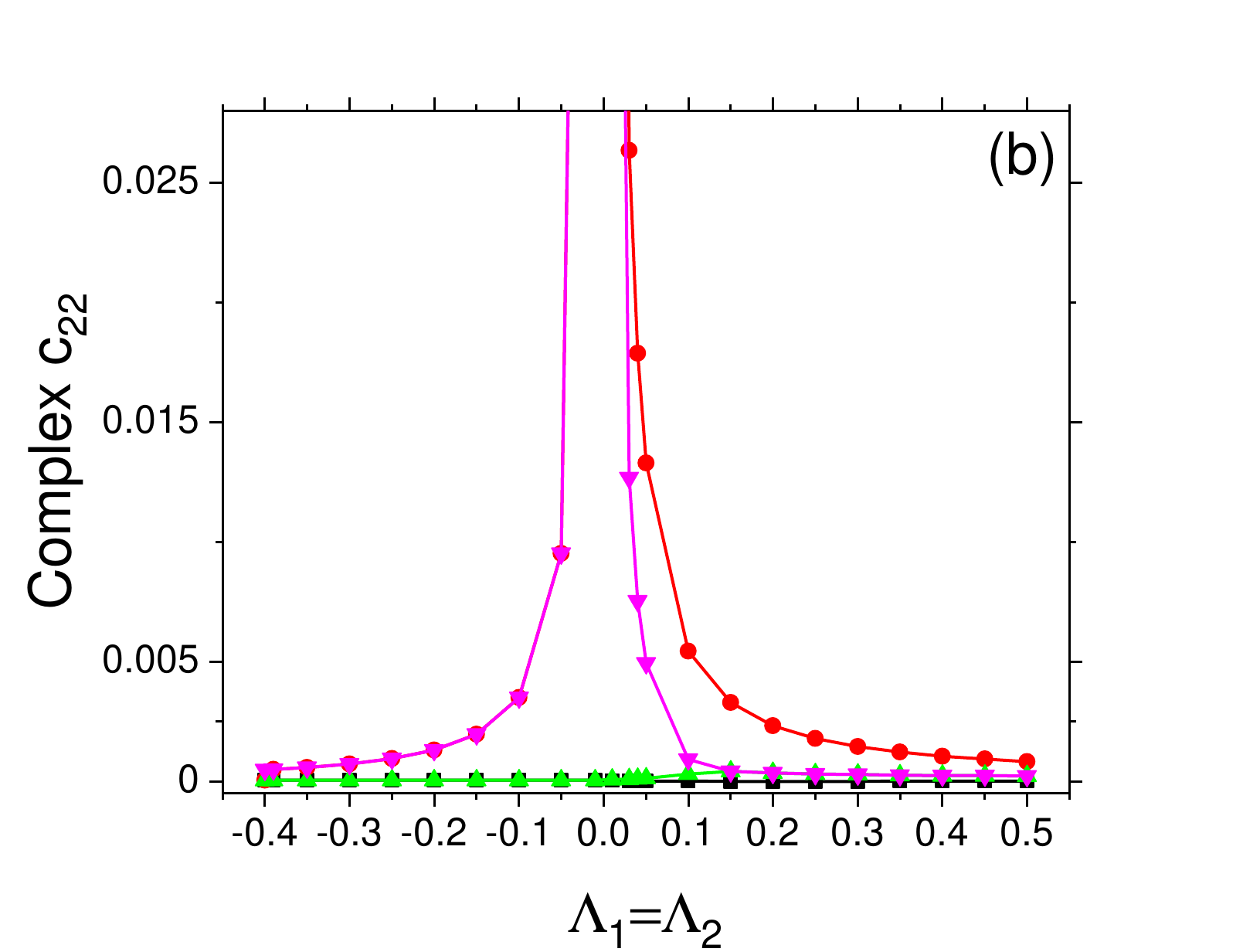}}\\
{\includegraphics[ scale=.30]{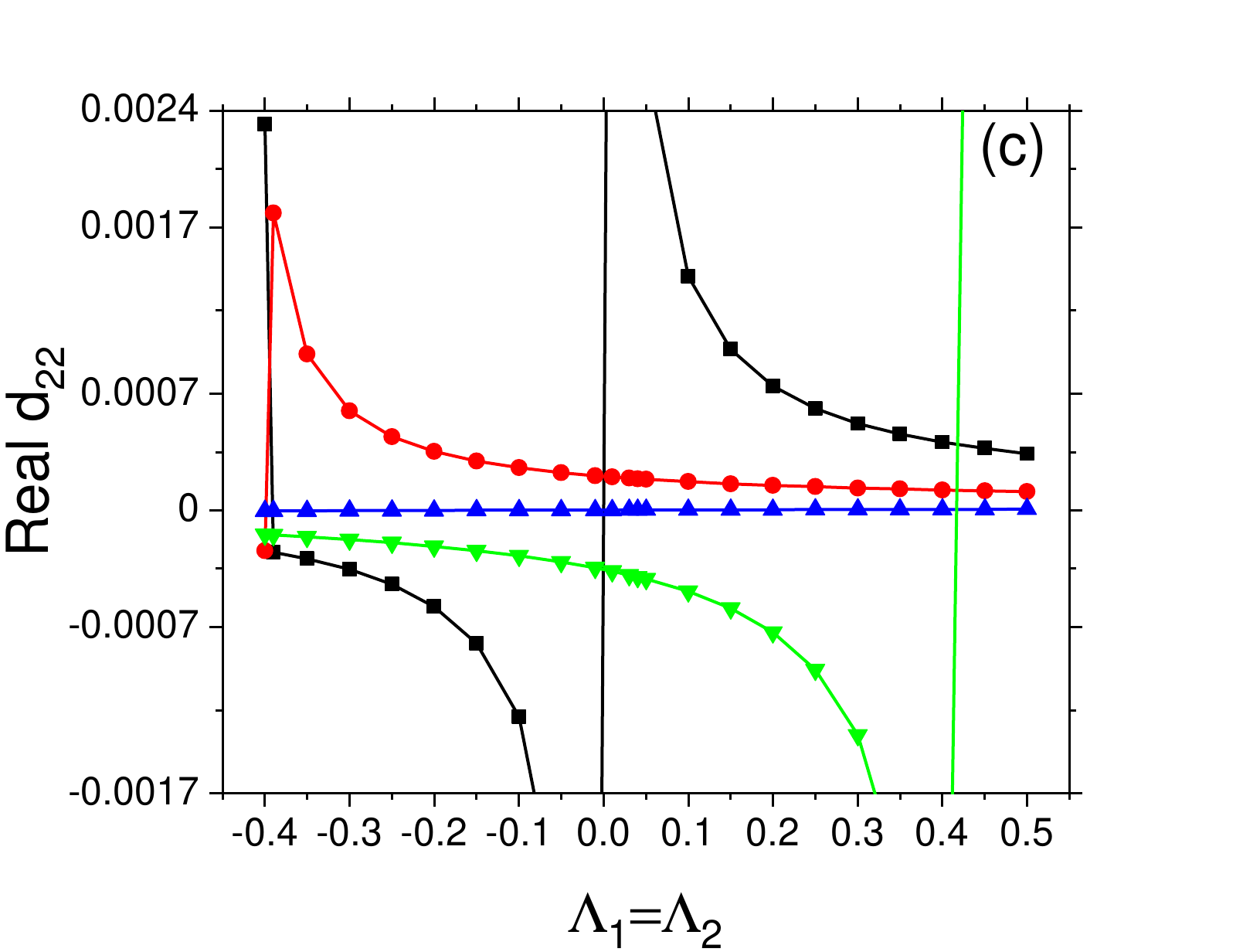}}
{\includegraphics[ scale=.30]{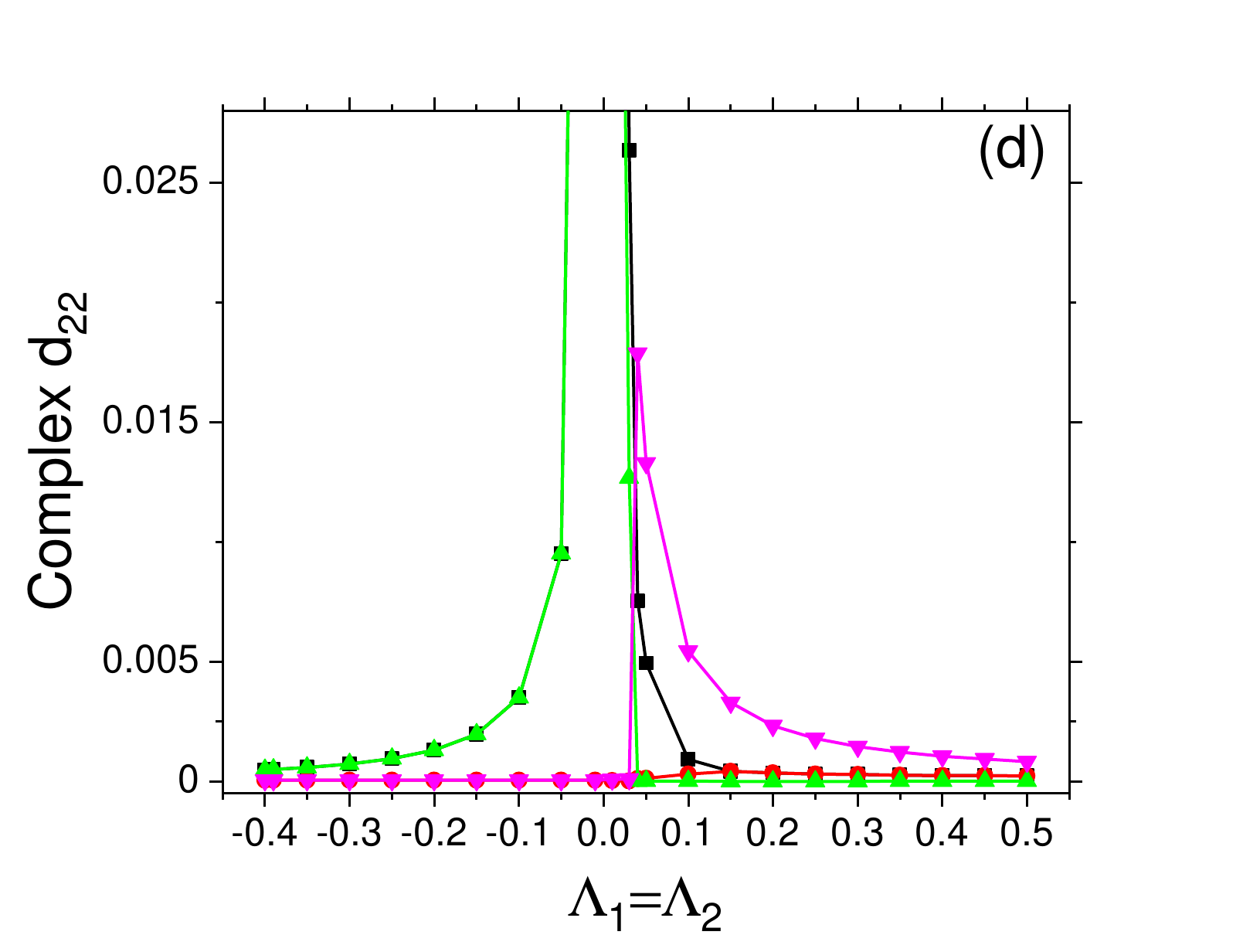}}\\
{\includegraphics[ scale=.30]{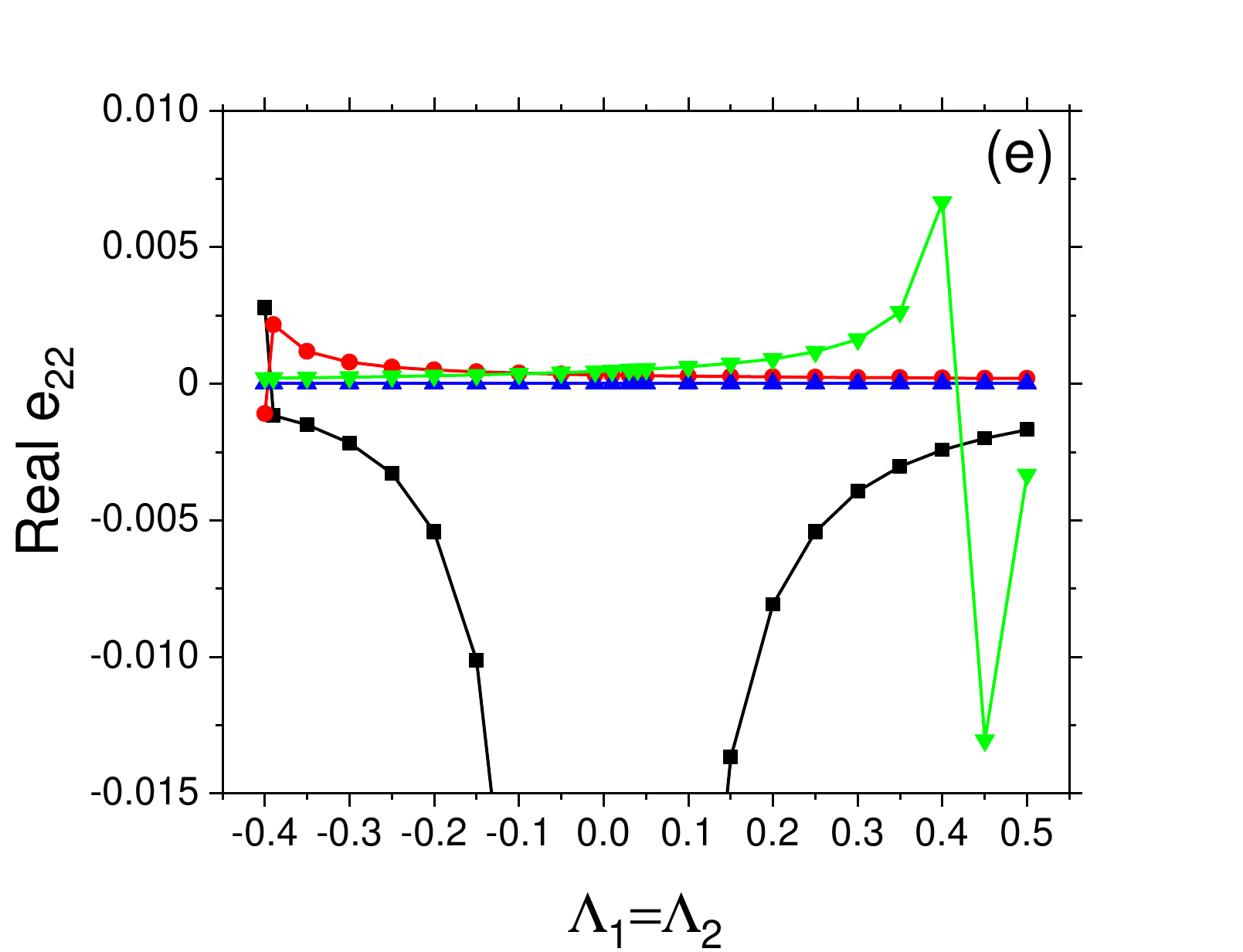}}
{\includegraphics[ scale=.30]{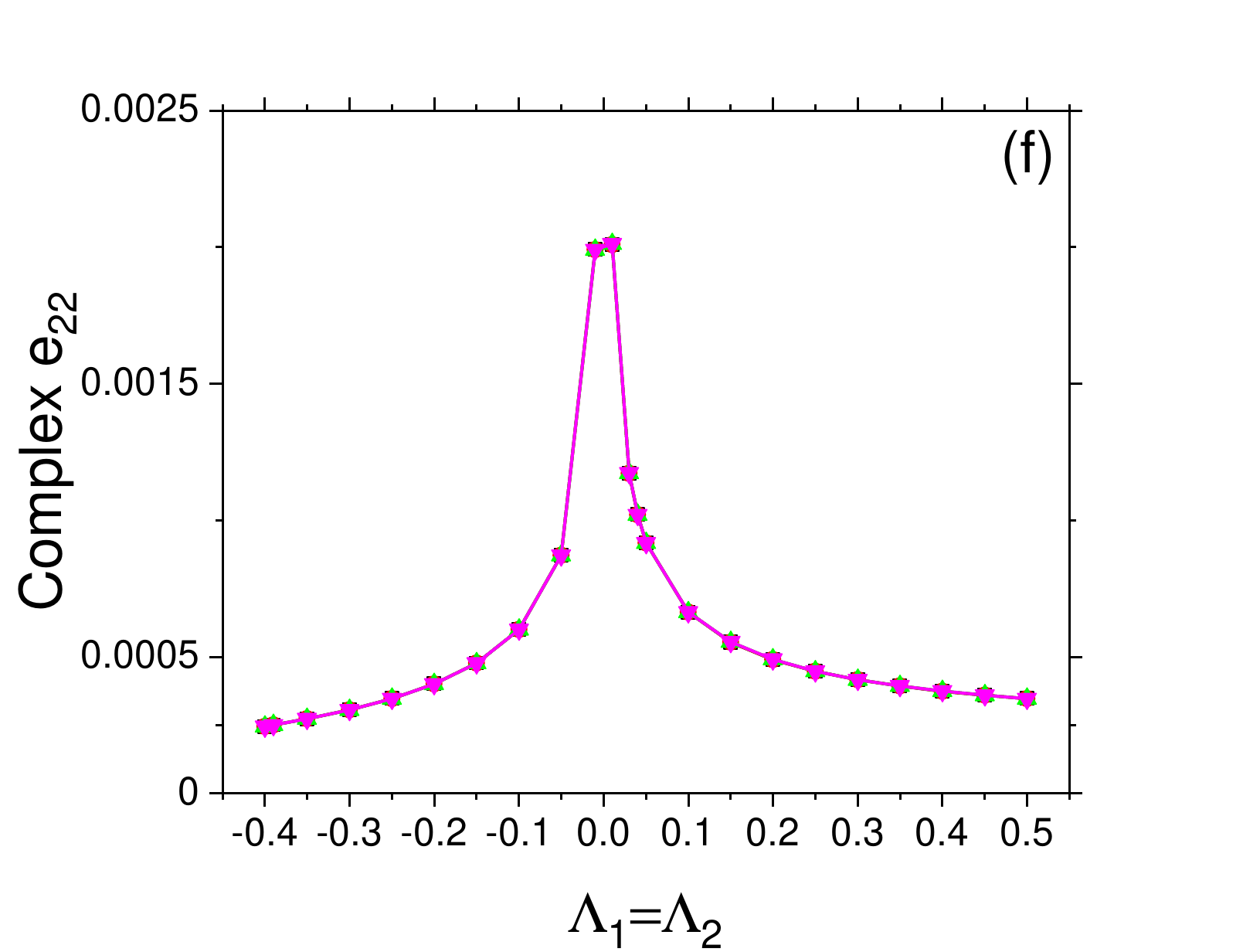}}\\
\caption{Variation of the coefficients, $c_{22}$,  $d_{22}$, and $e_{22}$ as a function of the intra-species interaction parameters. Here,  the numbers of bosons are  $N_1=N_2=10000$ and the inter-species interaction parameters  $\Lambda_{12}=\Lambda_{21}=0.5$. Eight solutions of  coefficients $c_{22}$,  $d_{22}$, and $e_{22}$ are shown. Among the eight solutions, four real and four complex-valued solutions are presented in the  left and right columns, respectively. Absolute values are plotted in panels (b), (d), and (f).  The blue color line  in panels (a), (c), and (e) corresponds to  the true values of the  coefficients   by which the coupled-cluster energy is calculated. The rest are discarded for lacking various properties, see the text for further discussion.   All the quantities are dimensionless. }
\label{FigA1}
\end{figure}

\begin{figure}[!h]
{\includegraphics[ scale=.30]{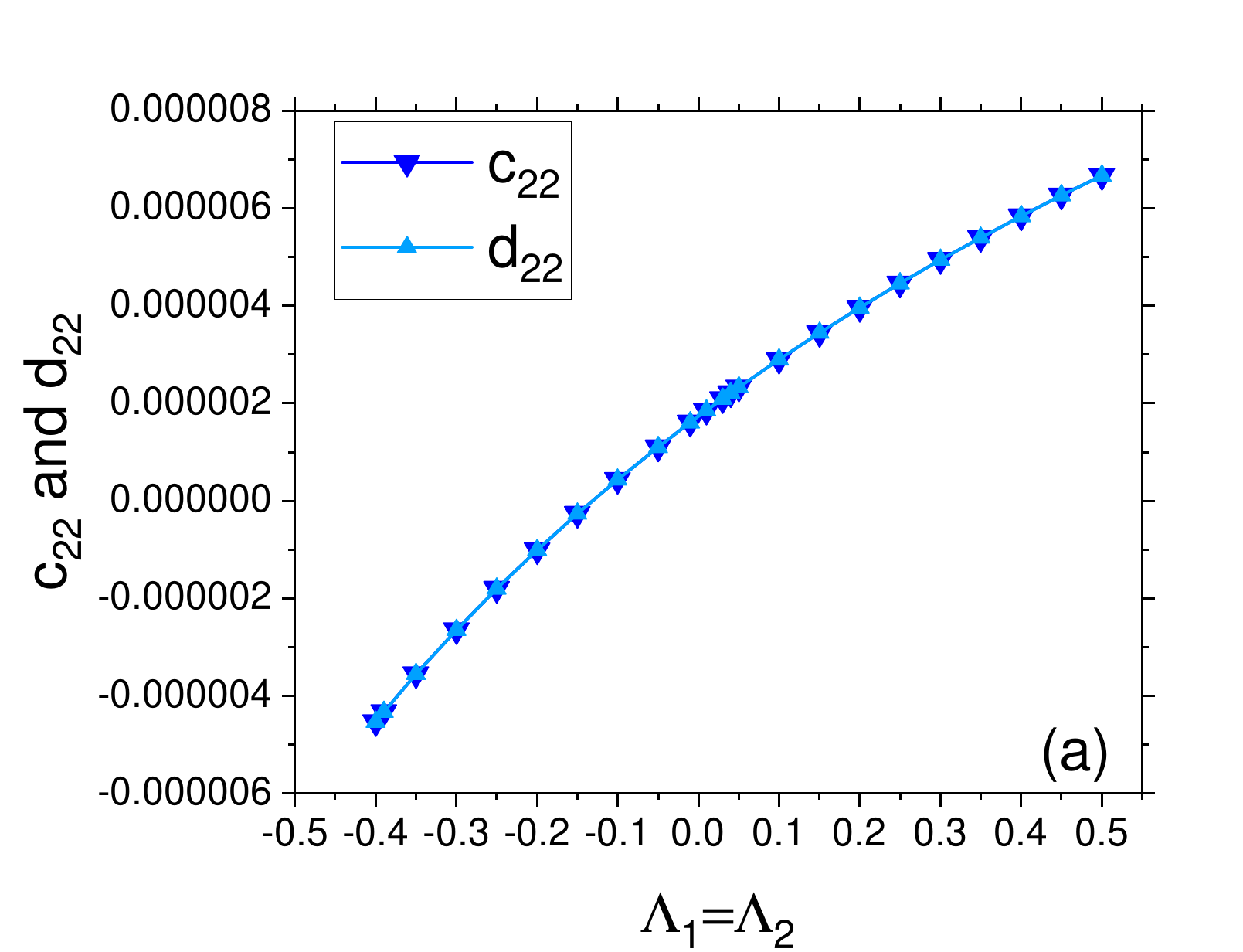}}
{\includegraphics[ scale=.30]{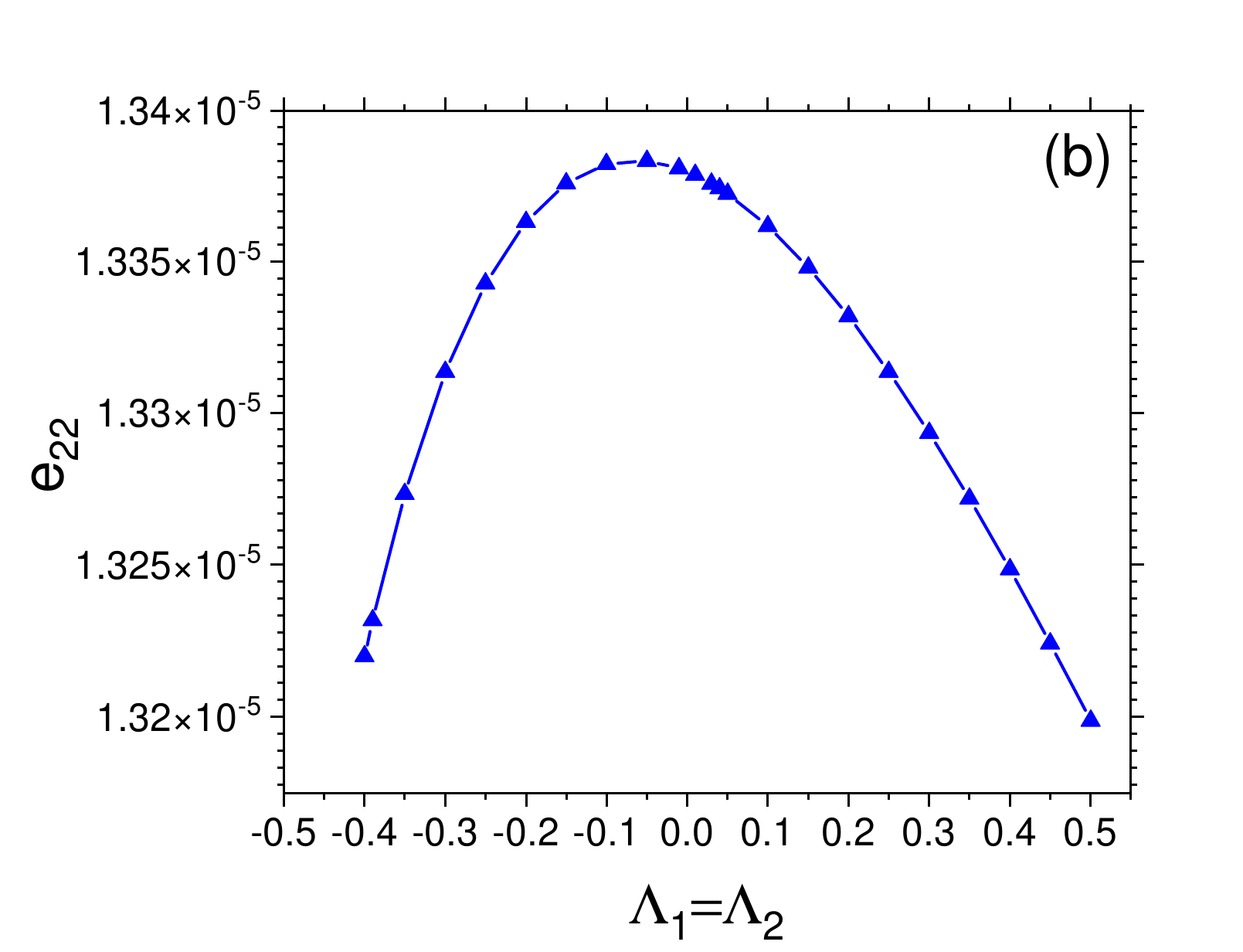}}\\
\caption{ Variation of the true values of the coefficients  $c_{22}$, $d_{22}$ and $e_{22}$ as a function of the intra-species interaction parameters. The numbers of bosons are  $N_1=N_2=10000$ and  the inter-species interaction parameters $\Lambda_{12}=\Lambda_{21}=0.5$.     All the quantities are dimensionless.  }
\label{FigA2}
\end{figure}

Here we discuss the details of calculating the coupled-cluster coefficients,  $c_{22}$, $d_{22}$, and $e_{22}$, which are required to determine the energy of the bosonic mixture for the illustrative examples employing  the harmonic interaction model. This section gives us the knowledge of how we deduce the unknown coefficients. To display the coefficients, we choose a particular set, among all the results presented in the main text,  where $N_1=N_2=10000$ and $\Lambda_{12}=\Lambda_{21}=0.5$, see Fig~\ref{Fig1} (e) and (f) of the main text. The coupled equations ~\ref{eqn_5.1} to ~\ref{eqn_5.3} derived in the main text give us eight families of solutions for the coefficients  $c_{22}$, $d_{22}$, and $e_{22}$. Among them, four families are real and the  other four are  complex-valued. The real and complex-valued solutions of $c_{22}$, $d_{22}$, and $e_{22}$ are presented in Fig.~\ref{FigA1}.  For the complex-valued coefficients, we plot the absolute values. It is obvious that the complex-valued families of the coefficients, right column of  Fig.~\ref{FigA1}, yield a complex-valued coupled-cluster energy and thereby they are discarded.  Among the four  real-valued families of the coefficients, three coefficients are fluctuating as a function of    the intra-species interaction parameters. Therefore, they are also discarded.  All in all,  the blue curve with triangles in Fig.~\ref{FigA1} (a), (c), and (e) is   the true solution for  the coefficients.  We take only the true solutions of $c_{22}$, $d_{22}$ and $e_{22}$ and present them in Fig.~\ref{FigA2}. It can be seen from  Fig.~\ref{FigA2} (a)  that $c_{22}$ and $d_{22}$ show the monotonous nature when moving from repulsive to attractive intra-species interactions and cross zero value at about $\Lambda_{1}=\Lambda_{2}=-0.05$. Since the two-species are interacting with each other having  $\Lambda_{12}=\Lambda_{21}=0.5$, $c_{22}$ and $d_{22}$  cross the zero value at a finite value of intra-species interaction parameter.   At the same value of  $\Lambda_{1}$ and $\Lambda_{2}$, i.e., -0.05, $e_{22}$ reaches   its maximal value. Therefore,  it is a very careful process to determine the true solution of the unknown coefficients.

\begin{figure}[!h]
{\includegraphics[ scale=.30]{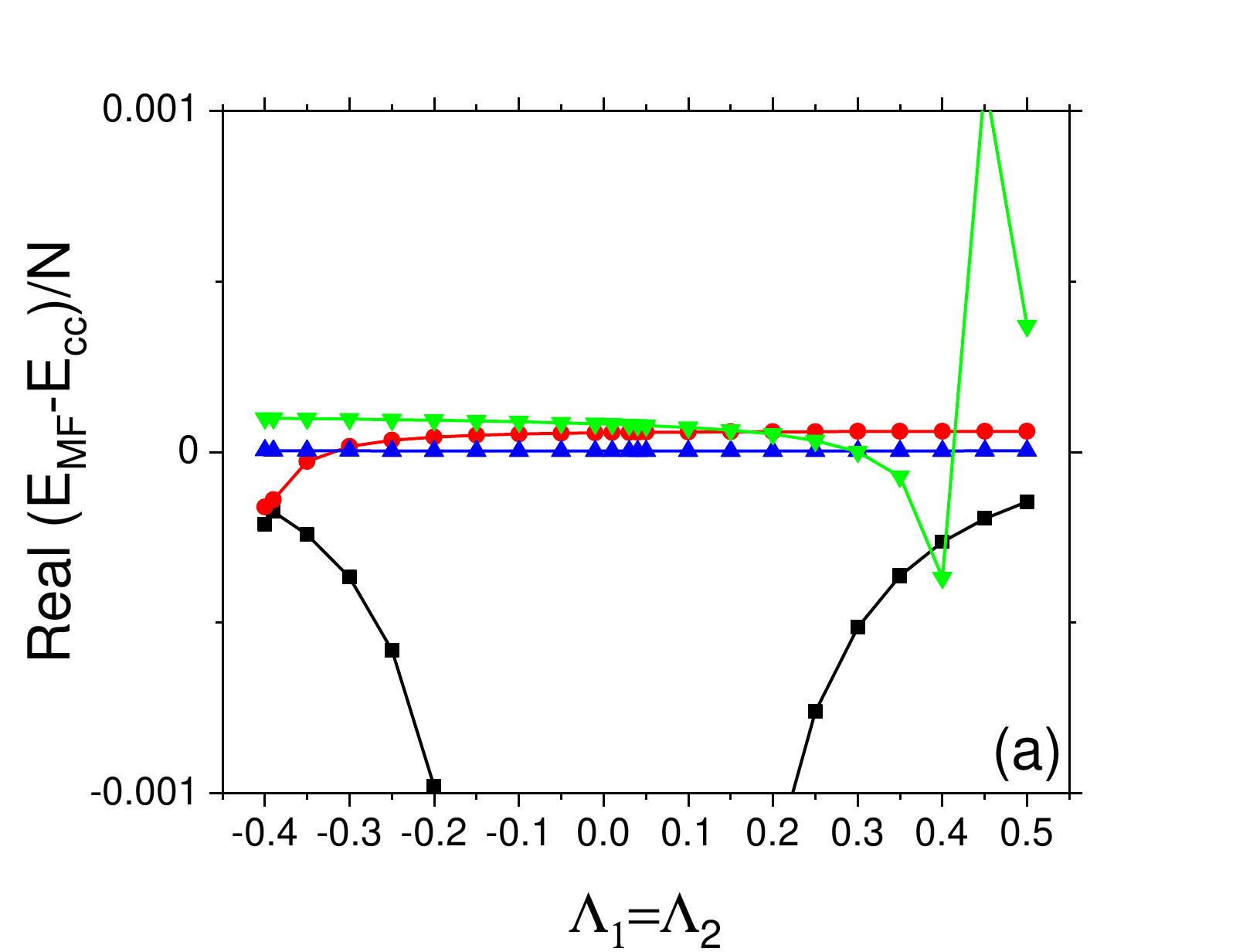}}
{\includegraphics[ scale=.30]{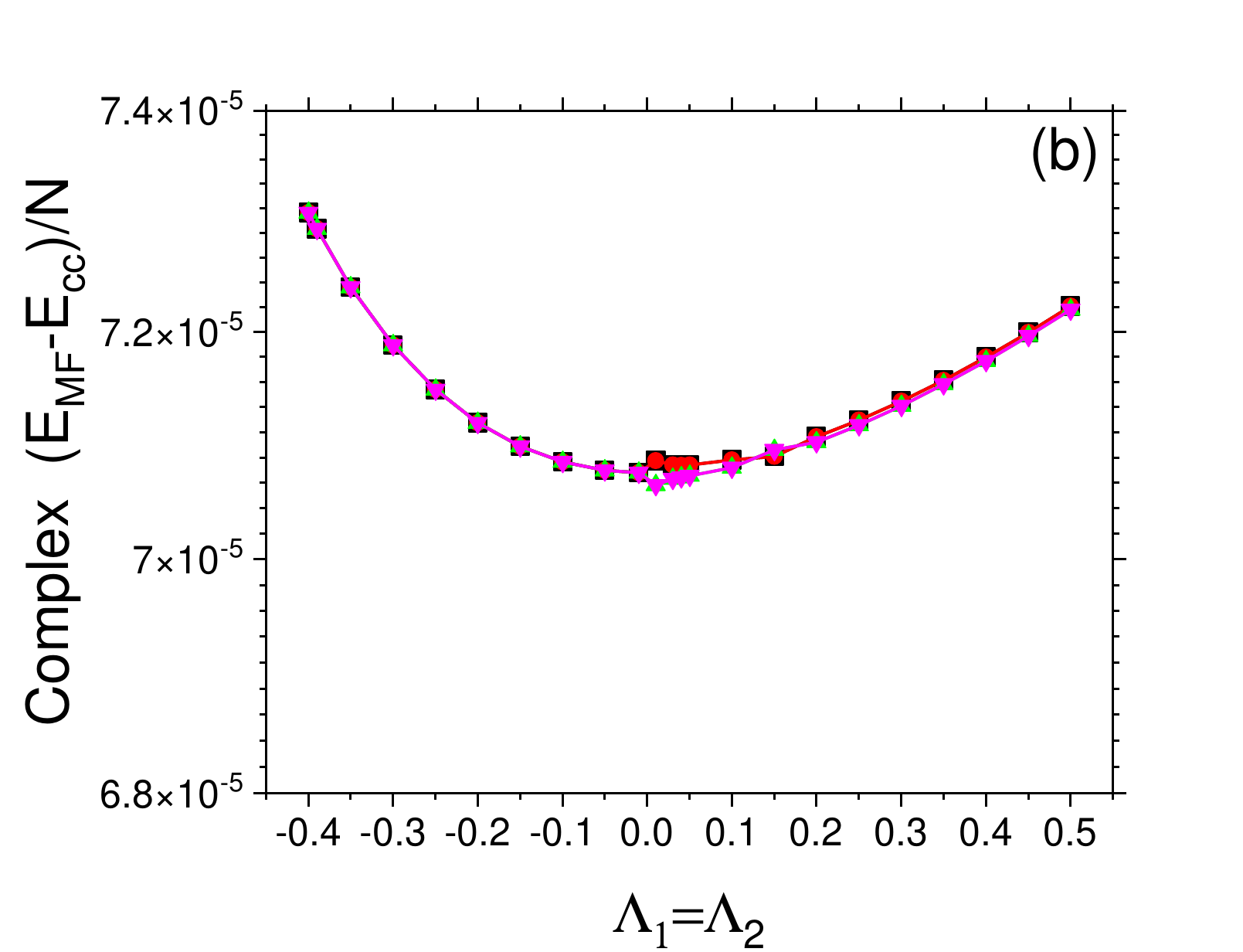}}\\
\caption{Variation of $(E_{\text{MF}}-E_{\text{cc}})/N$ as a function of the intra-species interaction parameters. The numbers of bosons are  $N_1=N_2=10000$ and  the inter-species interaction parameters $\Lambda_{12}=\Lambda_{21}=0.5$. The eight solutions of   the energy  are shown. The corresponding coefficients are presented in Fig.~\ref{FigA1}. Among the eight solutions, four real and four complex-valued  solutions are presented in  panels (a) and (b), respectively. See the text for further details of the determination of the true solutions [marked by blue triangles in panel (a)].    All the quantities are dimensionless.  }
\label{FigA3}
\end{figure}

Figs.~\ref{FigA3} (a) and (b) display   $(E_{\text{MF}}-E_{\text{cc}})/N$ for all the real and complex families, respectively, of the coefficients, $c_{22}$, $d_{22}$ and $e_{22}$.  As discussed before the energies are  complex in Fig.~\ref{FigA3} (b) and they are discarded.  Now, $(E_{\text{MF}}-E_{\text{cc}})/N$ in Fig.~\ref{FigA2} (a) presents  that, among the  four families of real energies, three families have more than one value of energy for which $(E_{\text{MF}}-E_{\text{cc}})/N$ is negative and therefore they could not be a solution for the coupled-cluster theory. Moreover, only one solution (blue color line) always gives a positive value for  $(E_{\text{MF}}-E_{\text{cc}})/N$ and  is closest to the zero line throughout the range of the intra-species interaction parameters.  Therefore, the actual coefficients will have the following properties: (i) the coefficients $c_{22}$ and $d_{22}$  have the monotonous nature as a function of intra-species interaction parameter and (ii) the coupled-cluster energy with those coefficients are always  smaller and closest  to the corresponding  mean-field energy.

\end{document}